\documentclass[fleqn,usenatbib]{mnras}

\usepackage[T1]{fontenc}

\DeclareRobustCommand{\VAN}[3]{#2}
\let\VANthebibliography\thebibliography
\def\thebibliography{\DeclareRobustCommand{\VAN}[3]{##3}\VANthebibliography}

\usepackage{graphicx}	
\usepackage{amsmath}	
\usepackage{makecell}
\usepackage{tabularx}
\usepackage{amssymb}	
\usepackage{orcidlink}
\usepackage{newtxtext,newtxmath}

\newcommand{\bs}{\boldsymbol}

\title[Secular evolution in the Galactic disc]{Orbital migration and heating history of the Galactic disc: a transition between the bimodal discs}

\author[Zhāng et al.]{HanYuan Zhāng,$^{1}$\thanks{hz420@cam.ac.uk (HZ)}\orcidlink{0009-0005-6898-0927}
Vasily Belokurov,$^{1}$\orcidlink{0000-0002-0038-9584}
Jason L. Sanders,$^{2}$\orcidlink{0000-0003-4593-6788}
N. Wyn Evans,$^{1}$\orcidlink{0000-0002-5981-7360}
David Chemaly,$^{1}$\orcidlink{0009-0001-4503-3071}
\newauthor
Daisuke Kawata,$^{3}$\orcidlink{0000-0001-8993-101X}
Natsuki Funakoshi,$^{3}$\orcidlink{0009-0001-0422-5093}
Neige Frankel,$^{4,5}$\orcidlink{0000-0002-6411-8695}
Sarah G. Kane,$^{1}$\orcidlink{0000-0001-8411-1012}
and Sergey E. Koposov$^{6,1}$\orcidlink{0000-0003-2644-135X}
\\
$^{1}$Institute of Astronomy, University of Cambridge, Madingley Road, Cambridge CB3 0HA, UK\\
$^{2}$Department of Physics and Astronomy, University College London, London WC1E 6BT, UK\\
$^{3}$Space and Climate Physics, Mullard Space Science Laboratory, University College London Holmbury St. Mary, Dorking, Surrey, RH5 6NT, UK\\
$^{4}$Canadian Institute for Theoretical Astrophysics, University of Toronto, 60 St. George Street, Toronto, ON M5S 3H8, Canada\\
$^{5}$David A. Dunlap Department of Astronomy and Astrophysics, University of Toronto, 50 St. George Street, Toronto, ON M5S 3H4, Canada\\
$^{6}$Institute for Astronomy, University of Edinburgh, Royal
Observatory, Blackford Hill, Edinburgh EH9 3HJ, UK
}

\date{Accepted XXX. Received YYY; in original form ZZZ}

\pubyear{2025}

\begin{document}
\label{firstpage}
\pagerange{\pageref{firstpage}--\pageref{lastpage}}
\maketitle

\begin{abstract}
Stellar orbits in the Galactic disc evolve from their birth to the current shape through both radial migration and dynamical heating. The history of their secular evolution is imprinted in the current kinematics and age-metallicity distribution. We construct a chrono-chemo-dynamical model of the disc, incorporating inside-out growth, metallicity evolution, radial migration, and heating to fit the observed age-metallicity-kinematics distribution of LAMOST subgiant stars in both the low and high-$\alpha$ disc. By modelling all distribution parameters with spline fitting, we present the first non-parametric stellar migration and heating history of the Galaxy. We determine the heating-to-migration ratio, the ratio of the root-mean-square changes in radial/vertical and azimuthal actions, to be $\approx0.075$ for radial to azimuthal actions and $\approx0.015$ for vertical to azimuthal actions, implying a highly anisotropic diffusion in the action space. Furthermore, we identify a transition in radial migration efficiency coinciding with the transition moment of the bimodal disc, for which the radial migration was more efficient for the high-$\alpha$ disc than for the low-$\alpha$ disc. 
This transition may be attributed to two correlated scenarios: 1) a bar formation epoch accompanied by violent outward migration, and 2) a drop in the gas mass fraction in the disc when the low-$\alpha$ disc began to form. These findings offer further constraints on the formation mechanisms of bimodal discs, favouring the downsizing scenario. We also briefly discuss the connection between our results and recent high-redshift observations. In addition to the secular evolution history, our model maps the Milky Way ISM metallicity gradient at different lookback times, which we find has only varied a little (in the range of $-0.07~\rm to~-0.10~dex/kpc$) since disc formation.
\end{abstract}




\section{Introduction}
Our Milky Way's disc formed early in time ($\sim 1-2 \rm\,Gyr$ after the Big Bang) at low metallicity \citep{Belokurov2022, Belokurov2024, Conroy2022, Rix2022}. Since its formation, the orbits of disc stars have experienced secular evolution (e.g. \citealt{Sellwood_Binney2002, Roskar2008, Schonrich2010}) and merger-induced heating (e.g. the Splash, \citealt{Bonaca2017, DiMatteo2019, Gallart2019,Belokurov2020}). The secular evolution of orbits is driven by internal processes, such as dynamical interactions with non-axisymmetric perturbations in the Galaxy (e.g. the bar, spiral arms, and giant molecular clouds; see \citealt{Sellwood_Binney2002, Roskar2008, Sellwood2014}). 

Secular evolution can be decomposed into two main components: changes in the ``size'' of orbits and changes in the ``shape'' of orbits. The former is usually described by drifts in the angular momentum or guiding radii of the orbits ($\Delta L_z$ or $\Delta R_g$), termed radial migration; the latter is described by orbital eccentricity and vertical extent or, canonically, changes in the radial and vertical actions of the orbits ($\Delta J_{R,z}$), termed heating \citep{Sellwood_Binney2002, Schonrich2010}. Radial migration processes were first discussed by \citet{Sellwood_Binney2002} and \citet{Roskar2008}, who illustrated that radial migration occurs through interaction with the resonances of transient spiral arms in simulated galaxies. The interaction with different resonances will cause varying degrees of heating of stars during migration. Perfect radial migration ($\Delta J_R/\Delta L_z=0$) only occurs around the corotation resonances, while other resonances cause additional heating during migration \citep{Sellwood_Binney2002, Minchev2006, Daniel2019}. In addition to transient spiral arms, many studies have identified other mechanisms that cause or enhance radial migration in simulated galaxies, such as bar-spiral resonance overlap \citep{Minchev_Famaey2010, Minchev2011_resonanceoverlap, Solway2012}, outward radial transport by the decelerating bar \citep{Khoperskov2020, ChibaSchonrich_2021, Haywood2024, Zhang2025}, and spatial overlap between the bar and spiral arms at the bar tips \citep{Marques2025}. Some studies also highlight the influence of giant molecular clouds and external galaxies \citep{Quillen2009, Bird2012, Carr2022}. 

Radial migration has a strong impact on the chemical distribution in galactic discs \citep{Minchev2013, Kubryk2013, Grand2015}. This has been explored intensively with simulated galaxies and also with theoretical modelling (e.g. \citealt{Schonrich2010, Loebman2016, Schonrich_MM2017}). Most notably, radial migration broadens the metallicity distribution function of a galaxy at different radii because it mixes stars born at different radii together \citep{Sellwood_Binney2002, Schonrich2010, Loebman2016}. As radial mixing occurs, it gradually flattens the radial metallicity gradient of the disc and also changes the vertical metallicity gradient \citep{Schonrich_MM2017, Kawata2017}. Interestingly, radial migration could also help produce the $\alpha$-bimodality in the Galactic disc by mixing stars born at different radii with different $\alpha$-knees \citep{Schonrich2010, Sharma2021_alphabimodality, Zhang2025}. Recent studies of simulations show the impact of radial migration on biasing the star formation history measurement of galaxies \citep{Minchev2025, Bernaldez2025, Ratcliffe2025}. 

Many studies suggest that radial migration was efficient in the Milky Way's past \citep{Frankel2018, Frankel2020}. Radial migration in the Milky Way is frequently studied via the age-metallicity-$R$ distribution of stars, as in \citet{Frankel2018, Frankel2020}. They revealed that stars in the low-$\alpha$ disc now in the solar vicinity have, on average, migrated more than $3$~kpc during the past $8$~Gyr. \citet{Haywood2024} used the breaks in the radial metallicity profile in different age bins to argue that stars born around $8-10$~Gyr ago experienced violent outward migration due to a bar formation epoch and migrated out to the outer Lindblad resonance radius of the bar, in agreement with the bar formation epoch derived by \citet{Sanders2024}. \citet{Zhang2025} also analysed the shape of the age-metallicity distributions at different radial bins to argue that a significant portion of stars currently located around the bar corotation radius migrated outwards due to the corotation resonance dragging of the decelerating bar from the very inner Galaxy to their present location. The amount of heat that accompanied this migration has recently been a topic of intense debate. \citet{Frankel2020} intriguingly revealed that stars experienced little heating compared to the amount of migration ($\Delta J_R/\Delta L_z\approx0.1$). \citet{Hamilton2024} compared this number with test particle simulations and transient spiral arm-driven radial migration therein and found a clear inconsistency between the model and observation, for which the observed heating is too little. However, \citet{Sellwood_Binney2025} re-ran the simulation originally presented in \citet{Sellwood_Binney2002} and found that the heating versus migration ratio is broadly consistent with that observed, but leaving no room for extra heating from other sources (such as mergers or giant molecular clouds). Recently, \citet{McCluskey2025} compared galaxies in the FIRE-2 cosmological simulation suite with external galaxies and found consistency, but noted that the Milky Way is about 2 times colder than the external and simulated galaxies. 

In this work, we construct a statistical model, similar to that in \citet{Frankel2020}, to fit the age-metallicity-kinematics distribution in the observation (the goal is similar to the chemodynamical model in \citealt{Binney2024} but with a different model). The main improvement is to extend the model to also include the high-$\alpha$ disc, and simultaneously fit all the model parameters (e.g. radial migration efficiency and birth metallicity gradient) non-parametrically with cubic splines so that we can fully explore the signature of radial migration and heating in both the time and radial domains. 

The paper is structured as follows. We illustrate the data that used in this work in Section~\ref{section:data}. We describe and verify our chrono-chemo-dynamical model in Section~\ref{section:model}. We present the best-fit model parameters in Section~\ref{section:results} and briefly discuss some implications. We discuss in detail the radial migration and heating signatures in the Galactic disc in Section~\ref{section:implication} and link them to high-redshift observations. We summarise and conclude the paper in Section~\ref{section:conclusion}.

\begin{figure}
    \centering
    \includegraphics[width=0.98\columnwidth]{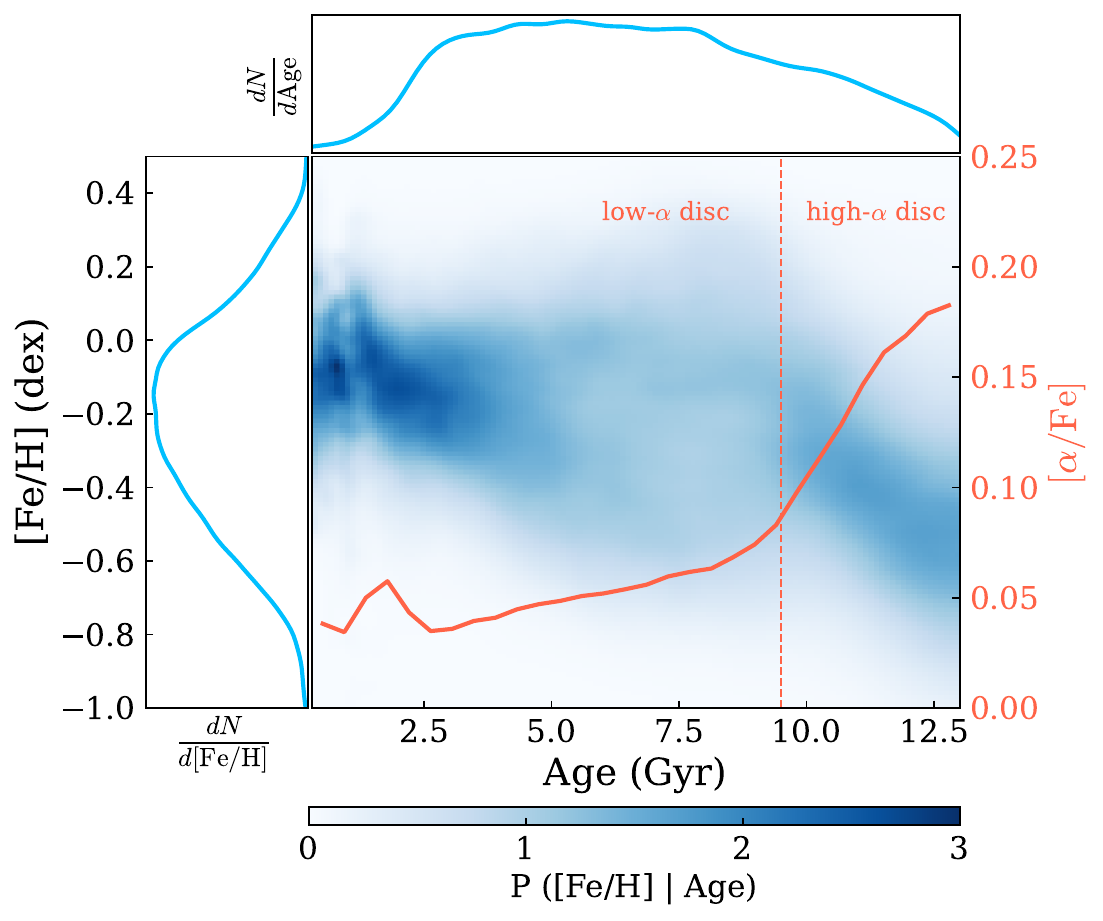}
    \caption{The column-normalised age-metallicity distribution of the LAMOST subgiant stars we used in this work \citep{Xiang_Rix2025}. The distribution of the age and metallicity are shown individually on the side panels next to the axes.}
    \label{fig:AMR_allstars}
\end{figure}

\section{Data} \label{section:data}
To unveil the orbital history of stars in the Milky Way's disc, accurate stellar age, metallicity, and kinematics measurements are required. We used the stellar parameter and isochrone age measurements of approximately $320,000$ subgiant stars derived in \citet{Xiang_Rix2025}, which is an update of \citet{Xiang_Rix2022}. Their subgiant sample utilises stellar parameter spectroscopic measurements from LAMOST DR7 \citep{Zhao2012_LAMOST, Xiang2019} and astrometric measurements from {\it Gaia} DR3 \citep{Gaia_DR3}. The subgiant stars are selected from the effective temperature--absolute magnitude ($T_{\rm eff}-M_K$) plane, following a series of quality cuts that remove binary and blue horizontal branch stars \citep[we refer the reader interested in the details to][]{Xiang_Rix2022, Xiang_Rix2025}. The median age uncertainty is $\sim 8\%$. For the kinematic measurements, we also adopt the astrometric measurements from {\it Gaia} DR3 and geometric distances from \citet{BJ2021}. The adopted radial velocity measurements are mainly from {\it Gaia} DR3, but for a small number of stars for which {\it Gaia} radial velocities are unavailable, we use the radial velocity measured by LAMOST \citep{Xiang2015_LAMOST}.

To further select high-quality disc stars, we apply the following cuts to the original sample in \citet{Xiang_Rix2025}. We select stars with metallicity uncertainty $\Sigma_{[\rm Fe/H]}<0.1$ and age uncertainty (on a logarithmic scale) $\Sigma_{\log_{10}\tau}<0.05$. We remove stars with large kinematics uncertainties by applying the parallax cut $\varpi/\Sigma_\varpi>5$ and the radial velocity uncertainty cut $\Sigma_{\rm RV}<10$~km~s$^{-1}$. We remove halo contamination by selecting stars with vertical height $|z|<3$~kpc, metallicity $[\rm Fe/H]>-1$ and ages $\tau<13$~Gyr, which is consistent with the spin-up epoch of the Milky Way considering a reasonable age uncertainty for old disc stars \citep{Miglio2021, Belokurov2022, Rix2022, Chandra2023, Queiroz2023, Belokurov2024, Zhang2023, Zhang2024}. All stars since the onset of the high-$\alpha$ disc's age-metallicity sequence are included \citep{Xiang_Rix2022}. We avoid any cuts in the velocity space so that the kinematic analyses in the next section are not biased. There are $182,465$ stars left after all selections. We compute the guiding radius, $R_g$, and the radial and vertical actions, $J_R$ and $J_z$, of these stars using the Galactic potential of \citet{McMillan2017} implemented in \textsc{AGAMA} \citep{Vasiliev2019}. The same potential is adopted throughout this work. The measurement uncertainties of astrometry, distances, and radial velocities are propagated to these dynamical quantities via Monte Carlo sampling.

The overall age-metallicity distributions are presented in Fig.~\ref{fig:AMR_allstars}. It shows rapid chemical enrichment in the high-$\alpha$ disc at early times \citep[][]{Ciuca2024}, and the bimodal "V-shaped" sequences for the low-$\alpha$ disc between $\sim 0-8$~Gyr \citep{Xiang_Rix2022}. The final sample covers a wide region around the solar vicinity of $6\lesssim R_g/\rm kpc \lesssim 13$. The age-metallicity distribution of $R_g<6~\rm kpc$ becomes noisy due to the low number of stars. Hence, we focus our analysis in the region of $R_g=6-13~\rm kpc$.


\section{Chrono-chemo-dynamical model}\label{section:model}

In this section, we present the model for the Milky Way's age-metallicity-kinematic distribution to decode its orbital history. We briefly summarise the assumptions of this model here, but for clarity, we present the detailed reasons and justifications in the corresponding sections. The main underlying assumptions are: 1) stars born  at the same time and radius have approximately the same metallicity within a tolerance; 2) stars are born on near circular orbits; 3) the change in stellar orbits is dominated by secular processes, such as radial migration and dynamical heating. Building on these assumptions, we construct a model describing $p({[\rm Fe/H]}, J_R, J_z |R_g,\tau)$ to disentangle radial migration (change in $R_g$ or angular momentum $L_z$) and dynamical heating (change in $J_R$ and $J_z$), similar to the model in \citet{Frankel2020}. We will now describe the ingredients of this model.

\subsection{Metallicity evolution}\label{model::metallicity}

The metallicity of stars, $[\rm Fe/H]$, is usually considered an imprint of their birthplace, known as chemical tagging \citep{Freeman_BH2002, Sellwood_Binney2002}. This technique has recently been revived to calculate the birth radius of stars in the Galactic disc \citep{Minchev2018, Lu2024, Ratcliffe2024, Ratcliffe2025}. We will adopt the same idea to decode the radial migration history in the Galactic disc by fitting a metallicity evolution model to the Milky Way. 

We assume that at a given age $\tau$, stellar metallicity follows a linear gradient with Galactocentric radial distance
\begin{equation*}
    \overline{[{\rm Fe/H}]} (\tau, R_b) = [{\rm Fe/H}]_{\rm max}(\tau) + R_b\nabla[{\rm Fe/H}](\tau) ,
\end{equation*}
where $R_b$ is the birth radius of stars, $[{\rm Fe/H}]_{\rm max}(\tau)$ is the central metallicity at that time or, equivalently, $\overline{[{\rm Fe/H}]}(\tau, R_b=0)$, and $\nabla[{\rm Fe/H}](\tau)$ is the radial metallicity gradient at time $\tau$ when the stars were born. We only consider the metallicity gradient to be negative in our model. One of the main assumptions of our model is that the metallicity of star-forming gas is azimuthally constant (within a tolerance, $\sigma_{[{\rm Fe/H}]}$) at all times and radii. This has been verified with simulations \citep[][]{Lu2022, Ratcliffe2024_tng50}, extragalactic observations \citep{Kreckel2019}, and, most relevantly, H$_{\rm II}$ regions in the present-day Milky Way \citep{Wenger2019}. Assuming stars are born on a near-circular orbit, we can easily write the probability distribution of stellar metallicity for stars born with age $\tau$ and angular momentum $L_b=R_b V_{\rm circ}$ as 
\begin{equation}
    p([{\rm Fe/H}]|L_b, \tau) = \mathcal{N}([{\rm Fe/H}]\,|\,\overline{[{\rm Fe/H}]} (\tau, L_b), \,\sigma_{[{\rm Fe/H}]} ),
    \label{eqn:met}
\end{equation}
where $\mathcal{N}$ denotes Gaussian distribution, and $\sigma_{[{\rm Fe/H}]}$ is the azimuthal metallicity dispersion, which we now set to a constant $\sigma_{[{\rm Fe/H}]}=0.05$ \citep{Wenger2019} throughout the age of the stellar disc, but we will revisit this assumption later in Section~\ref{limit::azimuthal}.

$\nabla[{\rm Fe/H}](\tau)$ is optimised to fit the data, whereas $[{\rm Fe/H}]_{\rm max}(\tau)$ is kept to be fixed.
We mainly adopt $[{\rm Fe/H}]_{\rm max}(\tau)$ derived in \citet{Lu2024}, which is obtained by taking the upper envelope of the age-metallicity distribution of stars with $R_g<5$~kpc in \citet{Xiang_Rix2022}. It has $[{\rm Fe/H}]_{\rm max}$ in the present day of $\sim0.6$~dex, which matches the gas-phase metallicity observed in the solar neighbourhood extrapolated to the Galactic centre \citep[][]{Wenger2019}. 
We fit the radial metallicity gradient over time $\nabla[{\rm Fe/H}](\tau)$ during the optimisation, while fixing its current value of $\nabla[{\rm Fe/H}](\tau=0)$ at $-0.07$, in agreement with current observations of young stars \citep[][]{Braganca2019}.

\subsection{Radial migration}\label{model::RM}
We use the same description of radial migration as in \citet{Sellwood_Binney2002}, characterising it as a series of stochastic scatterings in angular momentum. In this regime, the angular momentum evolution is modelled by the diffusion equation of angular momentum \citep{Sanders_Binney2015}:
\begin{equation*}
    \frac{\partial f}{\partial t} = \frac{\partial}{\partial L_z}\left(-D f+ \frac{\sigma_{L_z}^2}{2}\frac{\partial f}{\partial L_z} \right),
\end{equation*}
where $\sigma_{L_z}$ represents the diffusion of angular momentum, and $f$ denotes the angular momentum distribution function. To ensure that the total angular momentum of an exponential disc is approximately conserved, the drifting term $D$ can be constrained as $D=-\sigma_{L_z}^2/2V_{\rm circ}R_{d,0}$, where $R_{d,0}$ is the scale length of every mono-age stellar population at birth \citep{Sanders_Binney2015}. 

The solution of the diffusion equation allows us to link the present-day angular momentum distribution of stars to that of their birth by \citep{Sanders_Binney2015}:
\begin{equation}
    p(L_z|L_b, \tau) \propto \mathcal{N}\left(L_z\,|\,L_b - \frac{\sigma_{L_z}^2(R_g,\tau)}{2V_{\rm circ}R_{d,0}}, \,\sigma_{L_z}(R_g,\tau) \right),
\end{equation}
where $L_z>0$ for the disc, so an additional normalisation factor is needed. This radial migration kernel is also used in \citet{Frankel2018, Frankel2020} and \citet{Sharma2021_alphabimodality} to describe radial migration in the Milky Way's disc.
We will fit for the angular momentum diffusion $\sigma_{L_z}(R_g,\tau)$ as a function of time and guiding radius during the optimisation pipeline, and we fix $\sigma_{L_z}(R_g,\tau=0)=0$. Assuming stars are born on near circular orbits, the angular momentum diffusion term can be converted into the root-mean-square (RMS) change of angular momentum, $\delta L_z$, as $\delta L_z \approx \sqrt{\sigma_{L_z}^2 + D^{2}}\approx A\sigma_{L_z}$ \citep{Frankel2020}, where $A\sim1.2-1.5$ depends on $R_{d,0}$.

\begin{table*}
\caption{A summary of the model parameters and their priors}
\begin{center}
\begin{tabular}{lccccc}
\hline
\hline
 Models  & Functional form & Model parameters & Prior & Smoothing prior \\
  & & ($\Theta$) & & ($\sigma_{\ln{\theta}}$)\\
\hline
\\
Disc structure at birth & 
$p(L_{b}|\tau)=  \frac{L_b}{(V_{\rm circ}R_{d,0})^2}\exp\left(-\frac{L_b}{V_{\rm circ}R_{d,0}}\right)$ 
& $R_{d,0} (\tau)$ & $\mathcal{N}(\ln{R_{d,0}}|0.5, 0.5)$ & 0.2 \\

\\

Metallicity evolution & 
\makecell[c]{%
  $[{\rm Fe/H}]_\star = [{\rm Fe/H}]_{\rm max}- R_b\nabla[{\rm Fe/H}]$ \\
  $p([{\rm Fe/H}]|L_b, \tau) = \mathcal{N}([{\rm Fe/H}]\,|\,[{\rm Fe/H}]_\star, \,\sigma_{[{\rm Fe/H}]} )$
}
& $\nabla[{\rm Fe/H}](\tau)$ & $\mathcal{N}(\ln{(-\nabla[{\rm Fe/H}])}|-2.6, 0.5)$ & 0.2 \\

\\

Radial migration & 
\makecell[c]{
$p(L_z|L_b, \tau) \propto \mathcal{N}\left(L_z\,|\,L_b - \frac{\sigma_{L_z}^2}{2V_{\rm circ}R_{d,0}}, \,\sigma_{L_z} \right)$
}
& $\sigma_{L_z}(R_g,\tau)$ & $\mathcal{N}(\ln{(\sigma_{Lz}/\tau)}|4.7,0.5)$ & 0.5  \\

\\

Heating & 
\makecell[c]{
$p(J_R, J_z | L_z, \tau) \propto \exp\left(\frac{-\kappa J_R}{\sigma_R^2}\right)\exp\left(\frac{-\nu J_z}{\sigma_z^2}\right)$
}
& 
\makecell[c]{$\sigma_R(R_g, \tau)$\\$\sigma_z(R_g, \tau)$} & 
\makecell[c]{$\mathcal{N}(\ln{(\sigma_{R}/\tau)}|1.5,1.0)$\\
$\mathcal{N}(\ln{(\sigma_{z}/\tau)}|1.1,0.9)$}
& 0.3  \\
\\
\hline
\end{tabular}
\end{center}
\label{table:prior}
\end{table*}

\subsection{Radial and vertical heating} \label{model::heating}
We use the quasi-isothermal distribution function \citep{Binney2010, Sanders_Binney2015} to fit for the present-day radial and vertical kinematics, mapping the diffusion of the radial and vertical actions starting with:
\begin{equation}
    p(J_R, J_z | L_z, \tau) \propto \exp\left(\frac{-\kappa J_R}{\sigma_R^2(R_g, \tau)}\right)\exp\left(\frac{-\nu J_z}{\sigma_z^2(R_g, \tau)}\right),
\end{equation}
where $\kappa$ and $\nu$ are the radial and vertical epicyclic frequencies, and $\sigma_{R, z}(R_g, \tau)$ is the present-day radial and vertical velocity dispersion. In our pipeline, $\sigma_{R, z}(R_g, \tau)$ is optimised non-parametrically in both time and spatial domain. We fix $\sigma_{R, z}(R_g,\tau=0)=0$.


To convert the fitted velocity dispersion $\sigma_{R, z}(R_g, \tau)$ into the RMS of changes in the radial and vertical actions, $\delta J_{R, z}$, we can write 
\begin{equation*}
    \delta J_{i} = \sqrt{2\langle J_i\rangle^2 + 2J_{i,0}^2 - 2J_{i,0}\langle J_i\rangle}
\end{equation*}
for $i=R,z$, where $\langle J_{R(z)}\rangle\approx\sigma^2_{R(z)}/\kappa(\nu)$ in the epicyclic limit. Using the same approximation as before that stars are born on near-circular orbits, we can approximate $J_{i,0}\approx0$, and hence, $\delta J_{R}\approx\sqrt{2}\sigma^2_{R}/\kappa$ and $\delta J_{z}\approx\sqrt{2}\sigma^2_{z}/\nu$.

\subsection{Disc growth}\label{model::disc_growth}

The final ingredient of this model is the evolution of the surface density profile of the star-forming disc over time. The disc stars formed gradually in time, for which we model the birth angular momentum distribution of each stellar population with age as 
\begin{equation}
    p(L_b|\tau) = \frac{L_b}{L_{\rm disc}^2(\tau)}\exp\left(-\frac{L_b}{L_{\rm disc}(\tau)}\right),
\end{equation}
where $L_{\rm disc}(\tau)=R_{d,0}(\tau)V_{\rm circ}$ is the characteristic scale of the birth angular momentum, such that the surface density profile of the star-forming disc is roughly $\Sigma(R_b)\propto\exp(-R_b/R_{d,0})$. 

We fit for the evolution of the disc scale length at birth over time, $R_{d,0}(\tau)$. Our model, modulo age uncertainties, is insensitive to the star-formation rate in the Milky Way disc, so we do not model that in this architecture.

\subsection{Likelihood}\label{model::likelihood}

To find the best-fit model that can describe the observed age-metallicity-kinematics distribution of the Milky Way's disc, we combine the metallicity evolution, angular momentum diffusion, dynamical heating, and disc growth models described before to write the log-likelihood.

\begin{figure*}
    \centering
    \includegraphics[width=0.98\textwidth]{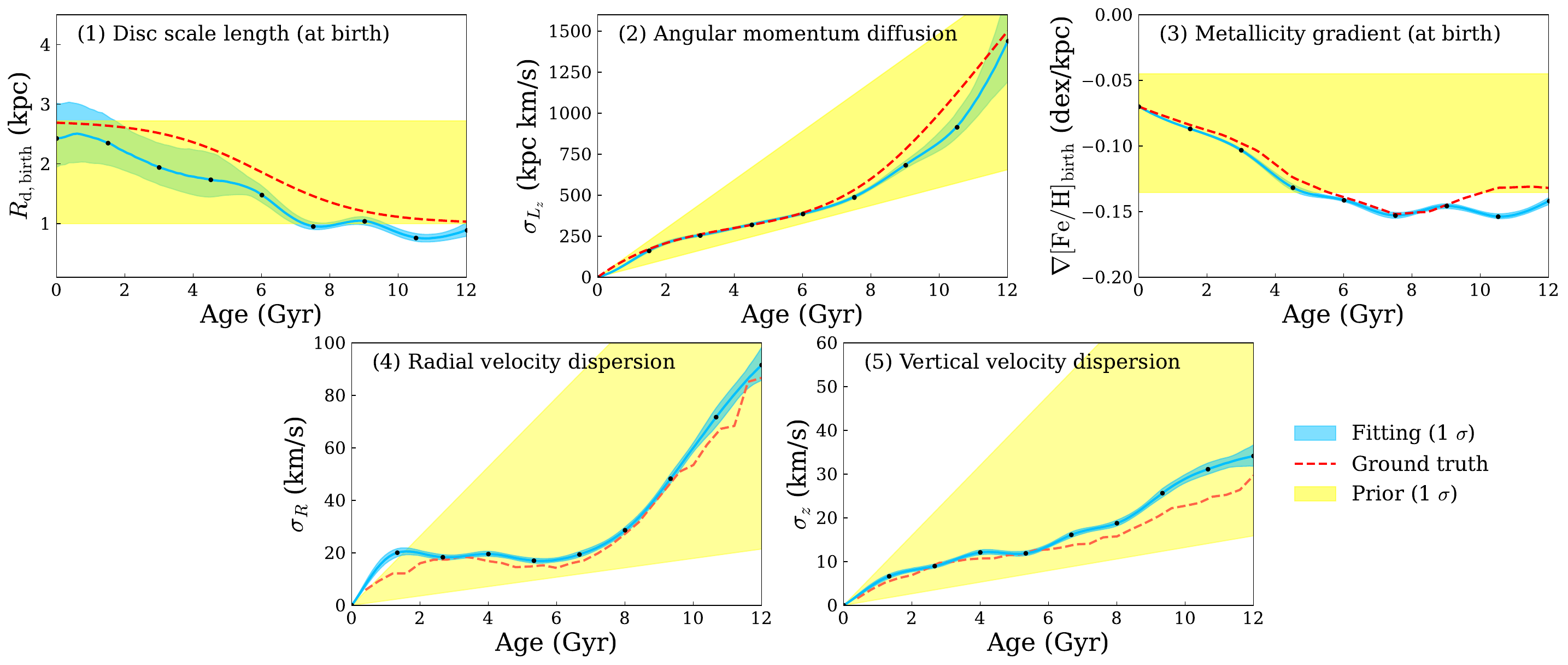}
    \caption{Results of the mock sample test of the model. The ground truth is shown in the red dashed line, and the recovered parameters are shown in the blue solid line and the blue shaded region. The black dots are the fitted knots in the cubic spline. The background yellow region denotes $1-\sigma$ of the Gaussian prior of these fitted parameters. Each different panel is for a different free parameter in the model, for which the {\it top left} panel is the initial scale length of the disc when the star was forming, {\it top middle} panel is for the angular momentum diffusion coefficient, {\it top right} panel is for the initial radial metallicity gradient when the star was forming (or the metallicity gradient of the Galactic ISM), {\it bottom} panels are the radial and vertical velocity dispersion.}
    \label{fig:mock_test}
\end{figure*}

We first write the measured distribution of the metallicity, present-day angular momentum, and age of a star as 
\begin{align*}
     p([{\rm Fe/H}], L_z, \log\tau|\bs \Theta_1) = & \int  {{\rm d} L_b}
    \,  p([{\rm Fe/H}]|L_b, \log\tau) \\ & p(L_z|L_b, \log\tau)p(L_b| \log\tau)p(\log\tau)\},
\end{align*}
where $\bs \Theta_1=\{R_{d,0}, \nabla[{\rm Fe/H}], \sigma_{L_z}\}$ are the model parameters.
We further convolve the measurement uncertainty to this distribution function as 
\begin{align*}
     p([{\rm Fe/H}], L_z, \log\tau|\bs \Theta_1) = & \int  {{\rm d}{\bs \chi'}} 
    \{\mathcal{N}(\bs \chi'|\bs \chi, \Sigma_{\chi}) \\
    &\, p([{\rm Fe/H}]', L_z', \log\tau'|\bs \Theta_1) \},
\end{align*}
where ${\bs \chi}=\{[{\rm Fe/H}], L_z, \log\tau\}$ are the observables, and $\Sigma_\chi$ is the error ellipse of these observables. 
Similarly, we convolve the measurement uncertainties to the kinematic distribution function
\begin{align*}
    p({\bf w}, \log\tau|& \bs \Theta_2) = \int {\rm d{\rm \bf  w'}d\log\tau'} \{\mathcal{N}(\log\tau'|\log\tau, \Sigma_{\log\tau}) \,\\
    & \mathcal{N}({\rm \bf  w'}|{\rm \bf w}, 
    \Sigma_{\rm \bf w}) p(J_R', J_z' | L_z', \log\tau') p(L_z'|\log\tau')p(\log\tau')\}, 
\end{align*}
where ${ \rm \bf w} = (\bs x, \bs v) = \bs J$, $\rm d{\bf w} = d{\bs x}d{\bs v} = (2\pi)^3 d\bs J$, and $\bs \Theta_2=\{\sigma_{R}, \sigma_{z}\}$. 

To eliminate the impact of age and spatial selection functions, we write 
\begin{equation}
    p([{\rm Fe/H}]| L_z, \log\tau) = \frac{p([{\rm Fe/H}], L_z, \tau)}{\int d[{\rm Fe/H}] p([{\rm Fe/H}], L_z, \tau)},
    \label{eqn::AMR}
\end{equation}
\begin{equation}
    p(v_R, v_z|R, z,L_z,\log\tau) = \frac{p({\bf w}, \tau)}{\int dv_R dv_z p({\bf w}, \tau)},
    \label{eqn::kine}
\end{equation}
where similar approach were successful in \citet{Zhang2023a} and \citet{ Funakoshi2025}.
These equations can be simplified by assuming the distribution of age and angular momentum are reasonably flat across the error ellipse of $L_z$ and $\tau$ after the quality cuts in Section~\ref{section:data}, which means $p(\tau)$ and $p(L_z|\tau)$ then cancel out in the denominators and numerators. We evaluate Equations~(\ref{eqn::AMR}) and (\ref{eqn::kine}) using Monte Carlo integration. We write the overall log-likelihoods as 
\begin{align*}
    & \ln L_1 = \sum_{j}^{N_{\rm star}}\ln p_j([{\rm Fe/H}]_j| L_{z,j}, \log\tau_j, \bs \Theta_1)\\
    & \ln L_2 = \sum_{j}^{N_{\rm star}}\ln p_j(v_{R,j}, v_{z,j}|R_j, z_j,L_{z,j},\log\tau_j, \bs \Theta_2)
\end{align*}
As these two log-likelihoods are independent of each other, we sample their posterior independently to save computational cost.

\subsection{Modelling pipeline}\label{model::pipeline}

We fit for $\Theta_1=\{R_{d,0}(\tau), \nabla[{\rm Fe/H}](\tau), \sigma_{L_z}(\tau, R_g)\}$ and $\Theta_2=\{\sigma_{R}(\tau, R_g), \sigma_{z}(\tau, R_g)\}$ non-parametrically, for which their time evolution is modelled with cubic spline functions (as the one in \citealt{Sanders2024} and \citealt{ Funakoshi2025}). We use 10 knots uniformly allocated across $0-13$~Gyr and draw cubic splines between the knots. Therefore, we have a total of $30$ and $20$ free parameters for each set, respectively. We sample the posterior of the parameters using the code in \citet{Sanders2024}, which adopted Hamiltonian Monte Carlo with the NUTS sampler implemented in \textsc{numpyro} \citep{numpyro, numpyro2}. We sample all the posterior from the natural logarithmic scale of these parameters. In addition, we apply a smoothing prior, $P_{\rm smooth}$, between knots to damp their oscillations:
\begin{equation*}
    \ln P_{\rm smooth} = \sum_{\theta}^{\bs \Theta}\sum_{k}^{N_{\rm knots}} \frac{(\ln \theta_{k+1} - \ln \theta_{k})^2}{2\sigma_{\ln\theta}^2}.
\end{equation*}
The choice of actual prior and smoothing prior $\sigma_{\ln\theta}$ for each set of parameters is summarised in Table~\ref{table:prior}.

As we want to model the radial dependence of angular momentum diffusion coefficient, radial and vertical velocity dispersion, but it is too computationally expensive to perform the spline fitting in both the radial and time domain, we divide the sample constructed in Section~\ref{section:data} into individual guiding radii bins, and we sample the posterior of $\bs \Theta_2$ directly at each guiding radii bin for $\sigma_{R}(\tau, R_g)$ and $ \sigma_{z}(\tau, R_g)$. The sampling routine for $\bs \Theta_1$ is more complicated because we require $R_{d,0}(\tau)$ and $\nabla[{\rm Fe/H}](\tau)$ to be the same for all guiding radii bins due to the model's assumption, while $\sigma_{L_z}(\tau, R_g)$ needs to be fitted at each guiding radii bin. For this purpose, we sample the posterior of $\bs \Theta_1$ iteratively. First, we apply the sampler to $\ln L_1$ computed using stars in all guiding radii bins to sample $\{R_{d,0}(\tau),\,\nabla[{\rm Fe/H}](\tau),\,\sigma_{L_z}(\tau, R_g)\}$ altogether. Secondly, we fix $R_{d,0}(\tau)$ and $\nabla[{\rm Fe/H}](\tau)$ using the fitted values and then sample the posterior of $\sigma_{L_z}(\tau, R_g)$ in each of the guiding radii bins. In the third step, we revisit $R_{d,0}(\tau)$ and $\nabla[{\rm Fe/H}](\tau)$ by fixing $\sigma_{L_z}(\tau, R_g)$ from the previous fitting, and then sample the posterior for $R_{d,0}(\tau)$ and $\nabla[{\rm Fe/H}](\tau)$ again using all the stars again. We iterate between the second and third steps until the results converge, which usually occurs reasonably within 3 iterations. In this way, we ensure that the angular momentum diffusion is modelled at different guiding radii, while keeping the birth scale length and metallicity gradient the same across different radius.

\begin{figure*}
    \centering
    \includegraphics[width=0.98\linewidth]{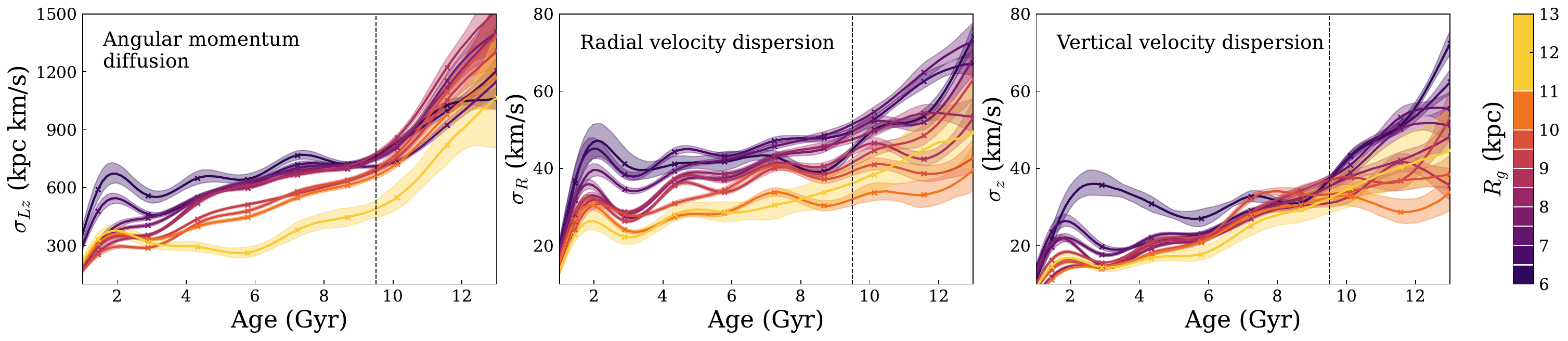}
    \caption{The parameters of the best fit model. {\it Left}: the angular momentum diffusion coefficient for stars at different ages. The results at different guiding radii are coloured differently. {\it Middle}: the radial velocity dispersion of the stars at different ages, coloured according to the guiding radius. {\it Right}: the vertical velocity dispersion of the stars at different ages, coloured according to the guiding radius. Vertical dotted lines label $\rm age = 9.5~Gyr$, around the transition time of $\sigma_{L_z}$.}
    \label{fig:fitting_actions}
\end{figure*}

\subsection{Mock data test} \label{model::mock_test}
We generate a mock sample to test our fitting routine. The purpose of this test is to ensure that our model is robust against the underlying selection functions and the measurement uncertainties of the sample. The systematic caveats of the model, e.g., model assumptions, are not tested in this section; they are instead discussed in Section~\ref{section:limitation}, and tested with a simulated galaxy presented in Appendix~\ref{appendix:TNG50}.

We postulate the free parameters in our model and their time dependence and use these as the ground truth to sample the metallicity and kinematics of the mock stars. First, we sample the ages of the stars from a uniform distribution between $0-12$~Gyr. Using the mock ages, we assign the position and velocity of the stars by sampling from the quasi-isothermal distribution function implemented in \textsc{AGAMA} using the \texttt{sampling routine} in \citet{Vasiliev2019}, assuming the \citet{McMillan2017} potential. We manually apply mock selection functions to the mock stars in age, Galactocentric radius, and angular momentum to mimic the distributions of the sample we construct in Section~\ref{section:data} in a similar approach as in \citet{Zhang2023a}. Then, we assign the metallicity to each star according to a probability distribution $p([{\rm Fe/H}]|\tau, L_z)$ in Eq.~(\ref{eqn::AMR}) without convolving the measurement uncertainties. After creating this sample of mock stars with all the observables $\{{\bf w}, \tau, {\rm [Fe/H]}\}$, we further add scatters to these observables to mock the measurement uncertainties, also on the same order of magnitude as the adopted sample. The mock error ellipse of this scattering is recorded for modelling.

We apply the model described above to this mock data sample, aiming to recover the postulated model parameters, i.e. $\{R_{d,0}, \nabla[{\rm Fe/H}], \sigma_{L_z}, \sigma_{R}, \sigma_{z}\}(\tau)$. We present our fitting result for stars with guiding radii between $9-11$~kpc in Fig.~\ref{fig:mock_test}. The red dashed line represents the ground truth parameters, and the blue shaded region shows the fitting results within $1\sigma$ error. 

We recover the angular momentum diffusion, metallicity gradient, radial and vertical velocity dispersion accurately as the ground truth with small fitting uncertainties for most of the knots. The birth scale length deviates slightly from the ground truth. As we do not consider spatial distribution of stars in our model, the constraints of the initial scale length is purely from the shape of the metallicity distribution of each mono-age stellar population, so the recovery of the initial scale length becomes worse when the age and metallicity uncertainty is large. The posterior's uncertainties also become larger for the younger population. This is because the constraining power on the birth scale length is inversely correlated to the angular momentum diffusion in our model, so the fitting becomes less accurate for younger populations which have had less time for radial migration to take place.

\section{Best fit model of the Milky Way}\label{section:results}

\begin{figure*}
    \centering
    \includegraphics[width = 0.98\linewidth]{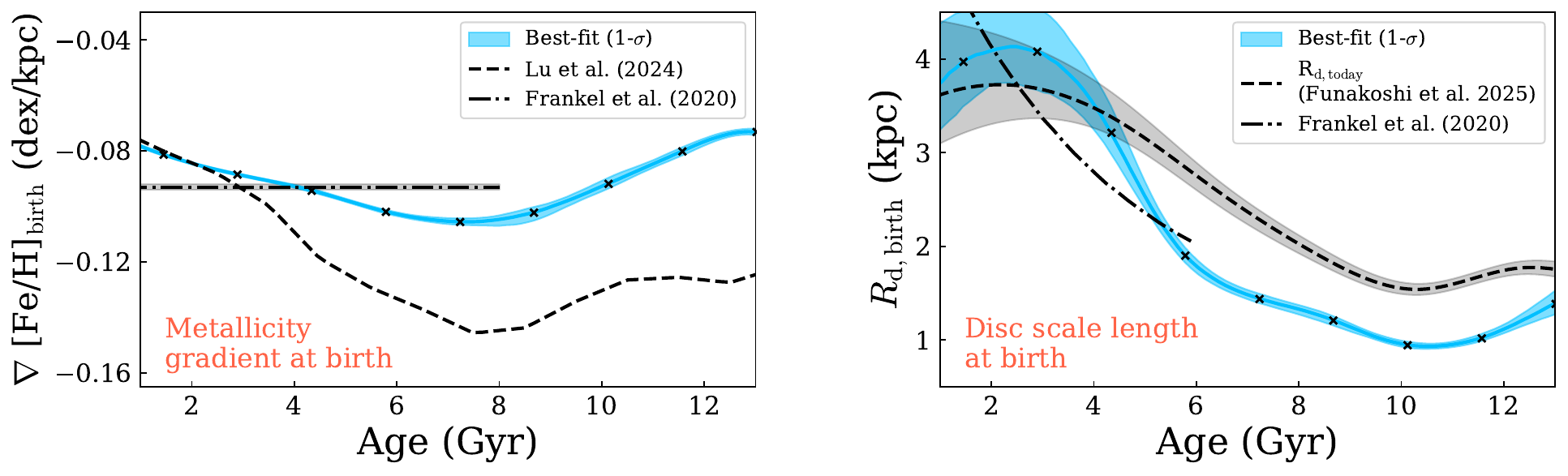}
    \caption{The parameters of the best fit model. 
    {\it Left}: The initial metallicity gradient at birth (or ISM metallicity gradient) at different lookback times. The blue line and blue bands are the fitting results within $1\sigma$ uncertainty. The black dot-dashed line is the fitting from \citet{Frankel2020} and the black dashed line is the inferred ISM metallicity in \citet{Lu2024}, obtained using a different method.
    {\it Right}: the initial scale length of the stars at birth. The blue line and blue bands are the fitting results within $1\sigma$ uncertainty. The black dotted line is the model results in \citet{Frankel2020} using low-$\alpha$ red clump stars, and the black dashed line is the present scale length of the disc fitted in \citet{Funakoshi2025}.}
    \label{fig:fitting_RdMHgrad}
\end{figure*}

In this section, we present the chrono-chemo-dynamical model of the Milky Way by applying the model described in section~\ref{section:model} to the subgiant star sample in section~\ref{section:data} with guiding radius from $6-13$~kpc. The best-fit model is shown in Fig.~\ref{fig:fitting_actions} and Fig.~\ref{fig:fitting_RdMHgrad}, in which the initial length of the disc scale and the metallicity gradient at birth are fitted globally for all radii, but the diffusion of angular momentum, the dispersion of radial and vertical velocity are fitted at different guiding radii.

\subsection{Diffusion of the orbital actions}\label{results::diffusions}

In the left panel of Fig.~\ref{fig:fitting_actions}, we show the best-fit angular momentum diffusion coefficient ($\sigma_{L_z}$) as a function of stellar age, coloured by the present-day guiding radius. We observe a smooth increase in radial migration strength with increasing stellar age for the young disc ($\tau\lesssim8$~Gyr), as expected. However, the model finds a transition around $\sim9-10$~Gyr, where old stars ($\tau\gtrsim10$~Gyr) in the Milky Way experienced much stronger radial migration at earlier times. This transition time matches that of the high-$\alpha$ to low-$\alpha$ disc on the timescale of this sample \citep{Xiang_Rix2022}. Therefore, the high radial migration efficiency in the model is required to explain the flat (or even slightly positive) radial metallicity distribution of the old, high-$\alpha$ disc \citep{Hayden2015, Chen2025}, implying that the Galactic high-$\alpha$ disc is fully radially mixed and no signature of the initial metallicity gradient remains. We leave the interpretation of this transition to Section~\ref{discussion::transition}. 

In the low-$\alpha$, young disc, $\sigma_{L_z}$ also increases with age, meaning that older stars are more radially mixed. Angular momentum diffusion can flatten the metallicity gradient \citep{Schonrich_MM2017, Kawata2018}, which is consistent with the observed gradual flattening of the metallicity gradient with increasing stellar age \citep[e.g.][]{Vickers2021, Chen2025}. In Fig.~\ref{fig:fitting_actions}, curves corresponding to larger $R_g$ (orange/yellow) consistently lie below the curves corresponding the smaller $R_g$ (red/purple). This means that the radial migration efficiency decreases radially outwards, which is frequently observed in simulations due to the decrease of bar and spiral perturbations (\citealt{Roskar2008, Bird2012, Kubryk2015}, Zhāng et al. in prep). In addition to the general radial decline of the radial migration strength, we also find that $\sigma_{L_z}(\tau)$ has similar behaviour in the $R_g$ bins of $\approx6.5-9$~kpc, $9-11$~kpc, and $\geq11$~kpc. This could be due to different radial migration mechanisms that dominate different radiuses (e.g. \citealt{Minchev_Famaey2010, Halle2015, Khoperskov2020, Haywood2024, Zhang2025, Marques2025}), which implies that the radial migration behaviour in the low-$\alpha$ disc is more structured than a simple stochastic diffusion with a smoothly varying diffusion coefficient across different radii. Bar-driven radial migration could be crucial for understanding this behaviour \citep{Halle2015, Khoperskov2020, Zhang2025, Cerqui2025, Marques2025}. We will discuss this in more detail in Section~\ref{discussion::radial_regimes}. 

The best-fit radial and vertical velocity dispersions are shown in the middle and right panels of Fig.~\ref{fig:fitting_actions}. Both velocity dispersions increase with increasing stellar ages, consistent with the age-velocity dispersion relation of the Milky Way \citep{Aumer2016, SandersDas2018, Sharma2021_avr}. The radial velocity dispersion exhibits a clear decline towards the outer radius, which is compatible with past observations, too. The vertical velocity dispersion also declines as the radius becomes larger, but this decline is much weaker than that in the radial velocity dispersion.


\subsection{Evolution of the metallicity gradient}\label{results::metallicity}

\begin{figure}
    \centering
    \includegraphics[width=0.98\linewidth]{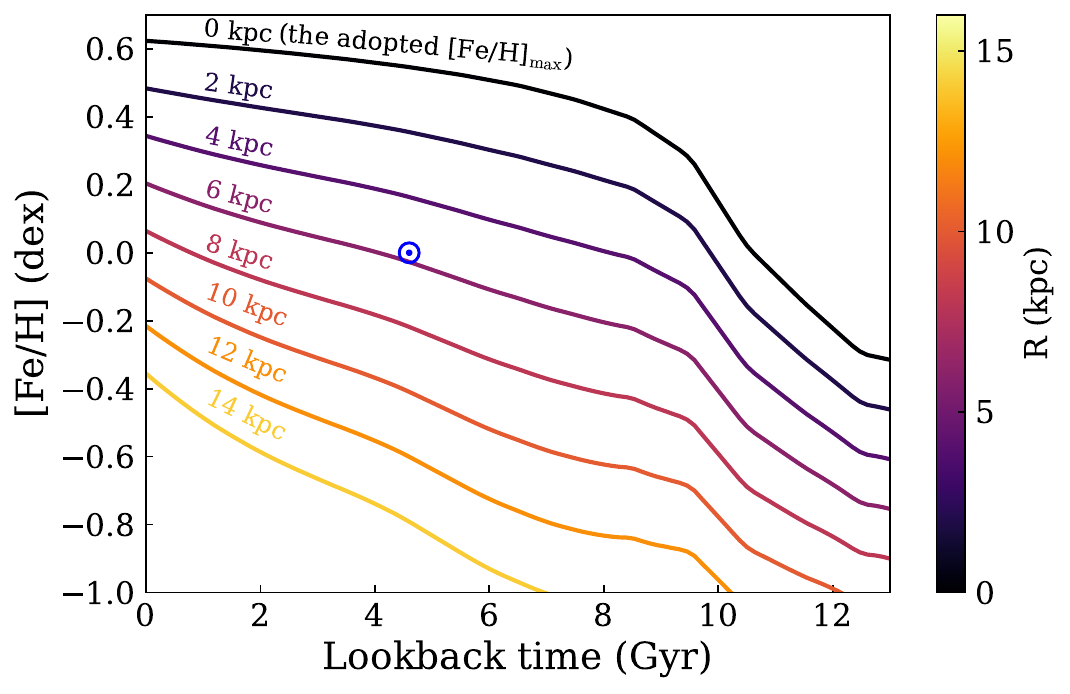}
    \caption{The best fit metallicity evolution model of the Milky Way. The coloured lines show the time evolution of ISM metallicity at different radii. The blue dot labels the Sun, implying that the Sun was born $\sim5.7$~kpc and migrated to its present location.}
    \label{fig:metallicity_model}
\end{figure}

\begin{figure*}
    \centering
    \includegraphics[width=0.98\linewidth]{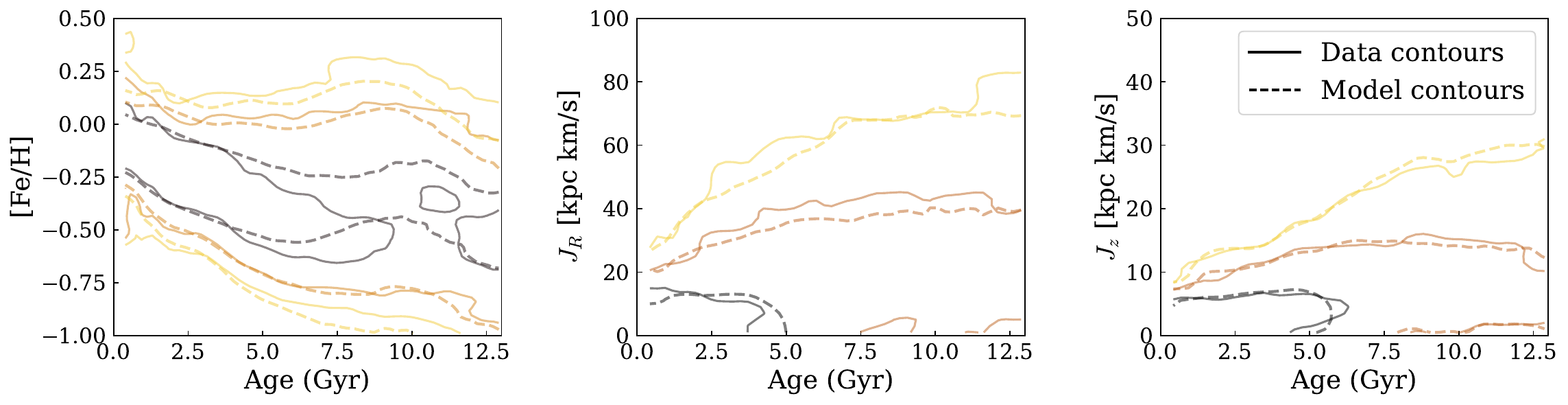}
    \caption{The comparison between the best fit model and the fitted data, where the model is in the dashed line and the data is in the solid line. Lines with different colours are different contours of the column-normalised distribution in the corresponding space. {\it Left:} the column-normalised age-metallicity plane of the guiding radius bin $9.5-10$~kpc. {\it Middle:} the column-normalised age-$J_R$ distribution at the same guiding radius bin. {\it Right:} the column-normalised age-$J_z$ distribution at the same guiding radius bin. Our model fits the data well.}
    \label{fig:model_vs_data}
\end{figure*}

The model of the birth metallicity gradient of stars is shown in the left panel of Fig.~\ref{fig:fitting_RdMHgrad}, in which our model is represented by the blue line and shaded region. We find a time-varying birth metallicity gradient for stars born at different ages. The metallicity gradient steepens at early times, reaching a minimum at approximately $8$~Gyr, and then starts to flatten to the present value of -$0.07$~dex/kpc \citep{Braganca2019}. The variation is much smaller than the ISM metallicity gradient calibrated using other methods directly from the stellar metallicity dispersion at fixed age bins (e.g. \citealt{Minchev2018, Lu2024, Ratcliffe2024_tng50}). We claim that this metallicity gradient could be more reliable because we are modelling both the angular momentum diffusion rate and the metallicity gradient together (see also a verification of this metallicity gradient recovery using a simulated galaxy in TNG50 in Appendix~\ref{appendix:TNG50}). Our result is consistent with \citet{Frankel2020}, who performed similar modelling for the low-$\alpha$ red clump stars in APOGEE. Their model assumed a time-invariant metallicity gradient and found a value of $\sim0.09$~dex that is consistent with ours if we consider a time-averaged metallicity gradient in our model.

The recovered metallicity evolution model of the Milky Way is shown in Fig.~\ref{fig:metallicity_model}, in which the time evolution of metallicity at different radii is coloured differently. The blue circle dot represents the age and metallicity of the Sun, which implies that the Sun was born at approximately $5.7$~kpc and migrated to its present location (see \citealt{Baba2024}). \citet{Zhang2025} argued that the Sun migrated outwards with the deceleration of the Galactic bar because it sits on the age-metallicity sequence that aligns with the evolution of the corotation radius. This metallicity model could also be used to trace the birth radius of stars in the Milky Way's disc for both the low- and high-$\alpha$ discs.

\subsection{Evolution of the disc scale length at birth}\label{results::scale_length}

The model for the initial scale length of the disc is presented in the right panel of Fig.~\ref{fig:fitting_RdMHgrad}, where the model result is shown by the blue line and blue shaded region. The constraint is tight for the early disc when the radial migration efficiency is high ($\sigma_{L_z}\gtrsim500$~kpc~km/s), enabling the model to learn the initial scale length of the disc from the shape of the metallicity distribution at the corresponding ages. The constraints on the late disc become weaker as the total angular momentum diffusion drops. Our results are broadly consistent with the initial scale length found in \citet{Frankel2020}. We also illustrate the \textit{current} disc scale length with the black dashed line, which is obtained by dynamical modelling of the disc kinematics \citep{Funakoshi2025}. \citet{Funakoshi2025} used the spectroscopic ages of red giant stars in \citet{Ciuca2024} to build a dynamical model of the Milky Way's disc from present-day kinematics and derived the current disc scale length from the velocity distributions in the Galactic disc. The initial scale length obtained is consistently shorter than the current scale length, as expected, because outward radial migration should flatten the surface profile of the disc (see \citealt{Ratcliffe2025} for similar results). 

The initial scale length of the disc shows a rapid increase for the Milky Way's low-$\alpha$ disc, consistent with the inside-out disc formation scenario \citep{Chiappini2001}. Interestingly, we observe a drop in the initial scale length at $\sim10$~Gyr. A similar drop is also observed in \citet{Funakoshi2025} at the same age for the current scale length, which was attributed to gas disc shrinkage caused by the Gaia-Sausage-Enceladus (GS/E) merger \citep{Grand2018, Funakoshi2025} at approximately $10$~Gyr ago \citep{Belokurov2020, Helmi2020}. Our result reinforces their argument.

\subsection{Model verification}\label{results::verification}

\begin{figure*}
    \centering
    \includegraphics[width=0.98\linewidth]{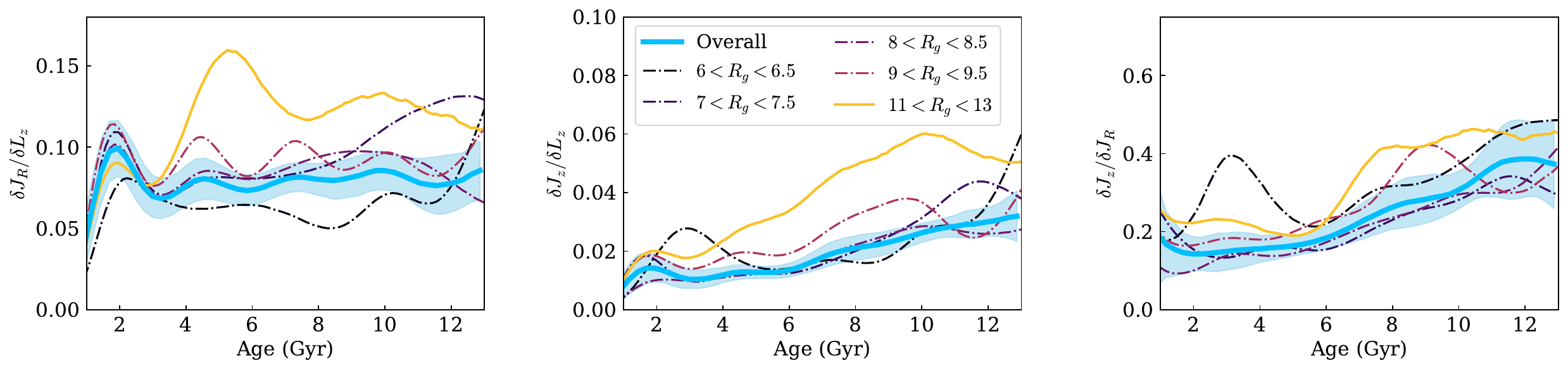}
    \caption{The ratio between the RMS changes in the three cylindrical actions. The lines are coloured according to the guiding radii, and the blue solid line and the shaded region is by marginalising these ratios over the stars in the sample in the solar vicinity. {\it Left:} the ratio between radial to azimuthal actions $\delta J_R/\delta L_z$. {\it Middle:} the ratio between vertical to azimuthal actions. {\it Right:} the ratio between vertical to radial actions. Overall, our results suggests the ratio of diffusions of the three actions are $\delta J_z:\delta J_R:\delta L_z\approx1:4:50$.}
    \label{fig:action_ratios}
\end{figure*}

To examine the goodness of fit of the model, we compare our best-fitting model to the original data distribution in the plane of age versus metallicity, $J_R$, and $J_z$, which together summarise all the observables used to constrain the model. To do so, we sample $[\rm Fe/H]$, $v_R$, and $v_z$ from Eq.~\ref{eqn::AMR} and Eq.~\ref{eqn::kine} using the best-fitting parameters at the given age, $R$, $R_g$, and $z$ of each observed star. Each observed star is oversampled ten times by drawing its observables from a Gaussian distribution centred on the observed value with a dispersion equal to the corresponding uncertainty. We further scatter the sampled $[\rm Fe/H]$, $v_R$, and $v_z$ values by the measurement uncertainty of the corresponding star. The samples of $J_R$ and $J_z$ are subsequently computed from the resulting sets of $(\bs x, \bs v)$. The comparison between the model and the data is shown in Fig.~\ref{fig:model_vs_data}, where the data contours are shown as solid lines and the model contours as dashed lines. Here, we only present the model-to-data distribution comparison for the guiding-radius bin of $9.5$--$10$~kpc. We refer the reader to Appendix~\ref{appendix:extra_figure} for the comparison across the full set of guiding-radius bins. The model and data show good overall consistency across all guiding radii. The slight discrepancies between the data and the model could be attributed to systematics in the model (e.g. non-exponential radial profile caused by bar quenching and the non-stochastic radial migration mechanism).

Changing the maximum metallicity curve, $[\rm Fe/H]_{\max}(\tau)$, could alter the fitted angular momentum dispersion (see Fig.~4 in \citealt{Frankel2020} for the degeneracy between $[\rm Fe/H]_{\max}$ and $\sigma_{L_z}$), such that increasing $[\rm Fe/H]_{\max}(\tau)$ decreases $\sigma_{L_z}$. To examine this effect, we repeat the analysis while shifting the $[\rm Fe/H]_{\max}$ 
curve up and down by approximately $\approx 0.15~\rm dex$, adjusting the enrichment rate in the young disc correspondingly to ensure that the present-day $[\rm Fe/H]_{\max}$ remains unchanged. We find that reasonable variations in 
$[\rm Fe/H]_{\max}$ do not affect the behaviour of $\sigma_{L_z}$, but instead systematically rescale $\sigma_{L_z}$ by $\sim15\%$--$25\%$ for a variation of $0.1$~dex in $[\rm Fe/H]_{\max}$. The value of $[\rm Fe/H]_{\max}$ also cannot be increased excessively; otherwise, the model yields an initial disc scale length longer than the present-day scale length of the high-$\alpha$, old disc. The choice of gravitational potential, as another possible source of systematic uncertainty, has only a limited 
effect on the model outputs, given that the Galactic potential is now well constrained (\citealt{McMillan2017, Hunter2024, Dillamore2025a}; see also the recent \citealt{Dillamore2025b}). Overall, we consider that the best-fitting model provides a good description of the observations and is robust against many potential sources of systematic uncertainty (see further discussion of model 
caveats in Section~\ref{section:limitation}).

\section{Implications for the Milky Way}\label{section:implication}

\begin{figure*}
    \centering
    \includegraphics[width=0.98\linewidth]{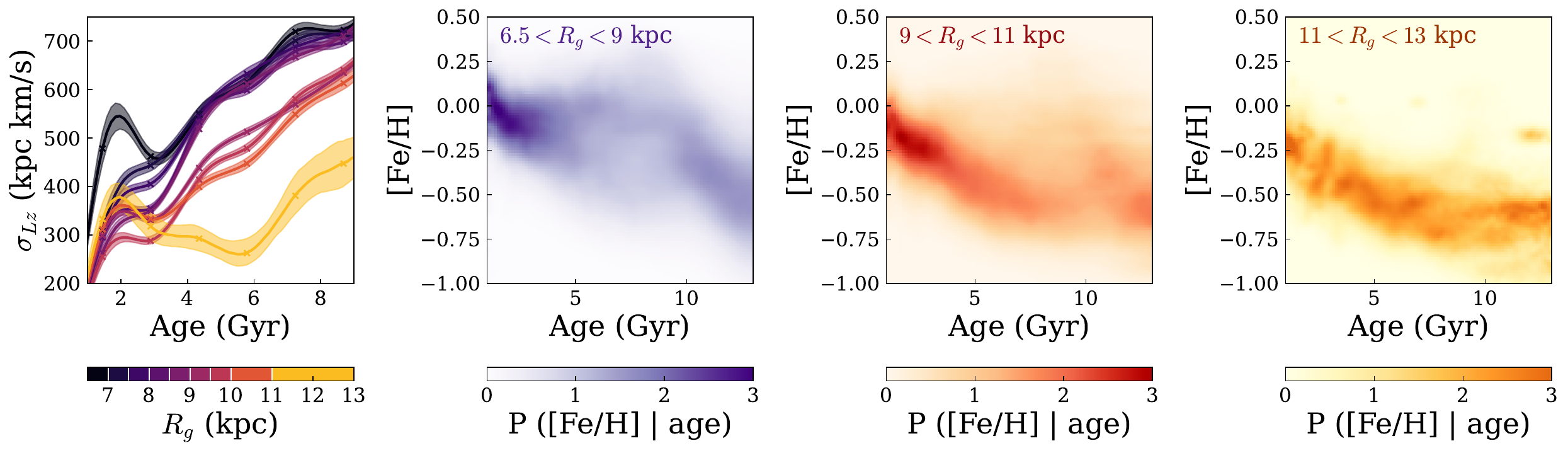}
    \caption{The three distinct radial regimes of radial migration in the low-$\alpha$ disc. {\it Left:} the same figure as the top right panel in Fig.~\ref{fig:fitting_actions}, but zoomed in to the low-$\alpha$ young disc ($\tau<9$~Gyr in the plot), where we see clearly three distinct radial groups have similar angular momentum diffusion coefficient within each group. {\it Middle left:} the observed column-normalised age-metallicity distribution of the first radial regime ($6.5<R_g/{\rm kpc}<9$), where a "$<$"-shaped structure appears in the low-$\alpha$ disc. {\it Middle right:} the column-normalised age-metallicity distribution of the second radial regime ($9<R_g/{\rm kpc}<11$), where the upper sequence in the previous regime disappears, and the metallicity distribution is similar to a Gaussian skewed to the metal-rich end at each ages. {\it Right:} the column-normalised age-metallicity distribution of the third, outermost radial regime ($R_g/{\rm kpc}>11$), where the age-metallicity sequences becomes much tighter.}
    \label{fig:SigmaLz_Rg_lowA}
\end{figure*}

We discuss the physical implications of the orbital diffusion measurement we showed in Section~\ref{section:results}. We link the features in $\sigma_{L_z}$, $\sigma_R$, and $\sigma_z$ measurements to the effect of the Galactic bar and the evolution of the gas content of the Milky Way. We also show our measurement of the Milky Way's heating-to-migration ratio in the context of the recent discussion about the coldness of the Milky Way's disc (see \citealt{Frankel2020, Hamilton2024, Sellwood_Binney2025, McCluskey2025}).

\subsection{Heating vs migration}\label{discussion::heating_churning}

We quantify the heating-to-migration ratio in the same way as in \citet{Frankel2020} and \citet{Hamilton2024}, but in this work, we extend it to three dimensions to map the diffusion of orbital action in cylindrical space $J_R$, $J_z$, $J_\phi$($L_z$). We compute the root mean square (RMS) of the change in actions (denoted $\delta J_i$) from $\sigma_{L_z}$, $\sigma_R$, and $\sigma_z$ using the derived equations in Sections~\ref{model::RM} and \ref{model::heating}, assuming stars are born on near-circular orbits. The ratios of the RMS of the three actions are shown in Fig.~\ref{fig:action_ratios}, coloured according to their guiding radii. The blue solid line illustrates the overall results by marginalising over the stars in the sample. We observe that the heating-to-migration ratio in the low-$\alpha$ young disc is close to a constant for both radial and vertical heating, with $\delta J_R/\delta L_z\approx0.075\pm0.022$ and $\delta J_z/\delta L_z\approx0.015\pm0.005$. The ratio of the three actions is $\delta J_z:\delta J_R:\delta L_z\approx1:4:50$. $\delta J_R/\delta L_z$ remains approximately constant with increasing age, and $\delta J_z/\delta L_z$ weakly increases with increasing age. The ratio of vertical-to-radial heating, shown in the right panel of Fig.~\ref{fig:action_ratios}, has a strong positive correlation with stellar age, implying a different power-law index of the age-velocity dispersion relation in the radial and vertical velocities. We also suggest not to overinterpret the ratio of the high-$\alpha$ disc because the underlying assumption that stars are born on near-circular orbits is more likely to be broken as the star-forming gas exhibits stronger perturbations at an early time \citep[e.g.][]{Brook2004, vanDonkelaar2022, BH2025}.

The radial heating-to-migration ratio we obtained is compatible with, but slightly lower than, the value in \citet{Frankel2020}. This ratio is particularly interesting because $\delta J_R/\delta L_z$ differs for different dynamical interactions. For example, migration driven by the corotation resonance alone experiences no radial heating \citep{Sellwood_Binney2002, Grand2012, Khoperskov2020}, but interactions with other resonances cause a change in $L_z$ and $J_R$ simultaneously \citep{Daniel2019}. If stars only experience an isotropic velocity kick, e.g., by interaction with giant molecular clouds or stellar clumps, we expect $\delta J_R/\delta L_z\approx2.5\sigma_R/V_{\rm circ}\approx0.3-0.6$
The Milky Way disc is much colder than this. Using test particle simulations with random transient spiral arms, \citet{Hamilton2024} argued that the observed $\delta J_R/\delta L_z$ ratio is much lower than the expectation from transient spiral arm-driven radial migration unless the spiral arms were loosely wound and concentrated around corotation and called for a revision of the radial migration mechanisms. The corotation resonance dragging of the decelerating bar could cause a kinematically cold migration \citep{Khoperskov2020, Chiba2021, Zhang2025} to solve this tension. However, \citet{Sellwood_Binney2025} subsequently examined this ratio in reruns of the N-body simulation originally presented in \citet{Sellwood_Binney2002} and reported that the observed low $\delta J_R/\delta L_z$ ratio is compatible with the self-consistently produced radial migration by the transient spiral arms in their simulations, but there is no room for extra heating by giant molecular clouds and mergers. \citet{McCluskey2025} also found that the Milky Way is kinematically colder than the galaxies in the Local Group and in cosmological simulations. In a parallel study (Zhāng et al. in prep.), we analyse the Nexus simulation \citep{TG24} to investigate the behaviour of radial migration as a function of the disc's gas mass fraction. The values $\delta J_R/\delta L_z\approx0.075$ and $\delta J_z/\delta L_z\approx0.015$ are compatible with the gas-poor galaxy in the simulations, for which the ratios become larger in a more gas-rich disc. The heating-to-migration ratio of the Milky Way is compatible with a gas-poor galaxy with a gas fraction $<20\%$.

There are no clear or monotonic radial trends in the heating-to-migration ratio for both radial and vertical heating, except for the outstandingly high ratios in the outermost radial bins in our analysis, where $\delta J_R/\delta L_z\approx0.12-0.15$ for stars with $R_g>11$~kpc. This could be due to the effect of the outer Lindblad resonance (OLR) barrier \citep{Halle2018, Khoperskov2020}, for which bar-driven migration diminishes around the OLR, and radial migration outside the OLR is dominated by perturbations other than the bar.

\subsection{Role of bar: the three distinct regimes in the low-$\mathrm{\alpha}$ disc}\label{discussion::radial_regimes}

As shown in the left panels of Fig.~\ref{fig:SigmaLz_Rg_lowA}, we observe three distinct radial regimes in the radial migration signatures within the young, low-$\alpha$ disc. The fitted angular momentum diffusion coefficients are similar in the guiding radii ranges of: 1) $6.5-9$~kpc in the black-purple lines, 2) $9-11$~kpc in the purple-red lines and 3) $\geq 11$~kpc in the orange line. We argue that this is a physical result rather than model overfitting, as the age-metallicity distributions of these radial regimes are qualitatively different, as illustrated in the three right panels of Fig.~\ref{fig:SigmaLz_Rg_lowA} in the corresponding colours of the lines. The effect of the Galactic bar on radial migration is a natural way to create these distinct behaviours. Even though our model does not incorporate the Galactic bar-driven radial migration directly, the high flexibility of the model allows us to capture the effect of the bar to some extent. 
\begin{itemize}
    \item[1)] The age-metallicity distribution in $6.5<R_g \, [\rm kpc]<9$ exhibits a "V-shape" for $\rm age\lesssim9$~Gyr, similar to the overall age-metallicity plane in Fig.~\ref{fig:AMR_allstars} (see a clearer illustration in \citealt{Xiang_Rix2022}). \citet{Zhang2025} interpreted this bimodal sequence as a consequence of the co-existence of stars migrating via two different mechanisms: a) bar corotation resonance dragging and b) random angular momentum scattering by transient spiral arms. In the former case, stars born around the corotation radius of the Galactic bar are trapped at the corotation resonance. With the deceleration of the Galactic bar, the corotation radius then expands with time, bringing these stars, born in the inner Galaxy and trapped at early times, outwards to be around the present corotation radius \citep{Khoperskov2020, Chiba2021, Zhang2025}. Measurements of the current pattern speed converge to $32-35$~km/s/kpc \citep{Hunt2025} through direct observation \citep{Sanders2019, ClarkeGerhard_2022, Zhang2024_patternspeed}, disc resonance \citep{Binney2020, Kawata_2021, Chiba2021}, and halo resonances \citep{Dillamore2024, Dillamore2025a}, which places the corotation radius $R_{\rm CR}\sim6.5-7.5$~kpc, and the OLR radius $R_{\rm OLR}\sim11-12$~kpc. Therefore, finding these inner Galaxy migrators around $6.5<R_g \, [\rm kpc]<9$ is expected. We argued in \citet{Zhang2025} that these stars lie on the upper sequence of the age-metallicity plane. The lower age-metallicity sequence comprises stars born locally around their birth radius and experienced radial migration due to interaction with transient spiral arms (e.g., \citealt{Roskar2008}).
    \\
    \item[2)] The angular momentum diffusion coefficients between $R_g\sim9-11$~kpc are grouped together, and their age-metallicity plane is demonstrated in the third panel of Fig.~\ref{fig:SigmaLz_Rg_lowA}. Compared with the previous regime, we notice that the bimodal sequences do not exist here, as the upper sequence is absent. This could be due to the bar corotation resonance dragging stars being contained within $\sim9$~kpc (corotation radius $+$ libration oscillation amplitude). The shape of the age-metallicity plane of the low-$\alpha$ disc in this regime is similar to a skewed Gaussian, which is compatible with the signature of radial migration driven by transient spiral arms. This radial migration mechanism is well described by the angular momentum diffusion equation adopted in this work. The relatively low radial migration efficiency in this regime suggests that the stars currently at these radii are not yet fully radially mixed. Radial migration broadened the metallicity distribution at each age, similar to a Gaussian but skewed towards the metal-rich end, but the peak of the distribution still reflects the local metallicity enrichment around $R\sim10$~kpc (e.g. \citealt{Hayden2015}).
    \\
    \item[3)] We observe a sharp drop in radial migration efficiency for stars currently outside $11$~kpc. The age-metallicity distribution of these stars is shown in the right panel of Fig.~\ref{fig:SigmaLz_Rg_lowA}. The age-metallicity distribution in this radial regime is much tighter than those in the two inner bins. As radial migration is inefficient here, this sequence traces the metallicity evolution at this local bin. This inefficient radial migration could be a consequence of the OLR barrier \citep{Halle2015, Halle2018, Khoperskov2020}, where the present OLR radius is $\sim11$~kpc (given bar pattern speed $32-34$~km/s/kpc). Stars born inside the OLR are likely to be retained inside the OLR for their orbital lifetime, so radial migration outside the OLR becomes inefficient, as stars born inside cannot migrate out, thereby forbidding angular momentum exchange with the inner Galaxy. This drop in angular momentum efficiency provides direct proof of the existence of the OLR barrier, which has already been suggested by the change in the chemical properties of the Galactic disc \citep{Haywood2024}. However, see also the analysis in \citet{Lian2022}.

\end{itemize}

\subsection{Transition phase between the young and old disc}\label{discussion::transition}
Another interesting feature in the $\sigma_{L_z}(\tau)$ fitting in Fig.~\ref{fig:fitting_actions} is the transition of radial migration efficiency around $\sim8-10$~Gyr ago, for which stars older than $\sim10$~Gyr witnessed a higher radial migration efficiency than younger stars, as shown in the left panel of Fig.~\ref{fig:fitting_actions}. By marginalising over the sample and taking the average of the time derivative of $\sigma_{L_z}$ over a window with a width of 1.5~Gyr, we find $\langle dR_g/dt\rangle\approx0.49\pm0.08$~kpc/Gyr for $\tau<8$~Gyr and $\langle dR_g/dt\rangle\approx0.92\pm0.11$~kpc/Gyr for $\tau>10$~Gyr. (It is noteworthy that the results in this paper are effective solutions that fit the present-day state of the Galactic disc, which does not necessarily describe the evolution of the Galactic disc, so these numbers should be taken with caution.) There are two ways to interpret this transition: 1) a high bar deceleration rate during the bar formation epoch (see \citealt{Haywood2024}); or 2) a transition of the gas content in the Galaxy (see Zhāng et al. in prep). The GS/E merger \citep{Belokurov2018, Helmi2018} could also induce some disc reshuffling, but the mass ratio of the Milky Way and the GS/E progenitor is unlikely to produce such a large effect (but see also \citealt{Lane2025} for recent updates on estimated GS/E's mass).

\begin{itemize}
    \item[1)] \citet{Haywood2024} demonstrated, using a hydrodynamical simulation, that there is an episode of outward radial migration during the onset of bar formation in that galaxy. They argued that the violent outward migration episode is driven by the early deceleration of the Galactic bar, where the bar decelerates from a high initial pattern speed of $\gtrsim100$~km/s/kpc at the moment that the quadrupole strength starts to increase to a final pattern speed of $\sim40$~km/s/kpc when the bar reaches its maximum strength. Many stars born inside the corotation radius $R_{\rm ini}<R_{\rm CR}(t)$ are dragged by the expansion of the corotation radius, while other stars are less affected. Stars experiencing this outward migration episode have their final destiny in a range from the innermost part of the Galaxy out to the OLR radius. Dating the star formation epoch in the nuclear stellar disc using Mira variables, \citet{Sanders2024} estimated the bar formation epoch of the Milky Way to be $\sim8-10$~Gyr, compatible with the time of transition of $\sigma_{L_z}$ as observed in this work. This implies that the Milky Way bar might experience an exceptionally strong deceleration episode during the bar formation epoch, instead of a constant pattern speed during bar formation that was conventionally adopted in previous models \citep[e.g.,][]{ChibaSchonrich_2021, Dillamore2023, Zhang2025}.
    \\
    \item[2)] In the Nexus simulation \citep{TG24}, it is demonstrated that the behaviour of radial migration (or orbital diffusion) varies as a function of the gas mass fraction ($f_{\rm gas}$) of the galaxies. A galaxy with a high gas fraction shows high efficiency of radial migration and simultaneously a higher heating-to-migration ratio. This is because stars in gas-rich galaxies experience stronger density perturbations with more clumpiness, mainly supported by the active gas in the discs. The angular momentum diffusion rate in the most gas-rich galaxy in the Nexus simulation suite, $f_{\rm gas}=80\%$, is $\approx2$ times greater than that in its gas-poor counterpart, $f_{\rm gas}=20\%$ (Zhāng et al. in prep.). Therefore, the transition phase of $\sigma_{L_z}$ in our model could also represent a transition from a gas-rich to a gas-poor disc when the low-$\alpha$ disc formed (see also relevant evidence and discussion of the gas disc shrinking in \citealt{Funakoshi2025}). Similar correlation between the radial migration strength and galaxies' gas fraction is also seen in galaxies in the TNG50 simulation suite, for which we present two examples in Appendix~\ref{appendix:TNG50}. 
    However, no obvious increase in the $\delta J_R/\delta L_z$ ratio is observed in our results, so it is unlikely that we can explain all results solely by the gas transition. 
    A careful description of the radial migration in Nexus simulations is beyond the scope of this work but is analysed by Zhāng et al. (in prep) in detail. We present some relevant Nexus results in Appendix~\ref{appendix:Nexus}.
\end{itemize}

It is likely that both mechanisms contributed together to the high radial mixing observed in the high-$\alpha$ old disc. Other simulation studies also support this, as the gas fraction of a galaxy also affects the formation and survival of the galactic bar and its deceleration rate \citep{Beane2023, BH2024}. \citet{BH2024} found that the galactic bar can form and survive more easily in a gas-poor disc than in a gas-rich disc (but see also counter-evidence in cosmological simulations \citealt{Fragkoudi2025}), and the galactic bar in a gas-poor disc decelerates faster \citep{Beane2023}. Therefore, these two factors are not mutually exclusive but are correlated with each other potentially.


Recent JWST observations revealed that single-disc systems at high redshift are more prevalently thick and compatible with the thick disc in the bimodal-disc system \citep{Tsukui2025}. This supports the bimodal disc formation scenario in which the old thick disc formed thick because of the high gas turbulence, and thin disc formation begins when the galaxy becomes gas-poor \citep{Brook2004, Genzel2011, Wisnioski2015, Yu2021}. This scenario is also compatible with our results, for which the gas-rich environments of the early Milky Way disc produced the high radial migration efficiency in the high-$\alpha$, old disc while the radial migration efficiency decreased as the galaxy becomes gas-poor and less turbulent when the low-$\alpha$ disc begins to form.

\section{Further discussions and model limitations}\label{section:limitation}

In this section, we discuss the systematic caveats of the model and examine the robustness of our results. 

\subsection{Azimuthal metallicity dispersion} \label{limit::azimuthal}

The azimuthal variation of the star-forming gaseous metallicity ($\sigma_{\rm [Fe/H]}$ in equation~(\ref{eqn:met})) is degenerate with the angular momentum diffusion coefficient because both can broaden the metallicity dispersion at a fixed age and guiding radius. In our implementation above, we fixed the metallicity dispersion to be $\sigma_{[\rm Fe/H]}=0.05$~dex, which is around the metallicity dispersion of H$_{\rm II}$ region of the Milky Way \citep{Wenger2019}, and a typical value of the azimuthal metallicity dispersion in external galaxies \citep{Kreckel2019}. However, the azimuthal metallicity dispersion in the past when the galaxy was more gas-rich could be different than today. 

\begin{figure}
    \centering
    \includegraphics[width=0.98\linewidth]{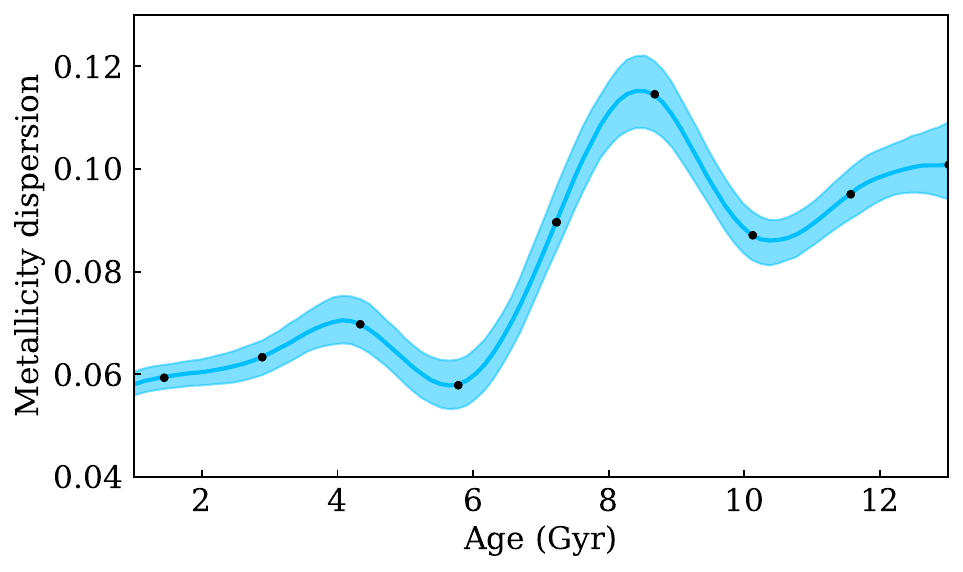}
    \caption{The recovered azimuthal metallicity dispersion as a function of the lookback time, assuming the radial heating-to-migration ratio remains a constant at all time $\delta J_R/\delta L_z\approx0.1$ \citep{Frankel2020}. Under such assumption, a transition between the azimuthal metallicity dispersion around the low and high-$\alpha$ disc transition is needed to explain the observation.}
    \label{fig:metallicity_dispersion}
\end{figure}

To examine this, we slightly change the model by fixing $\sigma_{L_z}(\tau, R_g)$ and fitting for $\sigma_{[\rm Fe/H]}(\tau)$ instead. For this exercise only, we choose a constant $\delta J_R/\delta L_z=0.1$ for stars of all ages and guiding radii \citep{Frankel2020}. Therefore, we can obtain the $\sigma_{L_z}(\tau, R_g)$ values purely from fitting the radial velocity distribution in the Galaxy by optimising the log-likelihood in equation~(\ref{eqn::kine}) without using information from the age-metallicity plane. We then also fix $R_{\rm d, 0}(\tau)$ and $\nabla[\rm Fe/H]$, which we obtained above in Fig.~\ref{fig:fitting_RdMHgrad}, leaving $\sigma_{[\rm Fe/H]}(\tau)$ as the only variable in the log-likelihood in equation~(\ref{eqn::AMR}). We sample the posterior of $\sigma_{[\rm Fe/H]}(\tau)$ and the result is shown in Fig.~\ref{fig:metallicity_dispersion}. The results also find two distinct phases, in which the metallicity dispersion for $\tau<6$~Gyr is $\sigma_{[\rm Fe/H]}\approx0.06$~dex and for $\tau>8$~Gyr is $\sigma_{[\rm Fe/H]}\approx0.10$~dex. The transition time is also similar to the transition time from the high-$\alpha$ to low-$\alpha$ disc, but slightly later. This suggests that we either need a high migration efficiency at an early time, or a high metallicity dispersion at an early time, both in line with the interpretation in this work that there could be a transition of gas content in the Milky Way when the low-$\alpha$ disc starts to form. 

\subsection{Missing of the bar radial transportation}\label{limit::Bar_RM}

Radial migration history of the Milky Way is modelled in this work by the 1D angular momentum diffusion equation \citep{Sanders_Binney2015}. It can describe stochastic random scattering of angular momentum, which is applicable for modelling the dynamical interaction between stars and transient spiral arms or density clumps. However, this model suffers from incorporation of bar-driven radial migration because it assigns an equal probability for a star to migrate inward or outward, but the Galactic bar might particularly favour the outward migration. For example, stars trapped by the bar corotation resonance at an early time will be transported outward by resonance dragging of the bar deceleration for violent deceleration during bar formation \citep{Haywood2024} or gradual deceleration after bar formation \citep{Khoperskov2020, Zhang2025}. Also, \citet{Marques2025} illustrated that stars at the tips of the bar would be transported outward when there are bar-spiral spatial overlaps. 

\citet{Zhang2025} demonstrated using a test particle with a decelerating bar that the stars transported outward by the expansion of the bar corotation resonance can create the "V-shaped" age-metallicity sequences as a signature we observe around the present corotation radius in the second panel of Fig.~\ref{fig:SigmaLz_Rg_lowA}. This bimodal "V-shaped" age-metallicity sequence cannot be easily fitted with the angular momentum diffusion kernel adopted in this work; instead, the model can only describe the unimodal metallicity distribution. Therefore, our model cannot be used as a perfect description of radial migration around bar resonances, but it finds an effective $\sigma_{L_z, {\rm eff}}$ that accounts for the change in angular momentum driven both by bar and transient perturbations. Building an accurate statistical description of the bar-driven radial migration is beyond the scope of this paper, and we leave it for future work.

\subsection{Selection effects}\label{limit::SF}

In this work, we model the distribution of metallicity conditioned on the observables that are subject to potential selection bias, e.g. $\tau$ and $R_g$, when fitting the angular momentum diffusion. However, the radial migration model that we adopted only describes the changes in the guiding radii of the stars, not their Galactocentric radius $R$. Hence, our $\sigma_{L_z}$ fitting would still be biased if there is an extra radial selection function on top of the guiding radius selection effect, which means that the orbital sets are not complete in the guiding radii bins. Incomplete orbital sets would bias the shape of the age-metallicity plane by favouring eccentric, non-locally born stars when the guiding radii bin is far away from the observation footprint. Therefore, we exclude the guiding radii bins inside 6~kpc, and outside 13~kpc, which are strongly affected by this selection effect. 

One way of solving this selection bias is to use the orbital superposition method to reconstruct the disc \citep{Khoperskov2025a, Khoperskov2025b}, and apply the model to the reconstructed age-metallicity plane. Recently, \citet{Khoperskov2025b} applied this method to APOGEE DR17 spectroscopic sample with spectroscopic ages, but the quality of the spectroscopic ages would make this radial migration analysis difficult. Repeating the analysis but on a subgiant isochronal ages sample would be beneficial to the ultimate goal of constructing a global model of the radial migration history in the Milky Way's disc. In the future, it is also worth including both the radial and vertical distributions of stars into the radial migration model, not only due to this selection effect, but also because the vertical metallicity gradient could provide further constraints on the radial migration history of the Galaxy \citep{Kawata2017}. 

\section{Conclusions}\label{section:conclusion}

We build a chrono-chemo-dynamical model for the Milky Way's discs over $13~\rm Gyr$ since disc formation, thereby describing the metallicity and secular evolution history of the Galactic disc. Similarly to \citet{Frankel2020}, the model involves inside-out growth, metallicity evolution, radial migration, and dynamical heating. All parameters in the model $\Theta=\{R_{d,0}(\tau), \nabla[{\rm Fe/H}](\tau), \sigma_{L_z}(\tau, R_g), \sigma_R(\tau, R_g), \sigma_z(\tau, R_g)\}$ are fitted non-parametrically with a cubic spline to fully explore the parameter space, allowing us to uncover the evolution history of the disc. The fitted parameters are presented in Fig.~\ref{fig:fitting_actions}and~\ref{fig:fitting_RdMHgrad}. We summarise the main results below:

\begin{itemize}
    \item[1)] As shown in Fig.~\ref{fig:fitting_actions}, the angular momentum diffusion coefficient gradually increases with age, and then there is a transition epoch of radial migration efficiency around $8-10$~Gyr ago, coinciding with the onset of the low-$\alpha$ thin disc. According to our model, the stars in the high-$\alpha$ old disc experienced a high radial migration efficiency with $\langle dR_g/dt\rangle\approx0.92\pm0.11$~kpc/Gyr, which then decreased to $\langle dR_g/dt\rangle\approx0.49\pm0.09$~kpc/Gyr for the low-$\alpha$ young disc. The high radial migration efficiency for the ancient, high-$\alpha$ disc implies that the high-$\alpha$ disc is fully radially mixed, which then explains the flat (or even positive) metallicity gradient there. This transition from a high to low radial migration efficiency could be caused by two distinct but interconnected mechanisms: a) the transition from a gas-rich to a gas-poor disc (Zhāng et al. in prep.); b) the onset of Galactic bar formation with a high initial pattern speed and the accompanied outward migration episode \citep{Haywood2024}. We conclude that both mechanisms are possible, and it is likely that both contributed to this transition.
    \\
    \item[2)] We observe a decline in radial migration efficiency with increasing guiding radius as shown in Fig.~\ref{fig:SigmaLz_Rg_lowA}. Interestingly, we detected three radial regimes of radial migration in the low-$\alpha$ disc with different radial migration histories. We interpret this as the effect of the Galactic bar. a) In the first regime, around the corotation radius, $6.5<R_g/{\rm kpc}<9$, radial migration is contributed both by the corotation resonance dragging of the decelerating bar and the interaction with transient spiral arms, leading to a "V-shaped" age-metallicity sequence. b) In the second regime, between the corotation and OLR radius, $9<R_g/{\rm kpc}<11$, the age-metallicity signature is consistent with migration driven by transient spiral arms. c) In the outermost radii, $R_g/{\rm kpc}\geq11$, a sharp decrease in radial migration efficiency is seen, reflecting the OLR radial migration barrier. These results emphasise the impact of the Galactic bar on the secular evolution history of the Galactic disc from the inner Galaxy out to the OLR radius.
    \\
    \item[3)] We map the diffusion in the 3D cylindrical action space. Our model suggests an approximately constant heating-to-migration ratio in terms of $\delta J_{R,z}/\delta L_z$ across all ages, with $\delta J_R/\delta L_z\approx0.075\pm0.022$ and $\delta J_z/\delta L_z\approx0.015\pm0.005$. The overall diffusion ratios of the three actions are highly anisotropic $\delta J_z:\delta J_R:\delta L_z\approx1:4:50$. There are no evident radial trends in $\delta J_{R,z}/\delta L_z$, except that the ratios are exceedingly high for the outermost radii $R_g>11$~kpc. This is consistent with simulation results that the heating-to-migration ratio becomes higher outside the OLR barrier. Comparing to the Nexus simulation, the measured low heating-to-migration ratios are consistent with a kinematically cold, gas-poor disc (Zhāng et al. in prep.). 
    \\
    \item[4)] We also produce a metallicity evolution model with high fidelity. This metallicity model could be used for weak chemical tagging, similar to the model in \citet{Lu2024, Ratcliffe2024}. The model predicts that the birth radius of the Sun is $R_{\odot, 0}\sim5.7$~kpc, which means that the Sun had to migrate for more than 2~kpc to its present location. It is difficult to determine the exact path of the radial migration, but an interesting possibility is that it migrated with the expansion of the bar corotation radius based on its location in the age-metallicity plane (see also the discussion in \citealt{Baba2024}).
\end{itemize}

Finally, in an interesting connection with high-redshift observations, our results are in line with and add further evidence to the bimodal disc formation scenario suggested in \citet{Tsukui2025} by the structural decomposition of high-redshift galaxies. The transition of radial migration efficiency and the heating-to-migration ratio all suggest a potential transition from a gas-rich to a gas-poor disc around the time of the low-$\alpha$ thin disc formation. These results, combined with recent studies of stellar birth kinematics (e.g. \citealt{Yu2021, vanDonkelaar2022, BH2025}), are compatible with the scenario in which the high-$\alpha$ thick disc was formed thick and the thin disc thinner because of the drop in the gas fraction. The radial extent of the thick disc today is much larger than that at birth due to the outward radial migration driven by the decelerating bar during and after bar formation.


\section*{Acknowledgements}

We thank the inspiring discussion with Ivan Minchev, Joss Bland-Hawthorn, and Zhao-Yu Li. We also appreciate the comments from Takafumi Tsukui on the draft of this paper.

HZ thanks the Science and Technology Facilities Council (STFC) for a PhD studentship (grant number 2888170). 
JLS acknowledge support from the Royal Society (URF\textbackslash R1\textbackslash191555).
VB and NWE acknowledge support from the Leverhulme Research Project Grant RPG-2021-205: "The Faint Universe Made Visible with Machine Learning". 
DC acknowledges funding from the Harding Distinguished Postgraduate Scholars Program.
SGK acknowledges PhD funding from the Marshall Scholarship, supported by the UK government and Trinity College, Cambridge. SK acknowledges support from the Science \& Technology Facilities Council (STFC) grant ST/Y001001/1.
DK and NF acknowledge MWGaiaDN, a Horizon Europe Marie Sk\l{}odowska-Curie Actions Doctoral Network funded under grant agreement no. 101072454 and also funded by UK Research and Innovation (EP/X031756/1). N. Frankel acknowledges the support of the Natural Sciences and Engineering Research Council of Canada (NSERC), [funding reference number 568580] through a CITA postdoctoral fellowship, and acknowledges partial support from an Arts \& Sciences Postdoctoral Fellowship at the University of Toronto.


\section*{Data availability}

The data used in this paper are all publicly available. The fitted chrono-chemo-dynamical model (results shown in Fig.~\ref{fig:fitting_actions}, \ref{fig:fitting_RdMHgrad}, and \ref{fig:metallicity_model}) is available at zenodo: 10.5281/zenodo.17882469



\bibliographystyle{mnras}
\bibliography{bibliography} 

@ARTICLE{Xiang_Rix2025,
       author = {{Xiang}, Maosheng and {Rix}, Hans-Walter and {Yang}, Hang and {Liu}, Jifeng and {Huang}, Yang and {Frankel}, Neige},
        title = "{The formation and survival of the Milky Way's oldest stellar disk}",
      journal = {Nature Astronomy},
         year = 2025,
        month = jan,
       volume = {9},
        pages = {101-110},
          doi = {10.1038/s41550-024-02382-w},
archivePrefix = {arXiv},
       eprint = {2410.09705},
 primaryClass = {astro-ph.GA},
       adsurl = {https://ui.adsabs.harvard.edu/abs/2025NatAs...9..101X}
}

@ARTICLE{Zhao2012_LAMOST,
       author = {{Zhao}, Gang and {Zhao}, Yong-Heng and {Chu}, Yao-Quan and {Jing}, Yi-Peng and {Deng}, Li-Cai},
        title = "{LAMOST spectral survey {\textemdash} An overview}",
      journal = {Research in Astronomy and Astrophysics},
         year = 2012,
        month = jul,
       volume = {12},
       number = {7},
        pages = {723-734},
          doi = {10.1088/1674-4527/12/7/002},
       adsurl = {https://ui.adsabs.harvard.edu/abs/2012RAA....12..723Z}
}

@ARTICLE{Xiang_Rix2022,
       author = {{Xiang}, Maosheng and {Rix}, Hans-Walter},
        title = "{A time-resolved picture of our Milky Way's early formation history}",
      journal = {\nat},
         year = 2022,
        month = mar,
       volume = {603},
       number = {7902},
        pages = {599-603},
          doi = {10.1038/s41586-022-04496-5},
archivePrefix = {arXiv},
       eprint = {2203.12110},
 primaryClass = {astro-ph.GA},
       adsurl = {https://ui.adsabs.harvard.edu/abs/2022Natur.603..599X}
}

@ARTICLE{BJ2021,
       author = {{Bailer-Jones}, C.~A.~L. and {Rybizki}, J. and {Fouesneau}, M. and {Demleitner}, M. and {Andrae}, R.},
        title = "{Estimating Distances from Parallaxes. V. Geometric and Photogeometric Distances to 1.47 Billion Stars in Gaia Early Data Release 3}",
      journal = {\aj},
         year = 2021,
        month = mar,
       volume = {161},
       number = {3},
          eid = {147},
        pages = {147},
          doi = {10.3847/1538-3881/abd806},
archivePrefix = {arXiv},
       eprint = {2012.05220},
 primaryClass = {astro-ph.SR},
       adsurl = {https://ui.adsabs.harvard.edu/abs/2021AJ....161..147B}
}

@ARTICLE{Gaia_DR3,
       author = {{Gaia Collaboration} and {Vallenari}, A. and {Brown}, A.~G.~A. and {Prusti}, T. and {de Bruijne}, J.~H.~J. and {Arenou}, F. and {Babusiaux}, C. and {Biermann}, M. and {Creevey}, O.~L. and {Ducourant}, C. and {Evans}, D.~W. and {Eyer}, L. and {Guerra}, R. and {Hutton}, A. and {Jordi}, C. and {Klioner}, S.~A. and {Lammers}, U.~L. and {Lindegren}, L. and {Luri}, X. and {Mignard}, F. and {Panem}, C. and {Pourbaix}, D. and {Randich}, S. and {Sartoretti}, P. and {Soubiran}, C. and {Tanga}, P. and {Walton}, N.~A. and {Bailer-Jones}, C.~A.~L. and {Bastian}, U. and {Drimmel}, R. and {Jansen}, F. and {Katz}, D. and {Lattanzi}, M.~G. and {van Leeuwen}, F. and {Bakker}, J. and {Cacciari}, C. and {Casta{\~n}eda}, J. and {De Angeli}, F. and {Fabricius}, C. and {Fouesneau}, M. and {Fr{\'e}mat}, Y. and {Galluccio}, L. and {Guerrier}, A. and {Heiter}, U. and {Masana}, E. and {Messineo}, R. and {Mowlavi}, N. and {Nicolas}, C. and {Nienartowicz}, K. and {Pailler}, F. and {Panuzzo}, P. and {Riclet}, F. and {Roux}, W. and {Seabroke}, G.~M. and {Sordo{\o}rcit}, R. and {Th{\'e}venin}, F. and {Gracia-Abril}, G. and {Portell}, J. and {Teyssier}, D. and {Altmann}, M. and {Andrae}, R. and {Audard}, M. and {Bellas-Velidis}, I. and {Benson}, K. and {Berthier}, J. and {Blomme}, R. and {Burgess}, P.~W. and {Busonero}, D. and {Busso}, G. and {C{\'a}novas}, H. and {Carry}, B. and {Cellino}, A. and {Cheek}, N. and {Clementini}, G. and {Damerdji}, Y. and {Davidson}, M. and {de Teodoro}, P. and {Nu{\~n}ez Campos}, M. and {Delchambre}, L. and {Dell'Oro}, A. and {Esquej}, P. and {Fern{\'a}ndez-Hern{\'a}ndez}, J. and {Fraile}, E. and {Garabato}, D. and {Garc{\'\i}a-Lario}, P. and {Gosset}, E. and {Haigron}, R. and {Halbwachs}, J. -L. and {Hambly}, N.~C. and {Harrison}, D.~L. and {Hern{\'a}ndez}, J. and {Hestroffer}, D. and {Hodgkin}, S.~T. and {Holl}, B. and {Jan{\ss}en}, K. and {Jevardat de Fombelle}, G. and {Jordan}, S. and {Krone-Martins}, A. and {Lanzafame}, A.~C. and {L{\"o}ffler}, W. and {Marchal}, O. and {Marrese}, P.~M. and {Moitinho}, A. and {Muinonen}, K. and {Osborne}, P. and {Pancino}, E. and {Pauwels}, T. and {Recio-Blanco}, A. and {Reyl{\'e}}, C. and {Riello}, M. and {Rimoldini}, L. and {Roegiers}, T. and {Rybizki}, J. and {Sarro}, L.~M. and {Siopis}, C. and {Smith}, M. and {Sozzetti}, A. and {Utrilla}, E. and {van Leeuwen}, M. and {Abbas}, U. and {{\'A}brah{\'a}m}, P. and {Abreu Aramburu}, A. and {Aerts}, C. and {Aguado}, J.~J. and {Ajaj}, M. and {Aldea-Montero}, F. and {Altavilla}, G. and {{\'A}lvarez}, M.~A. and {Alves}, J. and {Anders}, F. and {Anderson}, R.~I. and {Anglada Varela}, E. and {Antoja}, T. and {Baines}, D. and {Baker}, S.~G. and {Balaguer-N{\'u}{\~n}ez}, L. and {Balbinot}, E. and {Balog}, Z. and {Barache}, C. and {Barbato}, D. and {Barros}, M. and {Barstow}, M.~A. and {Bartolom{\'e}}, S. and {Bassilana}, J. -L. and {Bauchet}, N. and {Becciani}, U. and {Bellazzini}, M. and {Berihuete}, A. and {Bernet}, M. and {Bertone}, S. and {Bianchi}, L. and {Binnenfeld}, A. and {Blanco-Cuaresma}, S. and {Blazere}, A. and {Boch}, T. and {Bombrun}, A. and {Bossini}, D. and {Bouquillon}, S. and {Bragaglia}, A. and {Bramante}, L. and {Breedt}, E. and {Bressan}, A. and {Brouillet}, N. and {Brugaletta}, E. and {Bucciarelli}, B. and {Burlacu}, A. and {Butkevich}, A.~G. and {Buzzi}, R. and {Caffau}, E. and {Cancelliere}, R. and {Cantat-Gaudin}, T. and {Carballo}, R. and {Carlucci}, T. and {Carnerero}, M.~I. and {Carrasco}, J.~M. and {Casamiquela}, L. and {Castellani}, M. and {Castro-Ginard}, A. and {Chaoul}, L. and {Charlot}, P. and {Chemin}, L. and {Chiaramida}, V. and {Chiavassa}, A. and {Chornay}, N. and {Comoretto}, G. and {Contursi}, G. and {Cooper}, W.~J. and {Cornez}, T. and {Cowell}, S. and {Crifo}, F. and {Cropper}, M. and {Crosta}, M. and {Crowley}, C. and {Dafonte}, C. and {Dapergolas}, A. and {David}, M. and {David}, P. and {de Laverny}, P. and {De Luise}, F. and {De March}, R. and {De Ridder}, J. and {de Souza}, R. and {de Torres}, A. and {del Peloso}, E.~F. and {del Pozo}, E. and {Delbo}, M. and {Delgado}, A. and {Delisle}, J. -B. and {Demouchy}, C. and {Dharmawardena}, T.~E. and {Di Matteo}, P. and {Diakite}, S. and {Diener}, C. and {Distefano}, E. and {Dolding}, C. and {Edvardsson}, B. and {Enke}, H. and {Fabre}, C. and {Fabrizio}, M. and {Faigler}, S. and {Fedorets}, G. and {Fernique}, P. and {Fienga}, A. and {Figueras}, F. and {Fournier}, Y. and {Fouron}, C. and {Fragkoudi}, F. and {Gai}, M. and {Garcia-Gutierrez}, A. and {Garcia-Reinaldos}, M. and {Garc{\'\i}a-Torres}, M. and {Garofalo}, A. and {Gavel}, A. and {Gavras}, P. and {Gerlach}, E. and {Geyer}, R. and {Giacobbe}, P. and {Gilmore}, G. and {Girona}, S. and {Giuffrida}, G. and {Gomel}, R. and {Gomez}, A. and {Gonz{\'a}lez-N{\'u}{\~n}ez}, J. and {Gonz{\'a}lez-Santamar{\'\i}a}, I. and {Gonz{\'a}lez-Vidal}, J.~J. and {Granvik}, M. and {Guillout}, P. and {Guiraud}, J. and {Guti{\'e}rrez-S{\'a}nchez}, R. and {Guy}, L.~P. and {Hatzidimitriou}, D. and {Hauser}, M. and {Haywood}, M. and {Helmer}, A. and {Helmi}, A. and {Sarmiento}, M.~H. and {Hidalgo}, S.~L. and {Hilger}, T. and {H{\l}adczuk}, N. and {Hobbs}, D. and {Holland}, G. and {Huckle}, H.~E. and {Jardine}, K. and {Jasniewicz}, G. and {Jean-Antoine Piccolo}, A. and {Jim{\'e}nez-Arranz}, {\'O}. and {Jorissen}, A. and {Juaristi Campillo}, J. and {Julbe}, F. and {Karbevska}, L. and {Kervella}, P. and {Khanna}, S. and {Kontizas}, M. and {Kordopatis}, G. and {Korn}, A.~J. and {K{\'o}sp{\'a}l}, {\'A} and {Kostrzewa-Rutkowska}, Z. and {Kruszy{\'n}ska}, K. and {Kun}, M. and {Laizeau}, P. and {Lambert}, S. and {Lanza}, A.~F. and {Lasne}, Y. and {Le Campion}, J. -F. and {Lebreton}, Y. and {Lebzelter}, T. and {Leccia}, S. and {Leclerc}, N. and {Lecoeur-Taibi}, I. and {Liao}, S. and {Licata}, E.~L. and {Lindstr{\o}m}, H.~E.~P. and {Lister}, T.~A. and {Livanou}, E. and {Lobel}, A. and {Lorca}, A. and {Loup}, C. and {Madrero Pardo}, P. and {Magdaleno Romeo}, A. and {Managau}, S. and {Mann}, R.~G. and {Manteiga}, M. and {Marchant}, J.~M. and {Marconi}, M. and {Marcos}, J. and {Marcos Santos}, M.~M.~S. and {Mar{\'\i}n Pina}, D. and {Marinoni}, S. and {Marocco}, F. and {Marshall}, D.~J. and {Polo}, L. Martin and {Mart{\'\i}n-Fleitas}, J.~M. and {Marton}, G. and {Mary}, N. and {Masip}, A. and {Massari}, D. and {Mastrobuono-Battisti}, A. and {Mazeh}, T. and {McMillan}, P.~J. and {Messina}, S. and {Michalik}, D. and {Millar}, N.~R. and {Mints}, A. and {Molina}, D. and {Molinaro}, R. and {Moln{\'a}r}, L. and {Monari}, G. and {Mongui{\'o}}, M. and {Montegriffo}, P. and {Montero}, A. and {Mor}, R. and {Mora}, A. and {Morbidelli}, R. and {Morel}, T. and {Morris}, D. and {Muraveva}, T. and {Murphy}, C.~P. and {Musella}, I. and {Nagy}, Z. and {Noval}, L. and {Oca{\~n}a}, F. and {Ogden}, A. and {Ordenovic}, C. and {Osinde}, J.~O. and {Pagani}, C. and {Pagano}, I. and {Palaversa}, L. and {Palicio}, P.~A. and {Pallas-Quintela}, L. and {Panahi}, A. and {Payne-Wardenaar}, S. and {Pe{\~n}alosa Esteller}, X. and {Penttil{\"a}}, A. and {Pichon}, B. and {Piersimoni}, A.~M. and {Pineau}, F. -X. and {Plachy}, E. and {Plum}, G. and {Poggio}, E. and {Pr{\v{s}}a}, A. and {Pulone}, L. and {Racero}, E. and {Ragaini}, S. and {Rainer}, M. and {Raiteri}, C.~M. and {Rambaux}, N. and {Ramos}, P. and {Ramos-Lerate}, M. and {Re Fiorentin}, P. and {Regibo}, S. and {Richards}, P.~J. and {Rios Diaz}, C. and {Ripepi}, V. and {Riva}, A. and {Rix}, H. -W. and {Rixon}, G. and {Robichon}, N. and {Robin}, A.~C. and {Robin}, C. and {Roelens}, M. and {Rogues}, H.~R.~O. and {Rohrbasser}, L. and {Romero-G{\'o}mez}, M. and {Rowell}, N. and {Royer}, F. and {Ruz Mieres}, D. and {Rybicki}, K.~A. and {Sadowski}, G. and {S{\'a}ez N{\'u}{\~n}ez}, A. and {Sagrist{\`a} Sell{\'e}s}, A. and {Sahlmann}, J. and {Salguero}, E. and {Samaras}, N. and {Sanchez Gimenez}, V. and {Sanna}, N. and {Santove{\~n}a}, R. and {Sarasso}, M. and {Schultheis}, M. and {Sciacca}, E. and {Segol}, M. and {Segovia}, J.~C. and {S{\'e}gransan}, D. and {Semeux}, D. and {Shahaf}, S. and {Siddiqui}, H.~I. and {Siebert}, A. and {Siltala}, L. and {Silvelo}, A. and {Slezak}, E. and {Slezak}, I. and {Smart}, R.~L. and {Snaith}, O.~N. and {Solano}, E. and {Solitro}, F. and {Souami}, D. and {Souchay}, J. and {Spagna}, A. and {Spina}, L. and {Spoto}, F. and {Steele}, I.~A. and {Steidelm{\"u}ller}, H. and {Stephenson}, C.~A. and {S{\"u}veges}, M. and {Surdej}, J. and {Szabados}, L. and {Szegedi-Elek}, E. and {Taris}, F. and {Taylo}, M.~B. and {Teixeira}, R. and {Tolomei}, L. and {Tonello}, N. and {Torra}, F. and {Torra}, J. and {Torralba Elipe}, G. and {Trabucchi}, M. and {Tsounis}, A.~T. and {Turon}, C. and {Ulla}, A. and {Unger}, N. and {Vaillant}, M.~V. and {van Dillen}, E. and {van Reeven}, W. and {Vanel}, O. and {Vecchiato}, A. and {Viala}, Y. and {Vicente}, D. and {Voutsinas}, S. and {Weiler}, M. and {Wevers}, T. and {Wyrzykowski}, L. and {Yoldas}, A. and {Yvard}, P. and {Zhao}, H. and {Zorec}, J. and {Zucker}, S. and {Zwitter}, T.},
        title = "{Gaia Data Release 3: Summary of the content and survey properties}",
      journal = {arXiv e-prints},
         year = 2022,
        month = jul,
          eid = {arXiv:2208.00211},
        pages = {arXiv:2208.00211},
          doi = {10.48550/arXiv.2208.00211},
archivePrefix = {arXiv},
       eprint = {2208.00211},
 primaryClass = {astro-ph.GA},
       adsurl = {https://ui.adsabs.harvard.edu/abs/2022arXiv220800211G}
}

@ARTICLE{Belokurov2022,
       author = {{Belokurov}, Vasily and {Kravtsov}, Andrey},
        title = "{From dawn till disc: Milky Way's turbulent youth revealed by the APOGEE+Gaia data}",
      journal = {\mnras},
         year = 2022,
        month = jul,
       volume = {514},
       number = {1},
        pages = {689-714},
          doi = {10.1093/mnras/stac1267},
archivePrefix = {arXiv},
       eprint = {2203.04980},
 primaryClass = {astro-ph.GA},
       adsurl = {https://ui.adsabs.harvard.edu/abs/2022MNRAS.514..689B}
}

@ARTICLE{Rix2022,
       author = {{Rix}, Hans-Walter and {Chandra}, Vedant and {Andrae}, Ren{\'e} and {Price-Whelan}, Adrian M. and {Weinberg}, David H. and {Conroy}, Charlie and {Fouesneau}, Morgan and {Hogg}, David W. and {De Angeli}, Francesca and {Naidu}, Rohan P. and {Xiang}, Maosheng and {Ruz-Mieres}, Daniela},
        title = "{The Poor Old Heart of the Milky Way}",
      journal = {\apj},
         year = 2022,
        month = dec,
       volume = {941},
       number = {1},
          eid = {45},
        pages = {45},
          doi = {10.3847/1538-4357/ac9e01},
archivePrefix = {arXiv},
       eprint = {2209.02722},
 primaryClass = {astro-ph.GA},
       adsurl = {https://ui.adsabs.harvard.edu/abs/2022ApJ...941...45R}
}

@ARTICLE{Xiang2019,
       author = {{Xiang}, Maosheng and {Ting}, Yuan-Sen and {Rix}, Hans-Walter and {Sandford}, Nathan and {Buder}, Sven and {Lind}, Karin and {Liu}, Xiao-Wei and {Shi}, Jian-Rong and {Zhang}, Hua-Wei},
        title = "{Abundance Estimates for 16 Elements in 6 Million Stars from LAMOST DR5 Low-Resolution Spectra}",
      journal = {\apjs},
         year = 2019,
        month = dec,
       volume = {245},
       number = {2},
          eid = {34},
        pages = {34},
          doi = {10.3847/1538-4365/ab5364},
archivePrefix = {arXiv},
       eprint = {1908.09727},
 primaryClass = {astro-ph.SR},
       adsurl = {https://ui.adsabs.harvard.edu/abs/2019ApJS..245...34X}
}

@ARTICLE{Xiang2015_LAMOST,
       author = {{Xiang}, M.~S. and {Liu}, X.~W. and {Yuan}, H.~B. and {Huang}, Y. and {Huo}, Z.~Y. and {Zhang}, H.~W. and {Chen}, B.~Q. and {Zhang}, H.~H. and {Sun}, N.~C. and {Wang}, C. and {Zhao}, Y.~H. and {Shi}, J.~R. and {Luo}, A.~L. and {Li}, G.~P. and {Wu}, Y. and {Bai}, Z.~R. and {Zhang}, Y. and {Hou}, Y.~H. and {Yuan}, H.~L. and {Li}, G.~W. and {Wei}, Z.},
        title = "{The LAMOST stellar parameter pipeline at Peking University - LSP3}",
      journal = {\mnras},
         year = 2015,
        month = mar,
       volume = {448},
       number = {1},
        pages = {822-854},
          doi = {10.1093/mnras/stu2692},
archivePrefix = {arXiv},
       eprint = {1412.6627},
 primaryClass = {astro-ph.GA},
       adsurl = {https://ui.adsabs.harvard.edu/abs/2015MNRAS.448..822X}
}

@ARTICLE{Chandra2023,
       author = {{Chandra}, Vedant and {Semenov}, Vadim A. and {Rix}, Hans-Walter and {Conroy}, Charlie and {Bonaca}, Ana and {Naidu}, Rohan P. and {Andrae}, Rene and {Li}, Jiadong and {Hernquist}, Lars},
        title = "{The Three-Phase Evolution of the Milky Way}",
      journal = {arXiv e-prints},
         year = 2023,
        month = oct,
          eid = {arXiv:2310.13050},
        pages = {arXiv:2310.13050},
          doi = {10.48550/arXiv.2310.13050},
archivePrefix = {arXiv},
       eprint = {2310.13050},
 primaryClass = {astro-ph.GA},
       adsurl = {https://ui.adsabs.harvard.edu/abs/2023arXiv231013050C}
}

@ARTICLE{Conroy2022,
       author = {{Conroy}, Charlie and {Weinberg}, David H. and {Naidu}, Rohan P. and {Buck}, Tobias and {Johnson}, James W. and {Cargile}, Phillip and {Bonaca}, Ana and {Caldwell}, Nelson and {Chandra}, Vedant and {Han}, Jiwon Jesse and {Johnson}, Benjamin D. and {Speagle}, Joshua S. and {Ting}, Yuan-Sen and {Woody}, Turner and {Zaritsky}, Dennis},
        title = "{Birth of the Galactic Disk Revealed by the H3 Survey}",
      journal = {arXiv e-prints},
         year = 2022,
        month = apr,
          eid = {arXiv:2204.02989},
        pages = {arXiv:2204.02989},
          doi = {10.48550/arXiv.2204.02989},
archivePrefix = {arXiv},
       eprint = {2204.02989},
 primaryClass = {astro-ph.GA},
       adsurl = {https://ui.adsabs.harvard.edu/abs/2022arXiv220402989C}
}

@ARTICLE{Miglio2021,
       author = {{Miglio}, A. and {Chiappini}, C. and {Mackereth}, J.~T. and {Davies}, G.~R. and {Brogaard}, K. and {Casagrande}, L. and {Chaplin}, W.~J. and {Girardi}, L. and {Kawata}, D. and {Khan}, S. and {Izzard}, R. and {Montalb{\'a}n}, J. and {Mosser}, B. and {Vincenzo}, F. and {Bossini}, D. and {Noels}, A. and {Rodrigues}, T. and {Valentini}, M. and {Mandel}, I.},
        title = "{Age dissection of the Milky Way discs: Red giants in the Kepler field}",
      journal = {\aap},
         year = 2021,
        month = jan,
       volume = {645},
          eid = {A85},
        pages = {A85},
          doi = {10.1051/0004-6361/202038307},
archivePrefix = {arXiv},
       eprint = {2004.14806},
 primaryClass = {astro-ph.GA},
       adsurl = {https://ui.adsabs.harvard.edu/abs/2021A&A...645A..85M}
}

@ARTICLE{Queiroz2023,
       author = {{Queiroz}, A.~B.~A. and {Anders}, F. and {Chiappini}, C. and {Khalatyan}, A. and {Santiago}, B.~X. and {Nepal}, S. and {Steinmetz}, M. and {Gallart}, C. and {Valentini}, M. and {Dal Ponte}, M. and {Barbuy}, B. and {P{\'e}rez-Villegas}, A. and {Masseron}, T. and {Fern{\'a}ndez-Trincado}, J.~G. and {Khoperskov}, S. and {Minchev}, I. and {Fern{\'a}ndez-Alvar}, E. and {Lane}, R.~R. and {Nitschelm}, C.},
        title = "{StarHorse results for spectroscopic surveys and Gaia DR3: Chrono-chemical populations in the solar vicinity, the genuine thick disk, and young alpha-rich stars}",
      journal = {\aap},
         year = 2023,
        month = may,
       volume = {673},
          eid = {A155},
        pages = {A155},
          doi = {10.1051/0004-6361/202245399},
archivePrefix = {arXiv},
       eprint = {2303.09926},
 primaryClass = {astro-ph.GA},
       adsurl = {https://ui.adsabs.harvard.edu/abs/2023A&A...673A.155Q}
}

@ARTICLE{McMillan2017,
       author = {{McMillan}, Paul J.},
        title = "{The mass distribution and gravitational potential of the Milky Way}",
      journal = {\mnras},
         year = 2017,
        month = feb,
       volume = {465},
       number = {1},
        pages = {76-94},
          doi = {10.1093/mnras/stw2759},
archivePrefix = {arXiv},
       eprint = {1608.00971},
 primaryClass = {astro-ph.GA},
       adsurl = {https://ui.adsabs.harvard.edu/abs/2017MNRAS.465...76M}
}

@ARTICLE{Schonrich2010,
       author = {{Sch{\"o}nrich}, Ralph and {Binney}, James and {Dehnen}, Walter},
        title = "{Local kinematics and the local standard of rest}",
      journal = {\mnras},
         year = 2010,
        month = apr,
       volume = {403},
       number = {4},
        pages = {1829-1833},
          doi = {10.1111/j.1365-2966.2010.16253.x},
archivePrefix = {arXiv},
       eprint = {0912.3693},
 primaryClass = {astro-ph.GA},
       adsurl = {https://ui.adsabs.harvard.edu/abs/2010MNRAS.403.1829S}
}

@ARTICLE{Zhang2024_patternspeed,
       author = {{Zhang}, HanYuan and {Belokurov}, Vasily and {Evans}, N. Wyn and {Kane}, Sarah G. and {Sanders}, Jason L.},
        title = "{Kinematics and dynamics of the Galactic bar revealed by Gaia long-period variables}",
      journal = {\mnras},
         year = 2024,
        month = sep,
       volume = {533},
       number = {3},
        pages = {3395-3414},
          doi = {10.1093/mnras/stae2023},
archivePrefix = {arXiv},
       eprint = {2406.06678},
 primaryClass = {astro-ph.GA},
       adsurl = {https://ui.adsabs.harvard.edu/abs/2024MNRAS.533.3395Z}
}

@ARTICLE{Zhang2024,
       author = {{Zhang}, HanYuan and {Belokurov}, Vasily and {Evans}, N. Wyn and {Li}, Zhao-Yu and {Sanders}, Jason L. and {Ardern-Arentsen}, Anke},
        title = "{Deciphering the Milky Way disc formation time encrypted in the bar chrono-kinematics}",
      journal = {\mnras},
         year = 2024,
        month = dec,
       volume = {535},
       number = {3},
        pages = {2873-2888},
          doi = {10.1093/mnras/stae2546},
archivePrefix = {arXiv},
       eprint = {2408.16815},
 primaryClass = {astro-ph.GA},
       adsurl = {https://ui.adsabs.harvard.edu/abs/2024MNRAS.535.2873Z}
}

@ARTICLE{Zhang2023,
       author = {{Zhang}, HanYuan and {Ardern-Arentsen}, Anke and {Belokurov}, Vasily},
        title = "{On the existence of a very metal-poor disc in the Milky Way}",
      journal = {\mnras},
         year = 2024,
        month = sep,
       volume = {533},
       number = {1},
        pages = {889-907},
          doi = {10.1093/mnras/stae1887},
archivePrefix = {arXiv},
       eprint = {2311.09294},
 primaryClass = {astro-ph.GA},
       adsurl = {https://ui.adsabs.harvard.edu/abs/2024MNRAS.533..889Z}
}

@ARTICLE{Belokurov2024,
       author = {{Belokurov}, Vasily and {Kravtsov}, Andrey},
        title = "{In-situ versus accreted Milky Way globular clusters: a new classification method and implications for cluster formation}",
      journal = {\mnras},
         year = 2024,
        month = feb,
       volume = {528},
       number = {2},
        pages = {3198-3216},
          doi = {10.1093/mnras/stad3920},
archivePrefix = {arXiv},
       eprint = {2309.15902},
 primaryClass = {astro-ph.GA},
       adsurl = {https://ui.adsabs.harvard.edu/abs/2024MNRAS.528.3198B}
}

@ARTICLE{Vasiliev2019,
       author = {{Vasiliev}, Eugene},
        title = "{AGAMA: action-based galaxy modelling architecture}",
      journal = {\mnras},
         year = 2019,
        month = jan,
       volume = {482},
       number = {2},
        pages = {1525-1544},
          doi = {10.1093/mnras/sty2672},
archivePrefix = {arXiv},
       eprint = {1802.08239},
 primaryClass = {astro-ph.GA},
       adsurl = {https://ui.adsabs.harvard.edu/abs/2019MNRAS.482.1525V}
}

@ARTICLE{Frankel2020,
       author = {{Frankel}, Neige and {Sanders}, Jason and {Ting}, Yuan-Sen and {Rix}, Hans-Walter},
        title = "{Keeping It Cool: Much Orbit Migration, yet Little Heating, in the Galactic Disk}",
      journal = {\apj},
         year = 2020,
        month = jun,
       volume = {896},
       number = {1},
          eid = {15},
        pages = {15},
          doi = {10.3847/1538-4357/ab910c},
archivePrefix = {arXiv},
       eprint = {2002.04622},
 primaryClass = {astro-ph.GA},
       adsurl = {https://ui.adsabs.harvard.edu/abs/2020ApJ...896...15F}
}

@ARTICLE{Frankel2018,
       author = {{Frankel}, Neige and {Rix}, Hans-Walter and {Ting}, Yuan-Sen and {Ness}, Melissa and {Hogg}, David W.},
        title = "{Measuring Radial Orbit Migration in the Galactic Disk}",
      journal = {\apj},
         year = 2018,
        month = oct,
       volume = {865},
       number = {2},
          eid = {96},
        pages = {96},
          doi = {10.3847/1538-4357/aadba5},
archivePrefix = {arXiv},
       eprint = {1805.09198},
 primaryClass = {astro-ph.GA},
       adsurl = {https://ui.adsabs.harvard.edu/abs/2018ApJ...865...96F}
}

@ARTICLE{Lu2024,
       author = {{Lu}, Yuxi (Lucy) and {Minchev}, Ivan and {Buck}, Tobias and {Khoperskov}, Sergey and {Steinmetz}, Matthias and {Libeskind}, Noam and {Cescutti}, Gabriele and {Freeman}, Ken C. and {Ratcliffe}, Bridget},
        title = "{There is no place like home - finding birth radii of stars in the Milky Way}",
      journal = {\mnras},
         year = 2024,
        month = nov,
       volume = {535},
       number = {1},
        pages = {392-405},
          doi = {10.1093/mnras/stae2364},
archivePrefix = {arXiv},
       eprint = {2212.04515},
 primaryClass = {astro-ph.GA},
       adsurl = {https://ui.adsabs.harvard.edu/abs/2024MNRAS.535..392L}
}

@ARTICLE{Ratcliffe2024,
       author = {{Ratcliffe}, Bridget and {Khoperskov}, Sergey and {Minchev}, Ivan and {Lee}, Nathan D. and {Buck}, Tobias and {Marques}, L{\'e}a and {Lu}, Lucy and {Steinmetz}, Matthias},
        title = "{Evolution of the radial ISM metallicity gradient in the Milky Way disk since redshift $\approx 3$}",
      journal = {arXiv e-prints},
         year = 2024,
        month = oct,
          eid = {arXiv:2410.17326},
        pages = {arXiv:2410.17326},
          doi = {10.48550/arXiv.2410.17326},
archivePrefix = {arXiv},
       eprint = {2410.17326},
 primaryClass = {astro-ph.GA},
       adsurl = {https://ui.adsabs.harvard.edu/abs/2024arXiv241017326R}
}

@ARTICLE{Freeman_BH2002,
       author = {{Freeman}, Ken and {Bland-Hawthorn}, Joss},
        title = "{The New Galaxy: Signatures of Its Formation}",
      journal = {\araa},
         year = 2002,
        month = jan,
       volume = {40},
        pages = {487-537},
          doi = {10.1146/annurev.astro.40.060401.093840},
archivePrefix = {arXiv},
       eprint = {astro-ph/0208106},
 primaryClass = {astro-ph},
       adsurl = {https://ui.adsabs.harvard.edu/abs/2002ARA&A..40..487F}
}

@ARTICLE{Minchev2018,
       author = {{Minchev}, I. and {Anders}, F. and {Recio-Blanco}, A. and {Chiappini}, C. and {de Laverny}, P. and {Queiroz}, A. and {Steinmetz}, M. and {Adibekyan}, V. and {Carrillo}, I. and {Cescutti}, G. and {Guiglion}, G. and {Hayden}, M. and {de Jong}, R.~S. and {Kordopatis}, G. and {Majewski}, S.~R. and {Martig}, M. and {Santiago}, B.~X.},
        title = "{Estimating stellar birth radii and the time evolution of Milky Way's ISM metallicity gradient}",
      journal = {\mnras},
         year = 2018,
        month = dec,
       volume = {481},
       number = {2},
        pages = {1645-1657},
          doi = {10.1093/mnras/sty2033},
archivePrefix = {arXiv},
       eprint = {1804.06856},
 primaryClass = {astro-ph.GA},
       adsurl = {https://ui.adsabs.harvard.edu/abs/2018MNRAS.481.1645M}
}

@ARTICLE{Sellwood_Binney2002,
       author = {{Sellwood}, J.~A. and {Binney}, J.~J.},
        title = "{Radial mixing in galactic discs}",
      journal = {\mnras},
         year = 2002,
        month = nov,
       volume = {336},
       number = {3},
        pages = {785-796},
          doi = {10.1046/j.1365-8711.2002.05806.x},
archivePrefix = {arXiv},
       eprint = {astro-ph/0203510},
 primaryClass = {astro-ph},
       adsurl = {https://ui.adsabs.harvard.edu/abs/2002MNRAS.336..785S}
}

@ARTICLE{Ratcliffe2025,
       author = {{Ratcliffe}, Bridget and {Khoperskov}, Sergey and {Lee}, Nathan and {Minchev}, Ivan and {Di Matteo}, Paola and {van de Ven}, Glenn and {Haywood}, Misha and {Marques}, L{\'e}a and {Bernaldez}, John Paul and {Krajnovi{\'c}}, Davor and {Steinmetz}, Matthias},
        title = "{Rediscovering the Milky Way with an orbit superposition approach and APOGEE data V. The disc growth and history of star formation}",
      journal = {arXiv e-prints},
         year = 2025,
        month = sep,
          eid = {arXiv:2509.02691},
        pages = {arXiv:2509.02691},
          doi = {10.48550/arXiv.2509.02691},
archivePrefix = {arXiv},
       eprint = {2509.02691},
 primaryClass = {astro-ph.GA},
       adsurl = {https://ui.adsabs.harvard.edu/abs/2025arXiv250902691R}
}

@ARTICLE{Funakoshi2025,
       author = {{Funakoshi}, Natsuki and {Kawata}, Daisuke and {Sanders}, Jason L. and {Ciuc{\u{a}}}, Ioana and {Grand}, Robert J.~J. and {Zh{\={a}}ng}, HanYuan},
        title = "{Thick-to-thin disc transition and gas disc shrinking induced by the Gaia-Sausage-Enceladus merger}",
      journal = {arXiv e-prints},
         year = 2025,
        month = jul,
          eid = {arXiv:2507.22979},
        pages = {arXiv:2507.22979},
          doi = {10.48550/arXiv.2507.22979},
archivePrefix = {arXiv},
       eprint = {2507.22979},
 primaryClass = {astro-ph.GA},
       adsurl = {https://ui.adsabs.harvard.edu/abs/2025arXiv250722979F}
}

@ARTICLE{Zhang2025,
       author = {{Zhang}, HanYuan and {Belokurov}, Vasily and {Evans}, N. Wyn and {Sanders}, Jason L. and {Lu}, Yuxi(Lucy) and {Cao}, Chengye and {Myeong}, GyuChul and {Dillamore}, Adam M. and {Kane}, Sarah G. and {Li}, Zhao-Yu},
        title = "{Observational Constraints of Radial Migration in the Galactic Disk Driven by the Slowing Bar}",
      journal = {\apjl},
         year = 2025,
        month = apr,
       volume = {983},
       number = {1},
          eid = {L10},
        pages = {L10},
          doi = {10.3847/2041-8213/adc261},
archivePrefix = {arXiv},
       eprint = {2502.02642},
 primaryClass = {astro-ph.GA},
       adsurl = {https://ui.adsabs.harvard.edu/abs/2025ApJ...983L..10Z}
}

@ARTICLE{Chiba2021,
       author = {{Chiba}, Rimpei and {Friske}, Jennifer K.~S. and {Sch{\"o}nrich}, Ralph},
        title = "{Resonance sweeping by a decelerating Galactic bar}",
      journal = {\mnras},
         year = 2021,
        month = jan,
       volume = {500},
       number = {4},
        pages = {4710-4729},
          doi = {10.1093/mnras/staa3585},
archivePrefix = {arXiv},
       eprint = {1912.04304},
 primaryClass = {astro-ph.GA},
       adsurl = {https://ui.adsabs.harvard.edu/abs/2021MNRAS.500.4710C}
}

@ARTICLE{Wenger2019,
       author = {{Wenger}, Trey V. and {Balser}, Dana S. and {Anderson}, L.~D. and {Bania}, T.~M.},
        title = "{Metallicity Structure in the Milky Way Disk Revealed by Galactic H II Regions}",
      journal = {\apj},
         year = 2019,
        month = dec,
       volume = {887},
       number = {2},
          eid = {114},
        pages = {114},
          doi = {10.3847/1538-4357/ab53d3},
archivePrefix = {arXiv},
       eprint = {1910.14605},
 primaryClass = {astro-ph.GA},
       adsurl = {https://ui.adsabs.harvard.edu/abs/2019ApJ...887..114W}
}

@ARTICLE{Braganca2019,
       author = {{Bragan{\c{c}}a}, G.~A. and {Daflon}, S. and {Lanz}, T. and {Cunha}, K. and {Bensby}, T. and {McMillan}, P.~J. and {Garmany}, C.~D. and {Glaspey}, J.~W. and {Borges Fernandes}, M. and {Oey}, M.~S. and {Hubeny}, I.},
        title = "{Radial abundance gradients in the outer Galactic disk as traced by main-sequence OB stars}",
      journal = {\aap},
         year = 2019,
        month = may,
       volume = {625},
          eid = {A120},
        pages = {A120},
          doi = {10.1051/0004-6361/201834554},
archivePrefix = {arXiv},
       eprint = {1904.04340},
 primaryClass = {astro-ph.SR},
       adsurl = {https://ui.adsabs.harvard.edu/abs/2019A&A...625A.120B}
}

@ARTICLE{Lu2022,
       author = {{Lu}, Yuxi Lucy and {Ness}, Melissa K. and {Buck}, Tobias and {Carr}, Christopher},
        title = "{Turning points in the age-metallicity relations - created by late satellite infall and enhanced by radial migration}",
      journal = {\mnras},
         year = 2022,
        month = jun,
       volume = {512},
       number = {4},
        pages = {4697-4714},
          doi = {10.1093/mnras/stac780},
archivePrefix = {arXiv},
       eprint = {2112.05238},
 primaryClass = {astro-ph.GA},
       adsurl = {https://ui.adsabs.harvard.edu/abs/2022MNRAS.512.4697L}
}

@ARTICLE{Ratcliffe2024_tng50,
       author = {{Ratcliffe}, B. and {Khoperskov}, S. and {Minchev}, I. and {Lu}, L. and {de Jong}, R.~S. and {Steinmetz}, M.},
        title = "{Empirical derivation of the metallicity evolution with time and radius using TNG50 Milky Way and Andromeda analogues}",
      journal = {\aap},
         year = 2024,
        month = oct,
       volume = {690},
          eid = {A352},
        pages = {A352},
          doi = {10.1051/0004-6361/202449268},
archivePrefix = {arXiv},
       eprint = {2401.09260},
 primaryClass = {astro-ph.GA},
       adsurl = {https://ui.adsabs.harvard.edu/abs/2024A&A...690A.352R}
}

@ARTICLE{Kreckel2019,
       author = {{Kreckel}, K. and {Ho}, I. -T. and {Blanc}, G.~A. and {Groves}, B. and {Santoro}, F. and {Schinnerer}, E. and {Bigiel}, F. and {Chevance}, M. and {Congiu}, E. and {Emsellem}, E. and {Faesi}, C. and {Glover}, S.~C.~O. and {Grasha}, K. and {Kruijssen}, J.~M.~D. and {Lang}, P. and {Leroy}, A.~K. and {Meidt}, S.~E. and {McElroy}, R. and {Pety}, J. and {Rosolowsky}, E. and {Saito}, T. and {Sandstrom}, K. and {Sanchez-Blazquez}, P. and {Schruba}, A.},
        title = "{Mapping Metallicity Variations across Nearby Galaxy Disks}",
      journal = {\apj},
         year = 2019,
        month = dec,
       volume = {887},
       number = {1},
          eid = {80},
        pages = {80},
          doi = {10.3847/1538-4357/ab5115},
archivePrefix = {arXiv},
       eprint = {1910.07190},
 primaryClass = {astro-ph.GA},
       adsurl = {https://ui.adsabs.harvard.edu/abs/2019ApJ...887...80K}
}

@ARTICLE{Sharma2021_alphabimodality,
       author = {{Sharma}, Sanjib and {Hayden}, Michael R. and {Bland-Hawthorn}, Joss},
        title = "{Chemical enrichment and radial migration in the Galactic disc - the origin of the [{\ensuremath{\alpha}}Fe] double sequence}",
      journal = {\mnras},
         year = 2021,
        month = nov,
       volume = {507},
       number = {4},
        pages = {5882-5901},
          doi = {10.1093/mnras/stab2015},
archivePrefix = {arXiv},
       eprint = {2005.03646},
 primaryClass = {astro-ph.GA},
       adsurl = {https://ui.adsabs.harvard.edu/abs/2021MNRAS.507.5882S}
}

@ARTICLE{Sanders_Binney2015,
       author = {{Sanders}, Jason L. and {Binney}, James},
        title = "{Extended distribution functions for our Galaxy}",
      journal = {\mnras},
         year = 2015,
        month = jun,
       volume = {449},
       number = {4},
        pages = {3479-3502},
          doi = {10.1093/mnras/stv578},
archivePrefix = {arXiv},
       eprint = {1501.02227},
 primaryClass = {astro-ph.GA},
       adsurl = {https://ui.adsabs.harvard.edu/abs/2015MNRAS.449.3479S}
}

@ARTICLE{Binney2010,
       author = {{Binney}, James},
        title = "{Distribution functions for the Milky Way}",
      journal = {\mnras},
         year = 2010,
        month = feb,
       volume = {401},
       number = {4},
        pages = {2318-2330},
          doi = {10.1111/j.1365-2966.2009.15845.x},
archivePrefix = {arXiv},
       eprint = {0910.1512},
 primaryClass = {astro-ph.GA},
       adsurl = {https://ui.adsabs.harvard.edu/abs/2010MNRAS.401.2318B}
}

@ARTICLE{Zhang2023a,
       author = {{Zhang}, Hanyuan and {Sanders}, Jason L.},
        title = "{A kinematic calibration of the O-rich Mira variable period-age relation from Gaia}",
      journal = {\mnras},
         year = 2023,
        month = may,
       volume = {521},
       number = {1},
        pages = {1462-1478},
          doi = {10.1093/mnras/stad575},
archivePrefix = {arXiv},
       eprint = {2302.10024},
 primaryClass = {astro-ph.SR},
       adsurl = {https://ui.adsabs.harvard.edu/abs/2023MNRAS.521.1462Z}
}

@ARTICLE{Sanders2024,
       author = {{Sanders}, Jason L. and {Kawata}, Daisuke and {Matsunaga}, Noriyuki and {Sormani}, Mattia C. and {Smith}, Leigh C. and {Minniti}, Dante and {Gerhard}, Ortwin},
        title = "{The epoch of the Milky Way's bar formation: dynamical modelling of Mira variables in the nuclear stellar disc}",
      journal = {\mnras},
         year = 2024,
        month = may,
       volume = {530},
       number = {3},
        pages = {2972-2993},
          doi = {10.1093/mnras/stae711},
archivePrefix = {arXiv},
       eprint = {2311.00035},
 primaryClass = {astro-ph.GA},
       adsurl = {https://ui.adsabs.harvard.edu/abs/2024MNRAS.530.2972S}
}

@ARTICLE{Hayden2015,
       author = {{Hayden}, Michael R. and {Bovy}, Jo and {Holtzman}, Jon A. and {Nidever}, David L. and {Bird}, Jonathan C. and {Weinberg}, David H. and {Andrews}, Brett H. and {Majewski}, Steven R. and {Allende Prieto}, Carlos and {Anders}, Friedrich and {Beers}, Timothy C. and {Bizyaev}, Dmitry and {Chiappini}, Cristina and {Cunha}, Katia and {Frinchaboy}, Peter and {Garc{\'\i}a-Her{\'n}andez}, D.~A. and {Garc{\'\i}a P{\'e}rez}, Ana E. and {Girardi}, L{\'e}o and {Harding}, Paul and {Hearty}, Fred R. and {Johnson}, Jennifer A. and {M{\'e}sz{\'a}ros}, Szabolcs and {Minchev}, Ivan and {O'Connell}, Robert and {Pan}, Kaike and {Robin}, Annie C. and {Schiavon}, Ricardo P. and {Schneider}, Donald P. and {Schultheis}, Mathias and {Shetrone}, Matthew and {Skrutskie}, Michael and {Steinmetz}, Matthias and {Smith}, Verne and {Wilson}, John C. and {Zamora}, Olga and {Zasowski}, Gail},
        title = "{Chemical Cartography with APOGEE: Metallicity Distribution Functions and the Chemical Structure of the Milky Way Disk}",
      journal = {\apj},
         year = 2015,
        month = aug,
       volume = {808},
       number = {2},
          eid = {132},
        pages = {132},
          doi = {10.1088/0004-637X/808/2/132},
archivePrefix = {arXiv},
       eprint = {1503.02110},
 primaryClass = {astro-ph.GA},
       adsurl = {https://ui.adsabs.harvard.edu/abs/2015ApJ...808..132H}
}

@ARTICLE{Kubryk2015,
       author = {{Kubryk}, M. and {Prantzos}, N. and {Athanassoula}, E.},
        title = "{Evolution of the Milky Way with radial motions of stars and gas. I. The solar neighbourhood and the thin and thick disks}",
      journal = {\aap},
         year = 2015,
        month = aug,
       volume = {580},
          eid = {A126},
        pages = {A126},
          doi = {10.1051/0004-6361/201424171},
archivePrefix = {arXiv},
       eprint = {1412.0585},
 primaryClass = {astro-ph.GA},
       adsurl = {https://ui.adsabs.harvard.edu/abs/2015A&A...580A.126K}
}

@ARTICLE{Halle2015,
       author = {{Halle}, A. and {Di Matteo}, P. and {Haywood}, M. and {Combes}, F.},
        title = "{Quantifying stellar radial migration in an N-body simulation: blurring, churning, and the outer regions of galaxy discs}",
      journal = {\aap},
         year = 2015,
        month = jun,
       volume = {578},
          eid = {A58},
        pages = {A58},
          doi = {10.1051/0004-6361/201525612},
archivePrefix = {arXiv},
       eprint = {1501.00664},
 primaryClass = {astro-ph.GA},
       adsurl = {https://ui.adsabs.harvard.edu/abs/2015A&A...578A..58H}
}

@ARTICLE{Bird2012,
       author = {{Bird}, Jonathan C. and {Kazantzidis}, Stelios and {Weinberg}, David H.},
        title = "{Radial mixing in galactic discs: the effects of disc structure and satellite bombardment}",
      journal = {\mnras},
         year = 2012,
        month = feb,
       volume = {420},
       number = {2},
        pages = {913-925},
          doi = {10.1111/j.1365-2966.2011.19728.x},
archivePrefix = {arXiv},
       eprint = {1104.0933},
 primaryClass = {astro-ph.GA},
       adsurl = {https://ui.adsabs.harvard.edu/abs/2012MNRAS.420..913B}
}

@ARTICLE{Roskar2008,
       author = {{Ro{\v{s}}kar}, Rok and {Debattista}, Victor P. and {Quinn}, Thomas R. and {Stinson}, Gregory S. and {Wadsley}, James},
        title = "{Riding the Spiral Waves: Implications of Stellar Migration for the Properties of Galactic Disks}",
      journal = {\apjl},
         year = 2008,
        month = sep,
       volume = {684},
       number = {2},
        pages = {L79},
          doi = {10.1086/592231},
archivePrefix = {arXiv},
       eprint = {0808.0206},
 primaryClass = {astro-ph},
       adsurl = {https://ui.adsabs.harvard.edu/abs/2008ApJ...684L..79R}
}

@ARTICLE{Khoperskov2020,
       author = {{Khoperskov}, S. and {Di Matteo}, P. and {Haywood}, M. and {G{\'o}mez}, A. and {Snaith}, O.~N.},
        title = "{Escapees from the bar resonances. Presence of low-eccentricity metal-rich stars at the solar vicinity}",
      journal = {\aap},
         year = 2020,
        month = jun,
       volume = {638},
          eid = {A144},
        pages = {A144},
          doi = {10.1051/0004-6361/201937188},
archivePrefix = {arXiv},
       eprint = {1911.12424},
 primaryClass = {astro-ph.GA},
       adsurl = {https://ui.adsabs.harvard.edu/abs/2020A&A...638A.144K}
}

@ARTICLE{Haywood2024,
       author = {{Haywood}, Misha and {Khoperskov}, Sergey and {Cerqui}, Valeria and {Di Matteo}, Paola and {Katz}, David and {Snaith}, Owain},
        title = "{Timing the Milky Way bar formation and the accompanying radial migration episode}",
      journal = {\aap},
         year = 2024,
        month = oct,
       volume = {690},
          eid = {A147},
        pages = {A147},
          doi = {10.1051/0004-6361/202348767},
archivePrefix = {arXiv},
       eprint = {2403.08963},
 primaryClass = {astro-ph.GA},
       adsurl = {https://ui.adsabs.harvard.edu/abs/2024A&A...690A.147H}
}

@ARTICLE{Marques2025,
       author = {{Marques}, L. and {Minchev}, I. and {Ratcliffe}, B. and {Khoperskov}, S. and {Steinmetz}, M. and {Wenger}, T.~V. and {Buck}, T. and {Martig}, M. and {Kordopatis}, G. and {Schultheis}, M. and {Zucker}, D.~B.},
        title = "{Bar-spiral interaction induces radial migration and star formation bursts}",
      journal = {\aap},
         year = 2025,
        month = sep,
       volume = {701},
          eid = {A88},
        pages = {A88},
          doi = {10.1051/0004-6361/202554020},
archivePrefix = {arXiv},
       eprint = {2502.02651},
 primaryClass = {astro-ph.GA},
       adsurl = {https://ui.adsabs.harvard.edu/abs/2025A&A...701A..88M}
}

@ARTICLE{Minchev_Famaey2010,
       author = {{Minchev}, I. and {Famaey}, B.},
        title = "{A New Mechanism for Radial Migration in Galactic Disks: Spiral-Bar Resonance Overlap}",
      journal = {\apj},
         year = 2010,
        month = oct,
       volume = {722},
       number = {1},
        pages = {112-121},
          doi = {10.1088/0004-637X/722/1/112},
archivePrefix = {arXiv},
       eprint = {0911.1794},
 primaryClass = {astro-ph.GA},
       adsurl = {https://ui.adsabs.harvard.edu/abs/2010ApJ...722..112M}
}

@ARTICLE{Minchev2013,
       author = {{Minchev}, I. and {Chiappini}, C. and {Martig}, M.},
        title = "{Chemodynamical evolution of the Milky Way disk. I. The solar vicinity}",
      journal = {\aap},
         year = 2013,
        month = oct,
       volume = {558},
          eid = {A9},
        pages = {A9},
          doi = {10.1051/0004-6361/201220189},
archivePrefix = {arXiv},
       eprint = {1208.1506},
 primaryClass = {astro-ph.GA},
       adsurl = {https://ui.adsabs.harvard.edu/abs/2013A&A...558A...9M}
}

@ARTICLE{Minchev2011_resonanceoverlap,
       author = {{Minchev}, I. and {Famaey}, B. and {Combes}, F. and {Di Matteo}, P. and {Mouhcine}, M. and {Wozniak}, H.},
        title = "{Radial migration in galactic disks caused by resonance overlap of multiple patterns: Self-consistent simulations}",
      journal = {\aap},
         year = 2011,
        month = mar,
       volume = {527},
          eid = {A147},
        pages = {A147},
          doi = {10.1051/0004-6361/201015139},
archivePrefix = {arXiv},
       eprint = {1006.0484},
 primaryClass = {astro-ph.GA},
       adsurl = {https://ui.adsabs.harvard.edu/abs/2011A&A...527A.147M}
}

@ARTICLE{Aumer2016,
       author = {{Aumer}, Michael and {Binney}, James and {Sch{\"o}nrich}, Ralph},
        title = "{Age-velocity dispersion relations and heating histories in disc galaxies}",
      journal = {\mnras},
         year = 2016,
        month = oct,
       volume = {462},
       number = {2},
        pages = {1697-1713},
          doi = {10.1093/mnras/stw1639},
archivePrefix = {arXiv},
       eprint = {1607.01972},
 primaryClass = {astro-ph.GA},
       adsurl = {https://ui.adsabs.harvard.edu/abs/2016MNRAS.462.1697A}
}

@ARTICLE{Sharma2021_avr,
       author = {{Sharma}, Sanjib and {Hayden}, Michael R. and {Bland-Hawthorn}, Joss and {Stello}, Dennis and {Buder}, Sven and {Zinn}, Joel C. and {Kallinger}, Thomas and {Asplund}, Martin and {De Silva}, Gayandhi M. and {D'Orazi}, Valentina and {Freeman}, Ken and {Kos}, Janez and {Lewis}, Geraint F. and {Lin}, Jane and {Lind}, Karin and {Martell}, Sarah and {Simpson}, Jeffrey D. and {Wittenmyer}, Rob A. and {Zucker}, Daniel B. and {Zwitter}, Tomaz and {Chen}, Boquan and {Cotar}, Klemen and {Esdaile}, James and {Hon}, Marc and {Horner}, Jonathan and {Huber}, Daniel and {Kafle}, Prajwal R. and {Khanna}, Shourya and {Ting}, Yuan-Sen and {Nataf}, David M. and {Nordlander}, Thomas and {Saadon}, Mohd Hafiz Mohd and {Tepper-Garcia}, Thor and {Tinney}, C.~G. and {Traven}, Gregor and {Watson}, Fred and {Wright}, Duncan and {Wyse}, Rosemary F.~G.},
        title = "{Fundamental relations for the velocity dispersion of stars in the Milky Way}",
      journal = {\mnras},
         year = 2021,
        month = sep,
       volume = {506},
       number = {2},
        pages = {1761-1776},
          doi = {10.1093/mnras/stab1086},
archivePrefix = {arXiv},
       eprint = {2004.06556},
 primaryClass = {astro-ph.GA},
       adsurl = {https://ui.adsabs.harvard.edu/abs/2021MNRAS.506.1761S}
}

@ARTICLE{Cerqui2025,
       author = {{Cerqui}, Valeria and {Haywood}, Misha and {Snaith}, Owain and {Di Matteo}, Paola and {Casamiquela}, Laia},
        title = "{The chemical enrichment histories across the Milky Way disk}",
      journal = {\aap},
         year = 2025,
        month = jul,
       volume = {699},
          eid = {A277},
        pages = {A277},
          doi = {10.1051/0004-6361/202452448},
archivePrefix = {arXiv},
       eprint = {2504.20160},
 primaryClass = {astro-ph.GA},
       adsurl = {https://ui.adsabs.harvard.edu/abs/2025A&A...699A.277C}
}

@ARTICLE{Baba2024,
       author = {{Baba}, Junichi and {Tsujimoto}, Takuji and {Saitoh}, Takayuki R.},
        title = "{Solar System Migration Points to a Renewed Concept: Galactic Habitable Orbits}",
      journal = {\apjl},
         year = 2024,
        month = dec,
       volume = {976},
       number = {2},
          eid = {L29},
        pages = {L29},
          doi = {10.3847/2041-8213/ad9260},
archivePrefix = {arXiv},
       eprint = {2412.02963},
 primaryClass = {astro-ph.GA},
       adsurl = {https://ui.adsabs.harvard.edu/abs/2024ApJ...976L..29B}
}

@ARTICLE{Grand2018,
       author = {{Grand}, Robert J.~J. and {Bustamante}, Sebasti{\'a}n and {G{\'o}mez}, Facundo A. and {Kawata}, Daisuke and {Marinacci}, Federico and {Pakmor}, R{\"u}diger and {Rix}, Hans-Walter and {Simpson}, Christine M. and {Sparre}, Martin and {Springel}, Volker},
        title = "{Origin of chemically distinct discs in the Auriga cosmological simulations}",
      journal = {\mnras},
         year = 2018,
        month = mar,
       volume = {474},
       number = {3},
        pages = {3629-3639},
          doi = {10.1093/mnras/stx3025},
archivePrefix = {arXiv},
       eprint = {1708.07834},
 primaryClass = {astro-ph.GA},
       adsurl = {https://ui.adsabs.harvard.edu/abs/2018MNRAS.474.3629G}
}

@ARTICLE{Chiappini2001,
       author = {{Chiappini}, Cristina and {Matteucci}, Francesca and {Romano}, Donatella},
        title = "{Abundance Gradients and the Formation of the Milky Way}",
      journal = {\apj},
         year = 2001,
        month = jun,
       volume = {554},
       number = {2},
        pages = {1044-1058},
          doi = {10.1086/321427},
archivePrefix = {arXiv},
       eprint = {astro-ph/0102134},
 primaryClass = {astro-ph},
       adsurl = {https://ui.adsabs.harvard.edu/abs/2001ApJ...554.1044C}
}

@ARTICLE{Belokurov2020,
       author = {{Belokurov}, Vasily and {Sanders}, Jason L. and {Fattahi}, Azadeh and {Smith}, Martin C. and {Deason}, Alis J. and {Evans}, N. Wyn and {Grand}, Robert J.~J.},
        title = "{The biggest splash}",
      journal = {\mnras},
         year = 2020,
        month = may,
       volume = {494},
       number = {3},
        pages = {3880-3898},
          doi = {10.1093/mnras/staa876},
archivePrefix = {arXiv},
       eprint = {1909.04679},
 primaryClass = {astro-ph.GA},
       adsurl = {https://ui.adsabs.harvard.edu/abs/2020MNRAS.494.3880B}
}

@ARTICLE{Helmi2020,
       author = {{Helmi}, Amina},
        title = "{Streams, Substructures, and the Early History of the Milky Way}",
      journal = {\araa},
         year = 2020,
        month = aug,
       volume = {58},
        pages = {205-256},
          doi = {10.1146/annurev-astro-032620-021917},
archivePrefix = {arXiv},
       eprint = {2002.04340},
 primaryClass = {astro-ph.GA},
       adsurl = {https://ui.adsabs.harvard.edu/abs/2020ARA&A..58..205H}
}

@ARTICLE{Dillamore2025a,
       author = {{Dillamore}, Adam M. and {Sanders}, Jason L. and {Belokurov}, Vasily and {Zhang}, Hanyuan},
        title = "{Dynamical streams in the local stellar halo}",
      journal = {\mnras},
         year = 2025,
        month = jul,
       volume = {541},
       number = {1},
        pages = {214-233},
          doi = {10.1093/mnras/staf965},
archivePrefix = {arXiv},
       eprint = {2503.02926},
 primaryClass = {astro-ph.GA},
       adsurl = {https://ui.adsabs.harvard.edu/abs/2025MNRAS.541..214D}
}

@ARTICLE{Dillamore2025b,
       author = {{Dillamore}, Adam M. and {Sanders}, Jason L.},
        title = "{Geometry of the Milky Way's dark matter from dynamical models of the tilted stellar halo}",
      journal = {arXiv e-prints},
         year = 2025,
        month = sep,
          eid = {arXiv:2510.00095},
        pages = {arXiv:2510.00095},
          doi = {10.48550/arXiv.2510.00095},
archivePrefix = {arXiv},
       eprint = {2510.00095},
 primaryClass = {astro-ph.GA},
       adsurl = {https://ui.adsabs.harvard.edu/abs/2025arXiv251000095D}
}

@ARTICLE{Hunter2024,
       author = {{Hunter}, Glen H. and {Sormani}, Mattia C. and {Beckmann}, Jan P. and {Vasiliev}, Eugene and {Glover}, Simon C.~O. and {Klessen}, Ralf S. and {Soler}, Juan D. and {Brucy}, No{\'e} and {Girichidis}, Philipp and {G{\"o}ller}, Junia and {Ohlin}, Loke and {Tress}, Robin and {Molinari}, Sergio and {Gerhard}, Ortwin and {Benedettini}, Milena and {Smith}, Rowan and {Hennebelle}, Patrick and {Testi}, Leonardo},
        title = "{Testing kinematic distances under a realistic Galactic potential: Investigating systematic errors in the kinematic distance method arising from a non-axisymmetric potential}",
      journal = {\aap},
         year = 2024,
        month = dec,
       volume = {692},
          eid = {A216},
        pages = {A216},
          doi = {10.1051/0004-6361/202450000},
archivePrefix = {arXiv},
       eprint = {2403.18000},
 primaryClass = {astro-ph.GA},
       adsurl = {https://ui.adsabs.harvard.edu/abs/2024A&A...692A.216H}
}

@ARTICLE{Tsukui2025,
       author = {{Tsukui}, Takafumi and {Wisnioski}, Emily and {Bland-Hawthorn}, Joss and {Freeman}, Ken},
        title = "{The emergence of galactic thin and thick discs across cosmic history}",
      journal = {\mnras},
         year = 2025,
        month = jul,
       volume = {540},
       number = {4},
        pages = {3493-3522},
          doi = {10.1093/mnras/staf604},
archivePrefix = {arXiv},
       eprint = {2409.15909},
 primaryClass = {astro-ph.GA},
       adsurl = {https://ui.adsabs.harvard.edu/abs/2025MNRAS.540.3493T}
}

@ARTICLE{Hamilton2024,
       author = {{Hamilton}, Chris and {Modak}, Shaunak and {Tremaine}, Scott},
        title = "{Why is the Galactic disk so cool?}",
      journal = {arXiv e-prints},
         year = 2024,
        month = nov,
          eid = {arXiv:2411.08944},
        pages = {arXiv:2411.08944},
          doi = {10.48550/arXiv.2411.08944},
archivePrefix = {arXiv},
       eprint = {2411.08944},
 primaryClass = {astro-ph.GA},
       adsurl = {https://ui.adsabs.harvard.edu/abs/2024arXiv241108944H}
}

@ARTICLE{Sellwood_Binney2025,
       author = {{Sellwood}, J A and {Binney}, J},
        title = "{A comment on ``Why is the Galactic disk so cool?'', by Hamilton et al}",
      journal = {arXiv e-prints},
         year = 2025,
        month = jan,
          eid = {arXiv:2501.17907},
        pages = {arXiv:2501.17907},
          doi = {10.48550/arXiv.2501.17907},
archivePrefix = {arXiv},
       eprint = {2501.17907},
 primaryClass = {astro-ph.GA},
       adsurl = {https://ui.adsabs.harvard.edu/abs/2025arXiv250117907S}
}

@ARTICLE{McCluskey2025,
       author = {{McCluskey}, Fiona and {Wetzel}, Andrew and {Loebman}, Sarah and {Moreno}, Jorge},
        title = "{Stellar Velocity Dispersion versus Age: Consistency across Observations and Simulations, with the Milky Way as an Outlier}",
      journal = {arXiv e-prints},
         year = 2025,
        month = jun,
          eid = {arXiv:2506.11840},
        pages = {arXiv:2506.11840},
          doi = {10.48550/arXiv.2506.11840},
archivePrefix = {arXiv},
       eprint = {2506.11840},
 primaryClass = {astro-ph.GA},
       adsurl = {https://ui.adsabs.harvard.edu/abs/2025arXiv250611840M}
}

@ARTICLE{ChibaSchonrich_2021,
       author = {{Chiba}, Rimpei and {Sch{\"o}nrich}, Ralph},
        title = "{Tree-ring structure of Galactic bar resonance}",
      journal = {\mnras},
         year = 2021,
        month = aug,
       volume = {505},
       number = {2},
        pages = {2412-2426},
          doi = {10.1093/mnras/stab1094},
archivePrefix = {arXiv},
       eprint = {2102.08388},
 primaryClass = {astro-ph.GA},
       adsurl = {https://ui.adsabs.harvard.edu/abs/2021MNRAS.505.2412C}
}

@ARTICLE{ClarkeGerhard_2022,
       author = {{Clarke}, Jonathan P. and {Gerhard}, Ortwin},
        title = "{The pattern speed of the Milky Way bar/bulge from VIRAC and Gaia}",
      journal = {\mnras},
         year = 2022,
        month = may,
       volume = {512},
       number = {2},
        pages = {2171-2188},
          doi = {10.1093/mnras/stac603},
archivePrefix = {arXiv},
       eprint = {2107.10875},
 primaryClass = {astro-ph.GA},
       adsurl = {https://ui.adsabs.harvard.edu/abs/2022MNRAS.512.2171C}
}

@ARTICLE{Sanders2019,
       author = {{Sanders}, Jason L. and {Smith}, Leigh and {Evans}, N. Wyn},
        title = "{The pattern speed of the Milky Way bar from transverse velocities}",
      journal = {\mnras},
         year = 2019,
        month = oct,
       volume = {488},
       number = {4},
        pages = {4552-4564},
          doi = {10.1093/mnras/stz1827},
archivePrefix = {arXiv},
       eprint = {1903.02009},
 primaryClass = {astro-ph.GA},
       adsurl = {https://ui.adsabs.harvard.edu/abs/2019MNRAS.488.4552S}
}

@ARTICLE{Dillamore2024,
       author = {{Dillamore}, Adam M. and {Belokurov}, Vasily and {Evans}, N. Wyn},
        title = "{Radial halo substructure in harmony with the Galactic bar}",
      journal = {\mnras},
         year = 2024,
        month = aug,
       volume = {532},
       number = {4},
        pages = {4389-4407},
          doi = {10.1093/mnras/stae1789},
archivePrefix = {arXiv},
       eprint = {2402.14907},
 primaryClass = {astro-ph.GA},
       adsurl = {https://ui.adsabs.harvard.edu/abs/2024MNRAS.532.4389D}
}

@ARTICLE{Kawata_2021,
       author = {{Kawata}, Daisuke and {Baba}, Junichi and {Hunt}, Jason A.~S. and {Sch{\"o}nrich}, Ralph and {Ciuc{\u{a}}}, Ioana and {Friske}, Jennifer and {Seabroke}, George and {Cropper}, Mark},
        title = "{Galactic bar resonances inferred from kinematically hot stars in Gaia EDR3}",
      journal = {\mnras},
         year = 2021,
        month = nov,
       volume = {508},
       number = {1},
        pages = {728-736},
          doi = {10.1093/mnras/stab2582},
archivePrefix = {arXiv},
       eprint = {2012.05890},
 primaryClass = {astro-ph.GA},
       adsurl = {https://ui.adsabs.harvard.edu/abs/2021MNRAS.508..728K}
}

@ARTICLE{Binney2020,
       author = {{Binney}, James},
        title = "{Trapped orbits and solar-neighbourhood kinematics}",
      journal = {\mnras},
         year = 2020,
        month = jun,
       volume = {495},
       number = {1},
        pages = {895-904},
          doi = {10.1093/mnras/staa1103},
archivePrefix = {arXiv},
       eprint = {1912.07023},
 primaryClass = {astro-ph.GA},
       adsurl = {https://ui.adsabs.harvard.edu/abs/2020MNRAS.495..895B}
}

@ARTICLE{Halle2018,
       author = {{Halle}, A. and {Di Matteo}, P. and {Haywood}, M. and {Combes}, F.},
        title = "{Radial migration in a stellar galactic disc with thick components}",
      journal = {\aap},
         year = 2018,
        month = aug,
       volume = {616},
          eid = {A86},
        pages = {A86},
          doi = {10.1051/0004-6361/201832603},
archivePrefix = {arXiv},
       eprint = {1801.02403},
 primaryClass = {astro-ph.GA},
       adsurl = {https://ui.adsabs.harvard.edu/abs/2018A&A...616A..86H}
}

@ARTICLE{TG24,
       author = {{Tepper-Garc{\'\i}a}, Thor and {Bland-Hawthorn}, Joss and {Vasiliev}, Eugene and {Agertz}, Oscar and {Teyssier}, Romain and {Federrath}, Christoph},
        title = "{NEXUS: a framework for controlled simulations of idealized galaxies}",
      journal = {\mnras},
         year = 2024,
        month = nov,
       volume = {535},
       number = {1},
        pages = {187-206},
          doi = {10.1093/mnras/stae2372},
archivePrefix = {arXiv},
       eprint = {2406.00342},
 primaryClass = {astro-ph.GA},
       adsurl = {https://ui.adsabs.harvard.edu/abs/2024MNRAS.535..187T}
}

@ARTICLE{Vickers2021,
       author = {{Vickers}, John J. and {Shen}, Juntai and {Li}, Zhao-Yu},
        title = "{The Flattening Metallicity Gradient in the Milky Way's Thin Disk}",
      journal = {\apj},
         year = 2021,
        month = dec,
       volume = {922},
       number = {2},
          eid = {189},
        pages = {189},
          doi = {10.3847/1538-4357/ac27a9},
archivePrefix = {arXiv},
       eprint = {2109.09250},
 primaryClass = {astro-ph.GA},
       adsurl = {https://ui.adsabs.harvard.edu/abs/2021ApJ...922..189V}
}

@ARTICLE{Chen2025,
       author = {{Chen}, Ao and {Shen}, Juntai and {Wang}, Chun and {Huang}, Yang},
        title = "{Revisiting the Radial Metallicity Gradient-Age Relation in the Milky Way's Thin and Thick Disks}",
      journal = {arXiv e-prints},
         year = 2025,
        month = aug,
          eid = {arXiv:2508.20386},
        pages = {arXiv:2508.20386},
          doi = {10.48550/arXiv.2508.20386},
archivePrefix = {arXiv},
       eprint = {2508.20386},
 primaryClass = {astro-ph.GA},
       adsurl = {https://ui.adsabs.harvard.edu/abs/2025arXiv250820386C}
}

@ARTICLE{vanDonkelaar2022,
       author = {{van Donkelaar}, Floor and {Agertz}, Oscar and {Renaud}, Florent},
        title = "{From giant clumps to clouds - II. The emergence of thick disc kinematics from the conditions of star formation in high redshift gas rich galaxies}",
      journal = {\mnras},
         year = 2022,
        month = may,
       volume = {512},
       number = {3},
        pages = {3806-3814},
          doi = {10.1093/mnras/stac692},
archivePrefix = {arXiv},
       eprint = {2110.13165},
 primaryClass = {astro-ph.GA},
       adsurl = {https://ui.adsabs.harvard.edu/abs/2022MNRAS.512.3806V}
}

@ARTICLE{Dillamore2023,
       author = {{Dillamore}, Adam M. and {Belokurov}, Vasily and {Evans}, N. Wyn and {Davies}, Elliot Y.},
        title = "{Stellar halo substructure generated by bar resonances}",
      journal = {\mnras},
         year = 2023,
        month = sep,
       volume = {524},
       number = {3},
        pages = {3596-3608},
          doi = {10.1093/mnras/stad2136},
archivePrefix = {arXiv},
       eprint = {2303.00008},
 primaryClass = {astro-ph.GA},
       adsurl = {https://ui.adsabs.harvard.edu/abs/2023MNRAS.524.3596D}
}

@ARTICLE{BH2024,
       author = {{Bland-Hawthorn}, Joss and {Tepper-Garcia}, Thor and {Agertz}, Oscar and {Federrath}, Christoph},
        title = "{Turbulent Gas-rich Disks at High Redshift: Bars and Bulges in a Radial Shear Flow}",
      journal = {\apj},
         year = 2024,
        month = jun,
       volume = {968},
       number = {2},
          eid = {86},
        pages = {86},
          doi = {10.3847/1538-4357/ad4118},
archivePrefix = {arXiv},
       eprint = {2402.06060},
 primaryClass = {astro-ph.GA},
       adsurl = {https://ui.adsabs.harvard.edu/abs/2024ApJ...968...86B}
}

@ARTICLE{BH2025,
       author = {{Bland-Hawthorn}, Joss and {Tepper-Garcia}, Thor and {Agertz}, Oscar and {Federrath}, Christoph and {Haywood}, Misha and {di Matteo}, Paola and {Bedding}, Timothy R and {Tsukui}, Takafumi and {Wisnioski}, Emily and {Ness}, Melissa and {Freeman}, Ken},
        title = "{Turbulent gas-rich discs at high redshift: origin of thick stellar discs through 3D 'baryon sloshing'}",
      journal = {arXiv e-prints},
         year = 2025,
        month = feb,
          eid = {arXiv:2502.01895},
        pages = {arXiv:2502.01895},
          doi = {10.48550/arXiv.2502.01895},
archivePrefix = {arXiv},
       eprint = {2502.01895},
 primaryClass = {astro-ph.GA},
       adsurl = {https://ui.adsabs.harvard.edu/abs/2025arXiv250201895B}
}

@ARTICLE{Fragkoudi2025,
       author = {{Fragkoudi}, Francesca and {Grand}, Robert J.~J. and {Pakmor}, R{\"u}diger and {G{\'o}mez}, Facundo and {Marinacci}, Federico and {Springel}, Volker},
        title = "{Bar formation and evolution in the cosmological context: inputs from the Auriga simulations}",
      journal = {\mnras},
         year = 2025,
        month = apr,
       volume = {538},
       number = {3},
        pages = {1587-1608},
          doi = {10.1093/mnras/staf389},
archivePrefix = {arXiv},
       eprint = {2406.09453},
 primaryClass = {astro-ph.GA},
       adsurl = {https://ui.adsabs.harvard.edu/abs/2025MNRAS.538.1587F}
}

@ARTICLE{Beane2023,
       author = {{Beane}, Angus and {Hernquist}, Lars and {D'Onghia}, Elena and {Marinacci}, Federico and {Conroy}, Charlie and {Qi}, Jia and {Sales}, Laura V. and {Torrey}, Paul and {Vogelsberger}, Mark},
        title = "{Stellar Bars in Isolated Gas-rich Spiral Galaxies Do Not Slow Down}",
      journal = {\apj},
         year = 2023,
        month = aug,
       volume = {953},
       number = {2},
          eid = {173},
        pages = {173},
          doi = {10.3847/1538-4357/ace2b9},
archivePrefix = {arXiv},
       eprint = {2209.03364},
 primaryClass = {astro-ph.GA},
       adsurl = {https://ui.adsabs.harvard.edu/abs/2023ApJ...953..173B}
}

@ARTICLE{Kawata2017,
       author = {{Kawata}, Daisuke and {Grand}, Robert J.~J. and {Gibson}, Brad K. and {Casagrande}, Luca and {Hunt}, Jason A.~S. and {Brook}, Chris B.},
        title = "{Impacts of a flaring star-forming disc and stellar radial mixing on the vertical metallicity gradient}",
      journal = {\mnras},
         year = 2017,
        month = jan,
       volume = {464},
       number = {1},
        pages = {702-712},
          doi = {10.1093/mnras/stw2363},
archivePrefix = {arXiv},
       eprint = {1604.07412},
 primaryClass = {astro-ph.GA},
       adsurl = {https://ui.adsabs.harvard.edu/abs/2017MNRAS.464..702K}
}

@ARTICLE{Schonrich_MM2017,
       author = {{Sch{\"o}nrich}, Ralph and {McMillan}, Paul J.},
        title = "{Understanding inverse metallicity gradients in galactic discs as a consequence of inside-out formation}",
      journal = {\mnras},
         year = 2017,
        month = may,
       volume = {467},
       number = {1},
        pages = {1154-1174},
          doi = {10.1093/mnras/stx093},
archivePrefix = {arXiv},
       eprint = {1605.02338},
 primaryClass = {astro-ph.GA},
       adsurl = {https://ui.adsabs.harvard.edu/abs/2017MNRAS.467.1154S}
}

@ARTICLE{Khoperskov2025a,
       author = {{Khoperskov}, Sergey and {van de Ven}, Glenn and {Steinmetz}, Matthias and {Ratcliffe}, Bridget and {Minchev}, Ivan and {Krajnovi{\'c}}, Davor and {Haywood}, Misha and {Di Matteo}, Paola and {Kacharov}, Nikolay and {Marques}, L{\'e}a and {Valentini}, Marica and {de Jong}, Roelof S.},
        title = "{Rediscovering the Milky Way with an orbit superposition approach and APOGEE data: I. Method validation}",
      journal = {\aap},
         year = 2025,
        month = mar,
       volume = {695},
          eid = {A220},
        pages = {A220},
          doi = {10.1051/0004-6361/202453304},
archivePrefix = {arXiv},
       eprint = {2411.15062},
 primaryClass = {astro-ph.GA},
       adsurl = {https://ui.adsabs.harvard.edu/abs/2025A&A...695A.220K}
}

@ARTICLE{Khoperskov2025b,
       author = {{Khoperskov}, Sergey and {Steinmetz}, Matthias and {Haywood}, Misha and {van de Ven}, Glenn and {Krajnovi{\'c}}, Davor and {Ratcliffe}, Bridget and {Minchev}, Ivan and {Di Matteo}, Paola and {Kacharov}, Nikolay and {Marques}, L{\'e}a and {Valentini}, Marica and {de Jong}, Roelof S.},
        title = "{Rediscovering the Milky Way with an orbit superposition approach and APOGEE data: II. Chrono-chemo-kinematics of the disc}",
      journal = {\aap},
         year = 2025,
        month = aug,
       volume = {700},
          eid = {A89},
        pages = {A89},
          doi = {10.1051/0004-6361/202453305},
archivePrefix = {arXiv},
       eprint = {2411.16866},
 primaryClass = {astro-ph.GA},
       adsurl = {https://ui.adsabs.harvard.edu/abs/2025A&A...700A..89K}
}

@ARTICLE{Belokurov2018,
       author = {{Belokurov}, V. and {Erkal}, D. and {Evans}, N.~W. and {Koposov}, S.~E. and {Deason}, A.~J.},
        title = "{Co-formation of the disc and the stellar halo}",
      journal = {\mnras},
         year = 2018,
        month = jul,
       volume = {478},
       number = {1},
        pages = {611-619},
          doi = {10.1093/mnras/sty982},
archivePrefix = {arXiv},
       eprint = {1802.03414},
 primaryClass = {astro-ph.GA},
       adsurl = {https://ui.adsabs.harvard.edu/abs/2018MNRAS.478..611B}
}

@ARTICLE{Helmi2018,
       author = {{Helmi}, Amina and {Babusiaux}, Carine and {Koppelman}, Helmer H. and {Massari}, Davide and {Veljanoski}, Jovan and {Brown}, Anthony G.~A.},
        title = "{The merger that led to the formation of the Milky Way's inner stellar halo and thick disk}",
      journal = {\nat},
         year = 2018,
        month = oct,
       volume = {563},
       number = {7729},
        pages = {85-88},
          doi = {10.1038/s41586-018-0625-x},
archivePrefix = {arXiv},
       eprint = {1806.06038},
 primaryClass = {astro-ph.GA},
       adsurl = {https://ui.adsabs.harvard.edu/abs/2018Natur.563...85H}
}

@ARTICLE{Sellwood2014,
       author = {{Sellwood}, J.~A.},
        title = "{Secular evolution in disk galaxies}",
      journal = {Reviews of Modern Physics},
         year = 2014,
        month = jan,
       volume = {86},
       number = {1},
        pages = {1-46},
          doi = {10.1103/RevModPhys.86.1},
archivePrefix = {arXiv},
       eprint = {1310.0403},
 primaryClass = {astro-ph.GA},
       adsurl = {https://ui.adsabs.harvard.edu/abs/2014RvMP...86....1S}
}

@ARTICLE{Loebman2016,
       author = {{Loebman}, Sarah R. and {Debattista}, Victor P. and {Nidever}, David L. and {Hayden}, Michael R. and {Holtzman}, Jon A. and {Clarke}, Adam J. and {Ro{\v{s}}kar}, Rok and {Valluri}, Monica},
        title = "{Imprints of Radial Migration on the Milky Way{\textquoteright}s Metallicity Distribution Functions}",
      journal = {\apjl},
         year = 2016,
        month = feb,
       volume = {818},
       number = {1},
          eid = {L6},
        pages = {L6},
          doi = {10.3847/2041-8205/818/1/L6},
archivePrefix = {arXiv},
       eprint = {1511.06369},
 primaryClass = {astro-ph.GA},
       adsurl = {https://ui.adsabs.harvard.edu/abs/2016ApJ...818L...6L}
}

@ARTICLE{Minchev2006,
       author = {{Minchev}, I. and {Quillen}, A.~C.},
        title = "{Radial heating of a galactic disc by multiple spiral density waves}",
      journal = {\mnras},
         year = 2006,
        month = may,
       volume = {368},
       number = {2},
        pages = {623-636},
          doi = {10.1111/j.1365-2966.2006.10129.x},
archivePrefix = {arXiv},
       eprint = {astro-ph/0511037},
 primaryClass = {astro-ph},
       adsurl = {https://ui.adsabs.harvard.edu/abs/2006MNRAS.368..623M}
}

@ARTICLE{Solway2012,
       author = {{Solway}, Michael and {Sellwood}, J.~A. and {Sch{\"o}nrich}, Ralph},
        title = "{Radial migration in galactic thick discs}",
      journal = {\mnras},
         year = 2012,
        month = may,
       volume = {422},
       number = {2},
        pages = {1363-1383},
          doi = {10.1111/j.1365-2966.2012.20712.x},
archivePrefix = {arXiv},
       eprint = {1202.1418},
 primaryClass = {astro-ph.GA},
       adsurl = {https://ui.adsabs.harvard.edu/abs/2012MNRAS.422.1363S}
}

@ARTICLE{Daniel2019,
       author = {{Daniel}, Kathryne J. and {Schaffner}, David A. and {McCluskey}, Fiona and {Fiedler Kawaguchi}, Codie and {Loebman}, Sarah},
        title = "{When Cold Radial Migration is Hot: Constraints from Resonant Overlap}",
      journal = {\apj},
         year = 2019,
        month = sep,
       volume = {882},
       number = {2},
          eid = {111},
        pages = {111},
          doi = {10.3847/1538-4357/ab341a},
archivePrefix = {arXiv},
       eprint = {1907.10100},
 primaryClass = {astro-ph.GA},
       adsurl = {https://ui.adsabs.harvard.edu/abs/2019ApJ...882..111D}
}

@ARTICLE{Quillen2009,
       author = {{Quillen}, A.~C. and {Minchev}, Ivan and {Bland-Hawthorn}, Joss and {Haywood}, Misha},
        title = "{Radial mixing in the outer Milky Way disc caused by an orbiting satellite}",
      journal = {\mnras},
         year = 2009,
        month = aug,
       volume = {397},
       number = {3},
        pages = {1599-1606},
          doi = {10.1111/j.1365-2966.2009.15054.x},
archivePrefix = {arXiv},
       eprint = {0903.1851},
 primaryClass = {astro-ph.GA},
       adsurl = {https://ui.adsabs.harvard.edu/abs/2009MNRAS.397.1599Q}
}

@ARTICLE{Carr2022,
       author = {{Carr}, Christopher and {Johnston}, Kathryn V. and {Laporte}, Chervin F.~P. and {Ness}, Melissa K.},
        title = "{Migration and heating in the galactic disc from encounters between Sagittarius and the Milky Way}",
      journal = {\mnras},
         year = 2022,
        month = nov,
       volume = {516},
       number = {4},
        pages = {5067-5083},
          doi = {10.1093/mnras/stac2403},
archivePrefix = {arXiv},
       eprint = {2201.04133},
 primaryClass = {astro-ph.GA},
       adsurl = {https://ui.adsabs.harvard.edu/abs/2022MNRAS.516.5067C}
}

@ARTICLE{Minchev2025,
       author = {{Minchev}, I. and {Attard}, K. and {Ratcliffe}, B. and {Martig}, M. and {Walcher}, J. and {Khoperskov}, S. and {Bernaldez}, J.~P. and {Marques}, L. and {Sysoliatina}, K. and {Chiappini}, C. and {Steinmetz}, M. and {de Jong}, R.},
        title = "{The Impact of Radial Migration on Disk Galaxy SFHs: I. Biases in Spatially Resolved Estimates}",
      journal = {arXiv e-prints},
         year = 2025,
        month = aug,
          eid = {arXiv:2508.18367},
        pages = {arXiv:2508.18367},
          doi = {10.48550/arXiv.2508.18367},
archivePrefix = {arXiv},
       eprint = {2508.18367},
 primaryClass = {astro-ph.GA},
       adsurl = {https://ui.adsabs.harvard.edu/abs/2025arXiv250818367M}
}

@ARTICLE{Bernaldez2025,
       author = {{Bernaldez}, J.~P. and {Minchev}, I. and {Ratcliffe}, B. and {Marques}, L. and {Sysoliatina}, K. and {Walcher}, J. and {Khoperskov}, S. and {Martig}, M. and {de Jong}, R. and {Steinmetz}, M.},
        title = "{The Impact of Radial Migration on Disk Galaxy SFHs: II. The Role of bar strength, disk thickness, and merger history}",
      journal = {arXiv e-prints},
         year = 2025,
        month = aug,
          eid = {arXiv:2508.19340},
        pages = {arXiv:2508.19340},
          doi = {10.48550/arXiv.2508.19340},
archivePrefix = {arXiv},
       eprint = {2508.19340},
 primaryClass = {astro-ph.GA},
       adsurl = {https://ui.adsabs.harvard.edu/abs/2025arXiv250819340B}
}

@ARTICLE{Kubryk2013,
       author = {{Kubryk}, M. and {Prantzos}, N. and {Athanassoula}, E.},
        title = "{Radial migration in a bar-dominated disc galaxy - I. Impact on chemical evolution}",
      journal = {\mnras},
         year = 2013,
        month = dec,
       volume = {436},
       number = {2},
        pages = {1479-1491},
          doi = {10.1093/mnras/stt1667},
       adsurl = {https://ui.adsabs.harvard.edu/abs/2013MNRAS.436.1479K}
}

@ARTICLE{Kawata2018,
       author = {{Kawata}, Daisuke and {Allende Prieto}, Carlos and {Brook}, Chris B. and {Casagrande}, Luca and {Ciuc{\u{a}}}, Ioana and {Gibson}, Brad K. and {Grand}, Robert J.~J. and {Hayden}, Michael R. and {Hunt}, Jason A.~S.},
        title = "{Metallicity gradient of the thick disc progenitor at high redshift}",
      journal = {\mnras},
         year = 2018,
        month = jan,
       volume = {473},
       number = {1},
        pages = {867-878},
          doi = {10.1093/mnras/stx2464},
archivePrefix = {arXiv},
       eprint = {1706.01474},
 primaryClass = {astro-ph.GA},
       adsurl = {https://ui.adsabs.harvard.edu/abs/2018MNRAS.473..867K}
}

@ARTICLE{Lian2022,
       author = {{Lian}, Jianhui and {Zasowski}, Gail and {Hasselquist}, Sten and {Holtzman}, Jon A. and {Boardman}, Nicholas and {Cunha}, Katia and {Fern{\'a}ndez-Trincado}, Jos{\'e} G. and {Frinchaboy}, Peter M. and {Garcia-Hernandez}, D.~A. and {Nitschelm}, Christian and {Lane}, Richard R. and {Thomas}, Daniel and {Zhang}, Kai},
        title = "{Quantifying radial migration in the Milky Way: inefficient over short time-scales but essential to the very outer disc beyond  15 kpc}",
      journal = {\mnras},
         year = 2022,
        month = apr,
       volume = {511},
       number = {4},
        pages = {5639-5655},
          doi = {10.1093/mnras/stac479},
archivePrefix = {arXiv},
       eprint = {2202.08846},
 primaryClass = {astro-ph.GA},
       adsurl = {https://ui.adsabs.harvard.edu/abs/2022MNRAS.511.5639L}
}

@ARTICLE{Bonaca2017,
       author = {{Bonaca}, Ana and {Conroy}, Charlie and {Wetzel}, Andrew and {Hopkins}, Philip F. and {Kere{\v{s}}}, Du{\v{s}}an},
        title = "{Gaia Reveals a Metal-rich, in situ Component of the Local Stellar Halo}",
      journal = {\apj},
         year = 2017,
        month = aug,
       volume = {845},
       number = {2},
          eid = {101},
        pages = {101},
          doi = {10.3847/1538-4357/aa7d0c},
archivePrefix = {arXiv},
       eprint = {1704.05463},
 primaryClass = {astro-ph.GA},
       adsurl = {https://ui.adsabs.harvard.edu/abs/2017ApJ...845..101B}
}

@ARTICLE{DiMatteo2019,
       author = {{Di Matteo}, P. and {Haywood}, M. and {Lehnert}, M.~D. and {Katz}, D. and {Khoperskov}, S. and {Snaith}, O.~N. and {G{\'o}mez}, A. and {Robichon}, N.},
        title = "{The Milky Way has no in-situ halo other than the heated thick disc. Composition of the stellar halo and age-dating the last significant merger with Gaia DR2 and APOGEE}",
      journal = {\aap},
         year = 2019,
        month = dec,
       volume = {632},
          eid = {A4},
        pages = {A4},
          doi = {10.1051/0004-6361/201834929},
archivePrefix = {arXiv},
       eprint = {1812.08232},
 primaryClass = {astro-ph.GA},
       adsurl = {https://ui.adsabs.harvard.edu/abs/2019A&A...632A...4D}
}

@ARTICLE{Gallart2019,
       author = {{Gallart}, Carme and {Bernard}, Edouard J. and {Brook}, Chris B. and {Ruiz-Lara}, Tom{\'a}s and {Cassisi}, Santi and {Hill}, Vanessa and {Monelli}, Matteo},
        title = "{Uncovering the birth of the Milky Way through accurate stellar ages with Gaia}",
      journal = {Nature Astronomy},
         year = 2019,
        month = jul,
       volume = {3},
        pages = {932-939},
          doi = {10.1038/s41550-019-0829-5},
archivePrefix = {arXiv},
       eprint = {1901.02900},
 primaryClass = {astro-ph.GA},
       adsurl = {https://ui.adsabs.harvard.edu/abs/2019NatAs...3..932G}
}

@ARTICLE{Grand2015,
       author = {{Grand}, Robert J.~J. and {Kawata}, Daisuke and {Cropper}, Mark},
        title = "{Impact of radial migration on stellar and gas radial metallicity distribution}",
      journal = {\mnras},
         year = 2015,
        month = mar,
       volume = {447},
       number = {4},
        pages = {4018-4027},
          doi = {10.1093/mnras/stv016},
archivePrefix = {arXiv},
       eprint = {1410.3836},
 primaryClass = {astro-ph.GA},
       adsurl = {https://ui.adsabs.harvard.edu/abs/2015MNRAS.447.4018G}
}

@ARTICLE{Ciuca2024,
       author = {{Ciuc{\u{a}}}, Ioana and {Kawata}, Daisuke and {Ting}, Yuan-Sen and {Grand}, Robert J.~J. and {Miglio}, Andrea and {Hayden}, Michael and {Baba}, Junichi and {Fragkoudi}, Francesca and {Monty}, Stephanie and {Buder}, Sven and {Freeman}, Ken},
        title = "{Chasing the impact of the Gaia-Sausage-Enceladus merger on the formation of the Milky Way thick disc}",
      journal = {\mnras},
         year = 2024,
        month = feb,
       volume = {528},
       number = {1},
        pages = {L122-L126},
          doi = {10.1093/mnrasl/slad033},
archivePrefix = {arXiv},
       eprint = {2211.01006},
 primaryClass = {astro-ph.GA},
       adsurl = {https://ui.adsabs.harvard.edu/abs/2024MNRAS.528L.122C}
}

@article{numpyro,

  title={Composable Effects for Flexible and Accelerated Probabilistic Programming in NumPyro},

  author={Phan, Du and Pradhan, Neeraj and Jankowiak, Martin},

  journal={arXiv preprint arXiv:1912.11554},

  year={2019}

}

@article{numpyro2,

  author    = {Eli Bingham and

               Jonathan P. Chen and

               Martin Jankowiak and

               Fritz Obermeyer and

               Neeraj Pradhan and

               Theofanis Karaletsos and

               Rohit Singh and

               Paul A. Szerlip and

               Paul Horsfall and

               Noah D. Goodman},

  title     = {Pyro: Deep Universal Probabilistic Programming},

  journal   = {J. Mach. Learn. Res.},

  volume    = {20},

  pages     = {28:1--28:6},

  year      = {2019},

  url       = {http://jmlr.org/papers/v20/18-403.html}

}

@ARTICLE{SandersDas2018,
       author = {{Sanders}, Jason L. and {Das}, Payel},
        title = "{Isochrone ages for {\ensuremath{\sim}}3 million stars with the second Gaia data release}",
      journal = {\mnras},
         year = 2018,
        month = dec,
       volume = {481},
       number = {3},
        pages = {4093-4110},
          doi = {10.1093/mnras/sty2490},
archivePrefix = {arXiv},
       eprint = {1806.02324},
 primaryClass = {astro-ph.GA},
       adsurl = {https://ui.adsabs.harvard.edu/abs/2018MNRAS.481.4093S}
}

@ARTICLE{Brook2004,
       author = {{Brook}, Chris B. and {Kawata}, Daisuke and {Gibson}, Brad K. and {Freeman}, Ken C.},
        title = "{The Emergence of the Thick Disk in a Cold Dark Matter Universe}",
      journal = {\apj},
         year = 2004,
        month = sep,
       volume = {612},
       number = {2},
        pages = {894-899},
          doi = {10.1086/422709},
archivePrefix = {arXiv},
       eprint = {astro-ph/0405306},
 primaryClass = {astro-ph},
       adsurl = {https://ui.adsabs.harvard.edu/abs/2004ApJ...612..894B}
}

@ARTICLE{Grand2012,
       author = {{Grand}, Robert J.~J. and {Kawata}, Daisuke and {Cropper}, Mark},
        title = "{Dynamics of stars around spiral arms in an N-body/SPH simulated barred spiral galaxy}",
      journal = {\mnras},
         year = 2012,
        month = oct,
       volume = {426},
       number = {1},
        pages = {167-180},
          doi = {10.1111/j.1365-2966.2012.21733.x},
archivePrefix = {arXiv},
       eprint = {1202.6387},
 primaryClass = {astro-ph.GA},
       adsurl = {https://ui.adsabs.harvard.edu/abs/2012MNRAS.426..167G}
}

@ARTICLE{Hunt2025,
       author = {{Hunt}, Jason A.~S. and {Vasiliev}, Eugene},
        title = "{Milky Way dynamics in light of Gaia}",
      journal = {\nar},
         year = 2025,
        month = jun,
       volume = {100},
          eid = {101721},
        pages = {101721},
          doi = {10.1016/j.newar.2024.101721},
archivePrefix = {arXiv},
       eprint = {2501.04075},
 primaryClass = {astro-ph.GA},
       adsurl = {https://ui.adsabs.harvard.edu/abs/2025NewAR.10001721H}
}

@ARTICLE{Lane2025,
       author = {{Lane}, James M.~M. and {Bovy}, Jo},
        title = "{The orbital anisotropy profile of the Gaia-Sausage/Enceladus accretion remnant}",
      journal = {arXiv e-prints},
         year = 2025,
        month = sep,
          eid = {arXiv:2509.04557},
        pages = {arXiv:2509.04557},
          doi = {10.48550/arXiv.2509.04557},
archivePrefix = {arXiv},
       eprint = {2509.04557},
 primaryClass = {astro-ph.GA},
       adsurl = {https://ui.adsabs.harvard.edu/abs/2025arXiv250904557L}
}

@ARTICLE{Genzel2011,
       author = {{Genzel}, R. and {Newman}, S. and {Jones}, T. and {F{\"o}rster Schreiber}, N.~M. and {Shapiro}, K. and {Genel}, S. and {Lilly}, S.~J. and {Renzini}, A. and {Tacconi}, L.~J. and {Bouch{\'e}}, N. and {Burkert}, A. and {Cresci}, G. and {Buschkamp}, P. and {Carollo}, C.~M. and {Ceverino}, D. and {Davies}, R. and {Dekel}, A. and {Eisenhauer}, F. and {Hicks}, E. and {Kurk}, J. and {Lutz}, D. and {Mancini}, C. and {Naab}, T. and {Peng}, Y. and {Sternberg}, A. and {Vergani}, D. and {Zamorani}, G.},
        title = "{The Sins Survey of z \raisebox{-0.5ex}\textasciitilde 2 Galaxy Kinematics: Properties of the Giant Star-forming Clumps}",
      journal = {\apj},
         year = 2011,
        month = jun,
       volume = {733},
       number = {2},
          eid = {101},
        pages = {101},
          doi = {10.1088/0004-637X/733/2/101},
archivePrefix = {arXiv},
       eprint = {1011.5360},
 primaryClass = {astro-ph.CO},
       adsurl = {https://ui.adsabs.harvard.edu/abs/2011ApJ...733..101G}
}

@ARTICLE{Wisnioski2015,
       author = {{Wisnioski}, E. and {F{\"o}rster Schreiber}, N.~M. and {Wuyts}, S. and {Wuyts}, E. and {Bandara}, K. and {Wilman}, D. and {Genzel}, R. and {Bender}, R. and {Davies}, R. and {Fossati}, M. and {Lang}, P. and {Mendel}, J.~T. and {Beifiori}, A. and {Brammer}, G. and {Chan}, J. and {Fabricius}, M. and {Fudamoto}, Y. and {Kulkarni}, S. and {Kurk}, J. and {Lutz}, D. and {Nelson}, E.~J. and {Momcheva}, I. and {Rosario}, D. and {Saglia}, R. and {Seitz}, S. and {Tacconi}, L.~J. and {van Dokkum}, P.~G.},
        title = "{The KMOS$^{3D}$ Survey: Design, First Results, and the Evolution of Galaxy Kinematics from 0.7 <= z <= 2.7}",
      journal = {\apj},
         year = 2015,
        month = feb,
       volume = {799},
       number = {2},
          eid = {209},
        pages = {209},
          doi = {10.1088/0004-637X/799/2/209},
archivePrefix = {arXiv},
       eprint = {1409.6791},
 primaryClass = {astro-ph.GA},
       adsurl = {https://ui.adsabs.harvard.edu/abs/2015ApJ...799..209W}
}

@ARTICLE{Yu2021,
       author = {{Yu}, Sijie and {Bullock}, James S. and {Klein}, Courtney and {Stern}, Jonathan and {Wetzel}, Andrew and {Ma}, Xiangcheng and {Moreno}, Jorge and {Hafen}, Zachary and {Gurvich}, Alexander B. and {Hopkins}, Philip F. and {Kere{\v{s}}}, Du{\v{s}}an and {Faucher-Gigu{\`e}re}, Claude-Andr{\'e} and {Feldmann}, Robert and {Quataert}, Eliot},
        title = "{The bursty origin of the Milky Way thick disc}",
      journal = {\mnras},
         year = 2021,
        month = jul,
       volume = {505},
       number = {1},
        pages = {889-902},
          doi = {10.1093/mnras/stab1339},
archivePrefix = {arXiv},
       eprint = {2103.03888},
 primaryClass = {astro-ph.GA},
       adsurl = {https://ui.adsabs.harvard.edu/abs/2021MNRAS.505..889Y}
}

@ARTICLE{Nelson2019_TNG50,
       author = {{Nelson}, Dylan and {Springel}, Volker and {Pillepich}, Annalisa and {Rodriguez-Gomez}, Vicente and {Torrey}, Paul and {Genel}, Shy and {Vogelsberger}, Mark and {Pakmor}, Ruediger and {Marinacci}, Federico and {Weinberger}, Rainer and {Kelley}, Luke and {Lovell}, Mark and {Diemer}, Benedikt and {Hernquist}, Lars},
        title = "{The IllustrisTNG simulations: public data release}",
      journal = {Computational Astrophysics and Cosmology},
         year = 2019,
        month = may,
       volume = {6},
       number = {1},
          eid = {2},
        pages = {2},
          doi = {10.1186/s40668-019-0028-x},
archivePrefix = {arXiv},
       eprint = {1812.05609},
 primaryClass = {astro-ph.GA},
       adsurl = {https://ui.adsabs.harvard.edu/abs/2019ComAC...6....2N}
}

@ARTICLE{El-Badry2016,
       author = {{El-Badry}, Kareem and {Wetzel}, Andrew and {Geha}, Marla and {Hopkins}, Philip F. and {Kere{\v{s}}}, Dusan and {Chan}, T.~K. and {Faucher-Gigu{\`e}re}, Claude-Andr{\'e}},
        title = "{Breathing FIRE: How Stellar Feedback Drives Radial Migration, Rapid Size Fluctuations, and Population Gradients in Low-mass Galaxies}",
      journal = {\apj},
         year = 2016,
        month = apr,
       volume = {820},
       number = {2},
          eid = {131},
        pages = {131},
          doi = {10.3847/0004-637X/820/2/131},
archivePrefix = {arXiv},
       eprint = {1512.01235},
 primaryClass = {astro-ph.GA},
       adsurl = {https://ui.adsabs.harvard.edu/abs/2016ApJ...820..131E}
}

@ARTICLE{Binney2024,
       author = {{Binney}, James and {Vasiliev}, Eugene},
        title = "{Chemodynamical models of our Galaxy}",
      journal = {\mnras},
         year = 2024,
        month = jan,
       volume = {527},
       number = {2},
        pages = {1915-1934},
          doi = {10.1093/mnras/stad3312},
archivePrefix = {arXiv},
       eprint = {2306.11602},
 primaryClass = {astro-ph.GA},
       adsurl = {https://ui.adsabs.harvard.edu/abs/2024MNRAS.527.1915B}
}




\appendix

\section{Test on TNG50}\label{appendix:TNG50}

\subsection{Model verification with TNG50}
\begin{figure*}
    \centering
    \includegraphics[width=0.98\textwidth]{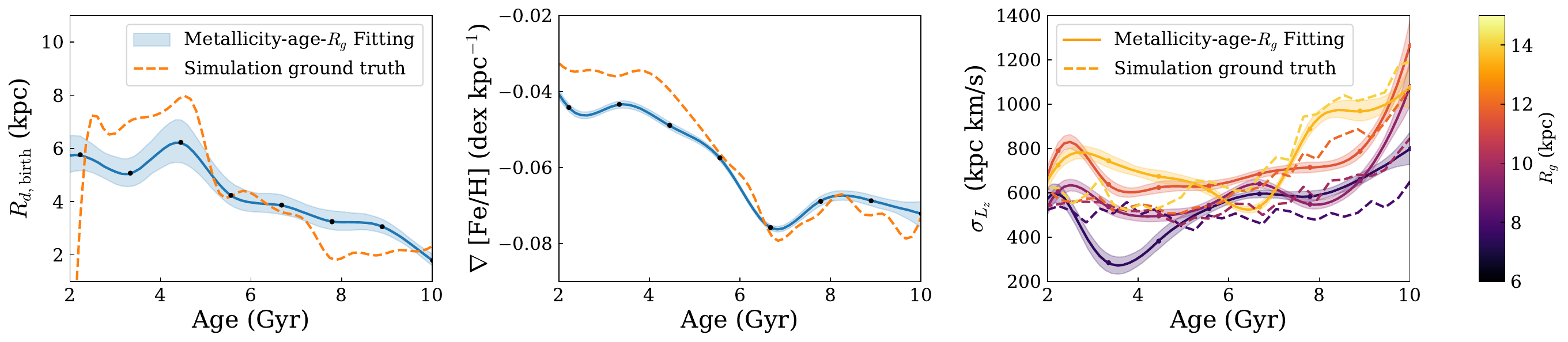}
    \caption{The chemo-chrono-dynamical fitting results of an example galaxy in TNG50 (id=547293). The left and middle panels show the fitted initial scale length and birth metallicity gradient, for which the ground truth is shown as the orange dashed lines, and the fitting results are shown as the blue solid line. The right panel shows the angular momentum diffusion coefficient for stars in different present-day guiding radii bins. The ground truth and fitting are shown as the dashed and solid lines with different colours corresponding to various guiding radius bins.}
    \label{fig:TNG50_fitting}
\end{figure*}

\begin{figure}
    \centering
    \includegraphics[width=0.98\columnwidth]{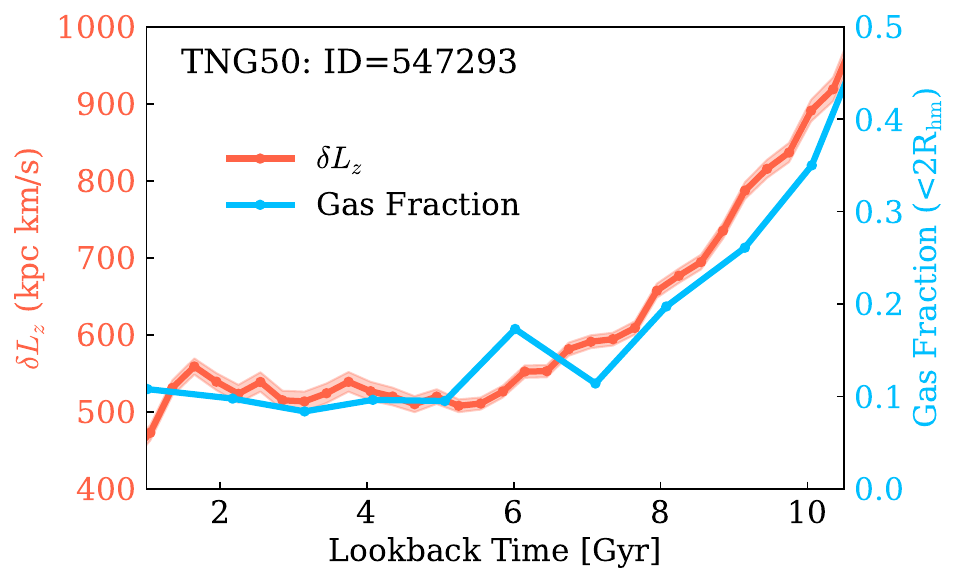}
    \includegraphics[width=0.98\columnwidth]{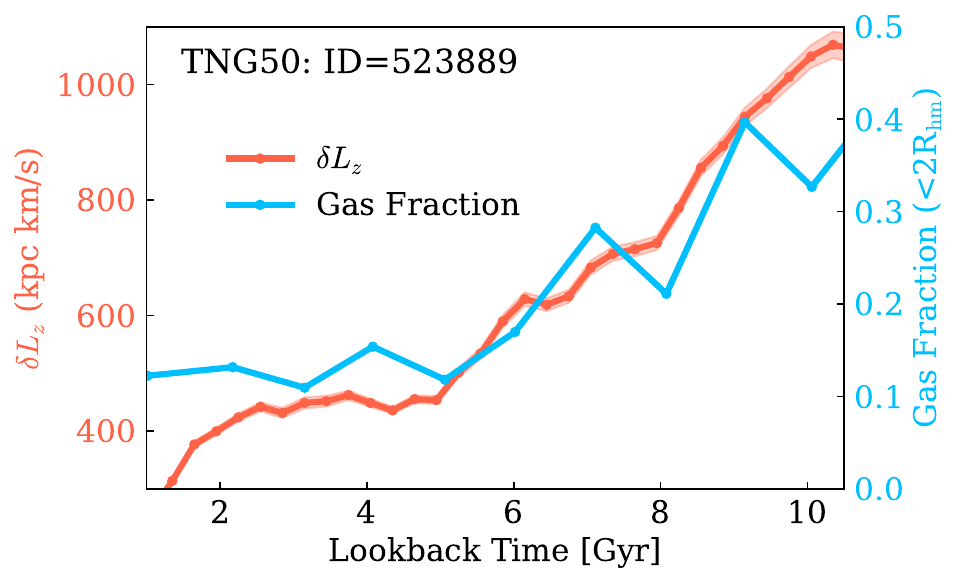}
    \caption{A demonstration of the correlation between the evolution of the gas fraction in a galaxy and the angular momentum diffusion strength. The RMS of variation of angular momentum is shown by the red lines, and the time evolution of the gas fraction is shown by the blue solid line. We show this for two example galaxies in TNG50 with id=547293 and id=523889 in the top and bottom panels, respectively. The angular momentum variation and gas fraction have a similar time dependence.}
    \label{fig:TNG50_gas_transition}
\end{figure}

To examine the chemo-chrono-dynamical model we built in this work, we apply the modelling procedure described in Section~\ref{section:model} to a galaxy with id=547293 in the TNG50 cosmological simulation suite \citep{Nelson2019_TNG50}. We select the disc star particles in this simulation with a simple age cut to select stars formed after the spin-up of the disc ($\rm age\approx10~Gyr$) in this galaxy \citep{Belokurov2022}, which has been demonstrated as a satisfactory selection of disc particles by \citet{Zhang2024}. We also remove stars with $\rm age<2~Gyr$. Our model cannot apply to the stellar population formed after $t_{\rm lb}\approx2~{\rm Gyr}$ because the AGN feedback creates a central hole in the star-forming gaseous disc, which deviates the radial profile of the star-forming disc far from the exponential profile, as assumed in our model. For stars with ages between 2 and 10 Gyr, we scatter the age and metallicity with uncertainties of $\rm \Sigma_{\rm log_{10}\, \tau}=0.02$, and $\Sigma_{[\rm Fe/H]}=0.1~\rm dex$. 

We implement the chemo-chrono-dynamical model described in Section~\ref{section:model} to the age-metallicity-$R_g$ distribution of the disc stellar particles in this simulated galaxy. The best-fit initial disc scale length and metallicity gradient are shown in the blue solid lines in the left and middle panels of Fig.~\ref{fig:TNG50_fitting}, and the best-fit angular momentum diffusion coefficient in the coloured solid lines in the right panel of Fig.~\ref{fig:TNG50_fitting}. 
The ground truth values of these three parameters are calculated using the recorded position and velocity distribution of these stars at birth and at the present day. The initial disc scale lengths are computed by fitting exponential profiles of the birth radii of the stars, binned according to their ages. Similarly, the initial radial metallicity gradients at different lookback times are obtained by fitting the linear model to the $[{\rm Fe/H}]\, {\rm v.s.}\, R_b$ sequence of stars with different ages. The ground truth angular momentum diffusion coefficients are computed as the RMS of the variation of the guiding radii of stars with various ages that are sitting at different guiding radii today. The ground truth values of these three parameters are shown in the dashed lines in Fig.~\ref{fig:TNG50_fitting}. 

Comparing the best-fit model as shown by the solid lines with the shaded regions with the ground truth as shown by the dashed lines, we find that our model recovers the initial disc scalelength, birth metallicity gradient, and angular momentum diffusion in this galaxy with good precision and satisfactory accuracy, especially for the initial metallicity gradient.

\subsection{Changing of radial migration strength as a consequence of gas-rich to gas-poor disc transition}

For the observed transition of angular momentum diffusion around $9.5~\rm Gyr$ (left panel in Fig.~\ref{fig:fitting_actions}), one of our proposed interpretations is a change in the gas fraction of the galaxy. We examined this by analysing the Nexus hydrodynamical simulation suite \citep{TG24} and TNG50 cosmological suite \citep{Nelson2019_TNG50}. We discuss the Nexus simulation results intensively in Appendix~\ref{appendix:Nexus}, whereas in this section, we discuss the results from the TNG50 simulation suite.

We select Milky Way-mass galaxies in TNG50 with an early forming disc ($t_{\rm disc~formation}\gtrsim10~\rm Gyr$). We calculate the gas mass fraction of these Milky Way-analogue galaxies by taking the ratio of the gas mass to total baryon mass within two times the half-mass radius of the disc, $R<2R_{hm}$. To compare the evolution of gas fraction with the radial migration, we again compute the RMS of the changes in the angular momentum of disc stars that are outside the bar/bulge regions. In Fig.~\ref{fig:TNG50_gas_transition}, we show the results of two galaxies with id=547293 and id=523889, in which the RMS angular momentum variation is shown in red lines with the axis on the left, and the gas fraction is shown in the blue lines with the axis on the right. We find that the evolution of gas fraction in these galaxies shares a similar trend to the strength of angular momentum variation, or radial migration. This could be because of the high turbulence of gas when the gas fraction is high, which can enhance radial migration in the stellar disc \citep{El-Badry2016}. Confirmed with TNG50 simulations, a transition of the gas fraction in the Milky Way could be a valid explanation for the transition of the angular momentum diffusion coefficient around the low-$\alpha$ disc formation time (see left panel of Fig.~\ref{fig:fitting_actions}).  


\section{Impact of gas on the secular evolution of the Milky Way}\label{appendix:Nexus}

\begin{figure}
    \centering
    \includegraphics[width=0.98\columnwidth]{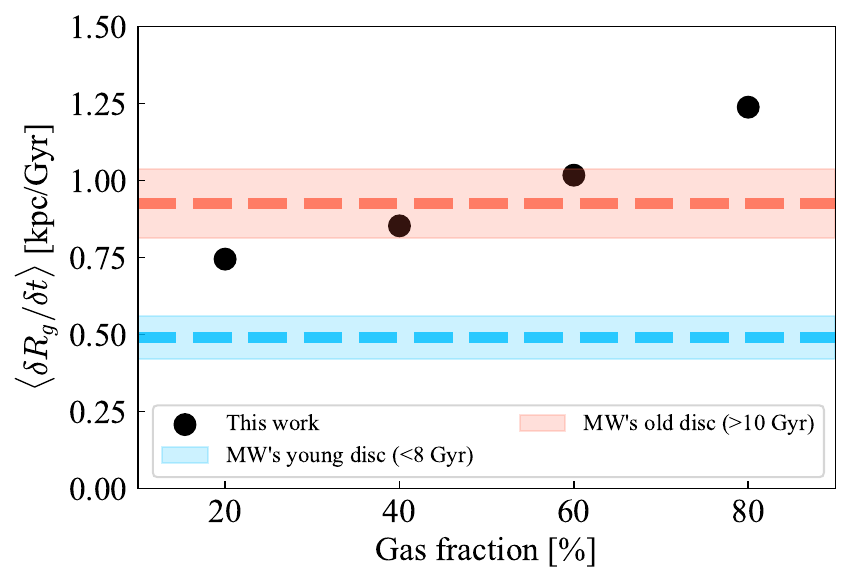}
    \caption{A figure adopted from Zhang et al. (in prep.), which illustrates the guiding radii variation of four galaxies with different gas fractions in the Nexus simulation suite. The averaged guiding radii variation rates of these four galaxies, as shown by the black dots, increase monotonically with increasing gas fraction. The blue and red lines show the same quantity but for the Milky Way from the fitting results we presented in Section~\ref{section:results}. We emphasise that the figure is for illustrative purposes, and the values should not be directly compared between isolated galaxy simulations and the Milky Way.}
    \label{fig:Nexus}
\end{figure}

To rigorously examine the impact of the gas fraction of the galaxy on the strength of radial migration of a galaxy, we analyse the galaxies in the \textsc{Nexus} simulations suite, a controlled hydrodynamical simulation. The details are presented in a separate paper (Zhang et al. in prep.), which kindly refers the audience to. Here, we quickly summarise the results that are relevant to the Section~\ref{section:implication} of this paper. 

We adopt four galaxies in the \textsc{Nexus} simulations suite, which all have total DM halo mass of $\approx 10^{11} M_\odot$, total (stellar+gas) disc mas of $\approx10^{10} M_\odot$. The initial gas fraction of these four galaxies are $f_{\rm gas}=\{20\%,~40\%,~60\%,~80\% \}$, respectively. Each galaxy model was run for $\approx 2~\rm Gyr$. We compute the cylindrical orbital actions of stellar particles in these galaxy models in all snapshots with a time cadence of $\approx 50~\rm Myr$ using \textsc{AGAMA} \citep{Vasiliev2019}, and we trace the variation of the orbital actions by calculating the RMS of changes in the three actions to study the scattering of stellar orbits. As these galaxy models share the same initial disc and halo masses, their stellar initial conditions are similar to each other, which provides an ideal controlled experiment laboratory. Comparing the RMS of changes of orbital actions in these four galaxies as a function of gas fraction, we can understand the impact of different gas fractions on the secular evolution of a galaxy. 

To summarise our main results, we find that the orbits of stars vary more significantly with increasing gas fraction. To ensure a better comparison with the observables in the Milky Way, we also compute the RMS of angular momentum variation, $\delta L_z$, and heating-to-migration ratio, $\delta J_{R,z} / \delta L_z$. As a linear approximation, we compute the mean gradient of guiding radii variation, $\langle \delta R_g/\delta t\rangle$, in each galaxy, for which we show them as the black solid dots in Fig.~\ref{fig:Nexus}. The RMS of angular momentum variation is systematically larger in the more gas-rich galaxies outside the bar region or $R>R_{\rm hm}$. In the same figure, we show $\langle \delta R_g/\delta t\rangle$ of the Milky Way's low-$\alpha$, young disc ($\rm age<8~Gyr$) and high-$\alpha$, old disc ($\rm age>10~Gyr$) from our modelling results in Fig.~\ref{fig:fitting_actions} with the blue and red dashed lines, respectively. It is noteworthy that the values of $\langle \delta R_g/\delta t\rangle$ in Fig.~\ref{fig:fitting_actions} should not be interpreted numerically, but the purpose of this result is to illustrate that the transition of the migration strength around the low-$\alpha$ disc formation time is compatible with a transition in the gas fraction in the Galactic disc. 
The heating-to-migration ratios in these galaxies also demonstrate a monotonic increasing trend with the increasing gas fraction. For the most gas-poor analogue of these galaxies ($f_{\rm gas}=20\%$), $\delta J_R / \delta L_z \approx 0.07$, and $\delta J_z / \delta L_z \approx 0.01$, while $\delta J_R / \delta L_z \approx 0.25$, and $\delta J_z / \delta L_z \approx 0.06$ for the most gas-rich analogue $f_{\rm gas}=80\%$. Both radial and vertical heating-to-migration ratio of the Milky Way's disc are compatible with the $f_{\rm gas}=20\%$ galaxy model, which suggests that the kinematics of the Milky Way's disc is compatible with a gas-poor disc.

\section{Extra figures}\label{appendix:extra_figure}

\begin{figure*}
    \centering
    \includegraphics[width=0.96\textwidth]{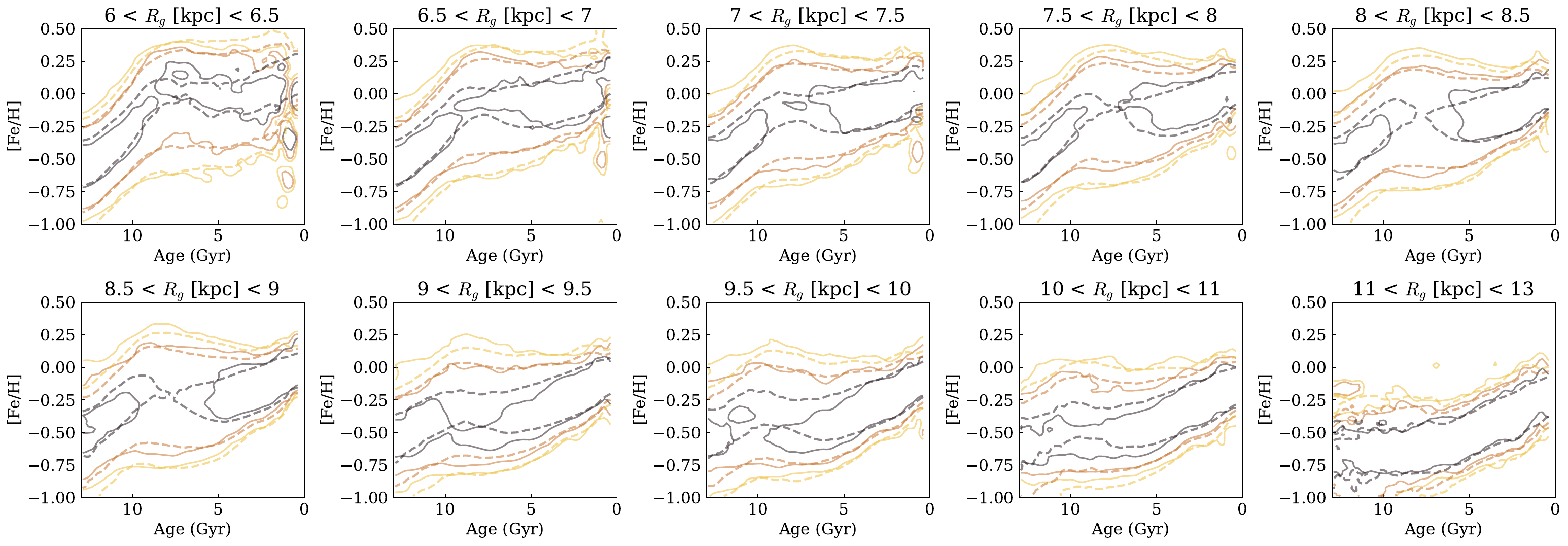}
    \caption{Model to data comparison of the best-fit model in Section~\ref{section:results} to the observation in the column-normalised age v.s. metallicity plane of various guiding radii. The model contour is shown by the dashed lines, and the data contour is shown by the solid line. }
    \label{fig:extra_AM_compare}
\end{figure*}

\begin{figure*}
    \centering
    \includegraphics[width=0.96\textwidth]{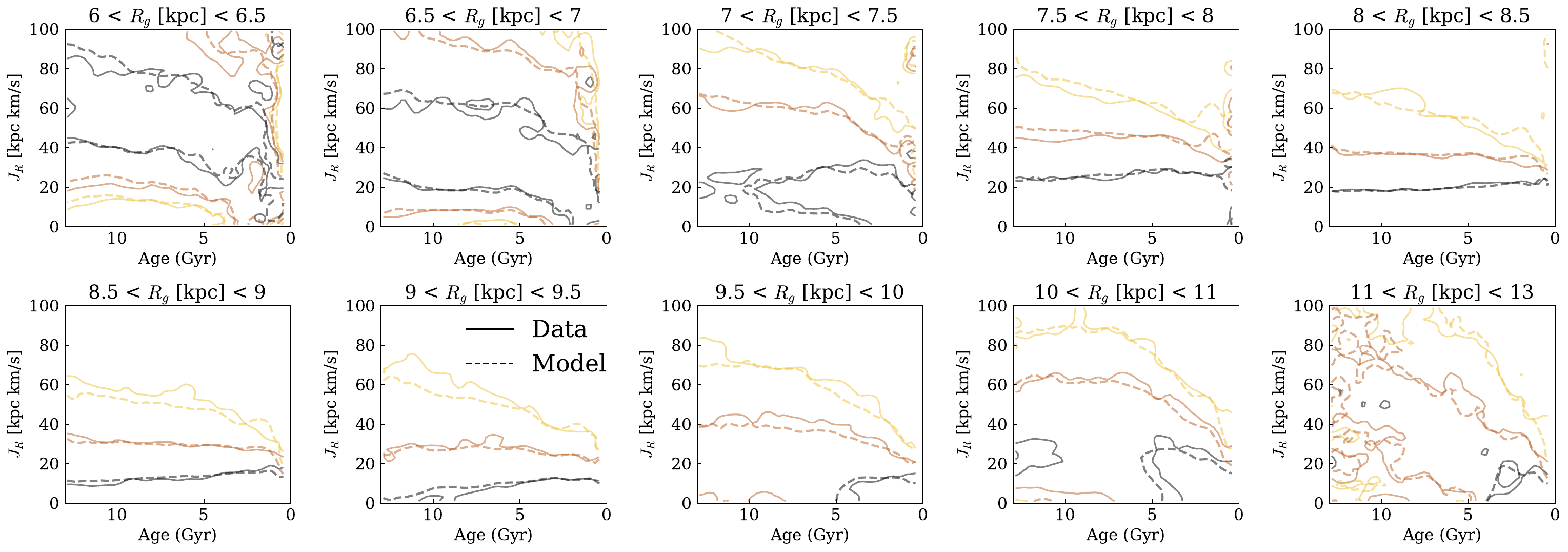}
    \caption{Same as Fig.~\ref{fig:extra_AM_compare} but for the column-normalised age v.s. $J_R$ plane.}
    \label{fig:extra_AJR_compare}
\end{figure*}

\begin{figure*}
    \centering
    \includegraphics[width=0.96\textwidth]{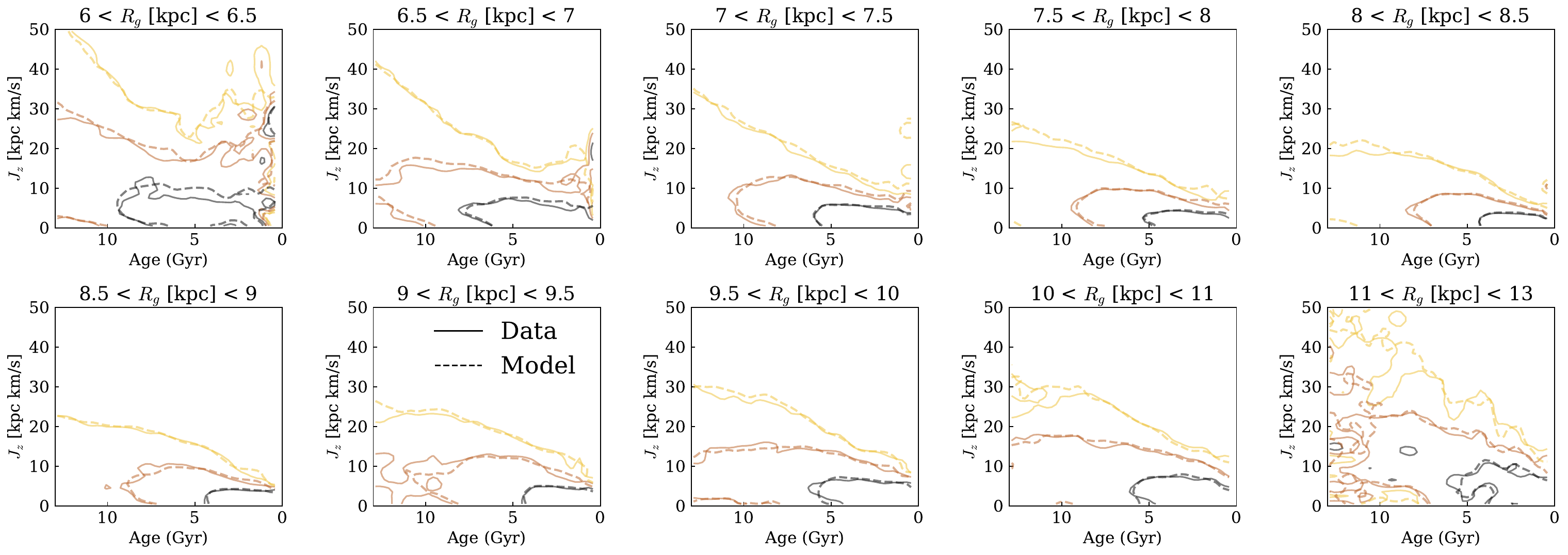}
    \caption{Same as Fig.~\ref{fig:extra_AM_compare} but for the column-normalised age v.s. $J_z$ plane.}
    \label{fig:extra_AJz_compare}
\end{figure*}

We demonstrate the goodness-of-fit in this section. We compare the best-fit chemo-chrono-dynamical model of the Milky Way's disc with the observed data. As the fitting is performed for each guiding radius bin, we show the column-normalised distributions in the age v.s. metallicity, age v.s. $J_R$, and age v.s. $J_z$ in Fig.~\ref{fig:extra_AM_compare},~\ref{fig:extra_AJR_compare},~and~\ref{fig:extra_AJz_compare}, in which the contour of data is shown by the solid lines, and the model contour is shown by the dashed lines. The model agrees well with the data, which reinforces the validity of our fitting.

\bsp	
\label{lastpage}
\end{document}